\documentclass[12pt,a4paper,twoside]{article}
\usepackage{latexsym,amsmath,amssymb,amsfonts,amsthm}
\usepackage[mathscr]{eucal}
\usepackage{makeidx}
\def\nabla{\bigtriangledown}
\newcommand{ \R} {\mbox{\rm I$\!$R}}
\newcommand{ \C} {\mbox{\rm I$\!$C}}
\newcommand{ \Z} {\mbox{\rm I$\!$Z}}
\makeindex
\begin{document}

\title{(Non) Commutative Finsler Geometry\\
from String/M--theory}
\author{Sergiu I. Vacaru \thanks{%
E-mail address:\ vacaru@fisica.ist.utl.pt, ~~
sergiu$_{-}$vacaru@yahoo.com,\
} \\
{\small \textit{Centro Multidisciplinar de Astrofisica - CENTRA,
Departamento de Fisica,}}\\
{\small \textit{Instituto Superior Tecnico, Av. Rovisco Pais 1,
Lisboa,
1049-001, Portugal}}\\
}
\date{November 8, 2002}
\maketitle

\begin{abstract}
We synthesize and extend the previous ideas about appearance of both
noncommutative and Finsler geometry in string theory with nonvanishing
B--field and/or anholonomic (super) frame structures \cite%
{vstring,vstr2,vnonc,vncf}. There are investigated the limits to the
Einstein gravity and string generalizations containing locally anisotropic
structures modeled by moving frames. The relation of anholonomic frames and
nonlinear connection geometry to M--theory and possible noncommutative
versions of locally anisotropic supergravity and D--brane physics is
discussed. We construct and analyze new classes of exact solutions with
noncommutative local anisotropy describing anholonomically deformed black
holes (black ellipsoids) in string gravity, embedded Finsler--string two
dimensional structures, solitonically moving black holes in extra dimensions
and wormholes with noncommutativity and anisotropy induced from string
theory.

\vskip5pt.

Pacs 11.10.Nx,\ 11.25.Uv,\ 02.40.Bw

\end{abstract}

\tableofcontents

\section{Introduction}

The idea that string/M--theory results in a noncommutative limit of field
theory and spacetime geometry is widely investigated by many authors both
from mathematical and physical perspectives \cite{strncg,sw,connes1} (see,
for instance, the reviews \cite{dn}). It is now generally accepted that
noncommutative geometry and quantum groups \cite{nc,qg,majid} play a
fundamental role in further developments of high energy particle physics and
gravity theory.

First of all we would like to give an exposition of some basic facts about
the geometry of anholonomic frames (vielbeins) and associated nonlinear
connection (N--connection) structures which emphasize surprisingly some new
results: We will consider N--connections in commutative geometry and we will
show that locally anisotropic spacetimes (anholonomic Riemannian, Finsler
like and their generalizations) can be obtained from the string/M--theory.
We shall discuss the related low energy limits to Einstein and gauge
gravity. Our second goal is to extend A. Connes' differential noncommutative
geometry as to include geometries with anholonomic frames and N--connections
and to prove that such 'noncommutative anisotropies' also arise very
naturally in the framework of strings and extra dimension gravity. We will
show that the anholonomic frame method is very useful in investigating of
new symmetries and nonperturbative states and for constructing new exact
solutions in string gravity with anholonomic and/or noncommutative
variables. We remember here that some variables are considered anholonomic
(equivalently, nonholonomic) if they are subjected to some constraints
(equivalently, anholonomy conditions).

Almost all of the physics paper dealing with the notion of (super) frame in
string theory do not use the well developed apparatus of E. Cartan's 'moving
frame' method \cite{cartan} which gave an unified approach to the Riemannian
and Finsler geometry, to bundle spaces and spinors, to the geometric theory
of systems of partial equations and to Einstein (and the so--called
Einstein--Cartan--Weyl) gravity. It is considered that very ''sophisticate''
geometries like the Finsler and Cartan ones, and theirs generalizations, are
less related to real physical theories. In particular, the bulk of frame
constructions in string and gravity theories are given by coefficients
defined with respect to coordinate frames or in abstract form with respect
to some general vielbein bases. It is completely disregarded the fact that
via anholonomic frames on (pseudo) Riemannian manifolds and on (co) tangent
and (co) vector bundles we can model different geometries and interactions
with local anisotropy even in the framework of generally accepted classical
and quantum theories. For instance, there were constructed a number of exact
solutions in general relativity and its lower/higher dimension extensions
with generic local anisotropy, which under certain conditions define Finsler
like geometries \cite{vexsol,vankin,vbel,vsingl,vsingl1,vsolsp}. It was
demonstrated that anholonomic geometric constructions are inevitable in the
theory of anisotropic stochastic, kinetic and thermodynamic processes in
curved spacetimes \cite{vankin} and proved that Finsler like (super)
geometries are contained alternatively in modern string theory \cite%
{vstring,vstr2,vmon1}.

We emphasize that we have not proposed any ''exotic'' locally
anisotropic modifications of string theory and general relativity
but demonstrated that such anisotropic structures, Finsler like
or another type ones, may appear alternatively to the Riemannian
geometry, or even can be modeled in the framework of a such
geometry, in the low energy limit of the string theory, if we are
dealing with frame (vielbein) constructions. One of our main goals
is to give an accessible exposition of some important notions and
results of N--connection geometry and to show how they can be
applied to concrete problems in string theory, noncommutative
geometry and gravity. We hope to convince a reader--physicist,
who knows that 'the B--field' in string theory may result in
noncommutative geometry, that the anholonomic (super) frames
could define nonlinear connections and Finsler like commutative
and/ or noncommutative geometries in string theory and (super)
gravity and this holds true in certain limits to general
relativity.

We address the present work to physicists who would like to learn about some
new geometrical methods and to apply them to mathematical problems arising
at the forefront of modern theoretical physics. We do not assume that such
readers have very deep knowledge in differential geometry and nonlinear
connection formalism (for convenience, we give an Appendix outlining the
basic results on the geometry of commutative spaces provided with
N--connection structures \cite{ma,vmon1,vmon2}) but consider that they are
familiar with some more geometric approaches to gravity \cite{haw,mtw} and
string theories \cite{deligne}.

Finally, we note that the first attempts to relate Riemann--Finsler spaces
(and spaces with anisotropy of another type) to noncommutative geometry and
physics were made in Refs. \cite{vnonc} where some models of noncommutative
gauge gravity (in the commutative limit being equivalent to the Einstein
gravity, or to different generalizations to de Sitter, affine, or Poincare
gauge gravity with, or not, nonlinear realization of the gauge groups) were
analyzed. \ Further developments of noncommutative geometries with
anholonomic/ anisotropic structures and their applications in modern
particle physics lead to a rigorous \ study of the geometry of
noncommutative anholonomic frames with associated N--connection structure %
\cite{vncf} (that work should be considered as the non--string partner of
the present paper).

The paper has the following structure:

In Section 2 we consider stings in general manifolds and bundles provided
with anholonomic frames and associated nonlinear connection structures and
analyze the low energy string anholonomic field equations. \ The conditions
when anholonomic Einstein or Finsler like gravity models can be derived from
string theory are stated.

Section 3 outlines the geometry of locally anisotropic supergravity models
contained in superstring theory. Superstring effective actions and
anisotropic toroidal compactifications are analyzed. The corresponding
anholonomic field equations with distinguishing of anholonomic
Riemannian--Finesler (super) gravities are derived.

In Section 4 we formulate the theory of noncommutative anisotropic scalar
and gauge fields interactions and examine their anholonomic symmetries.

In Section 5 we emphasize how noncommutative anisotropic structures are
embedded in string/M--theory and discuss their connection to anholonomic
geometry.

Section 6 is devoted to locally anisotropic gravity models generated on
noncommutative D--branes.

In Section 7 we construct four classes of exact solutions with
noncommutative and locally anisotropic structures. We analyze solutions
describing locally anisotropic black holes in string theory, define a class
of Finsler--string structures containing two dimensional Finsler metrics,
consider moving solitonic string--black hole configurations and give an
examples of anholonomic noncommutative wormhole solution induced from string
theory.

Finally, in Section 8, some additional comments and questions for further
developments are presented. The Appendix outlines the necessary results from
the geometry of nonlinear connections and generalized Finsler--Riemannian
spaces.

\section{String Theory and Commutative Riemann--Fins\-ler Gravity}

The string gravitational effects are computed from corresponding low--energy
effective actions and moving equations of stings in curved spacetimes (on
string theory, see monographs \cite{deligne}). The basic idea is to consider
propagation of a string not only of a flat 26--dimensional space with
Minkowski metric $\eta _{\mu \nu }$ but also its propagation in a background
more general manifold with metric tensor $g_{\mu \nu }$ from where one
derived string--theoretic corrections to general relativity when the vacuum
Einstein\ equations $R_{\mu \nu }=0$ correspond to vanishing of the
one--loop beta function in corresponding sigma model. More rigorous theories
were formulated by adding an antisymmetric tensor field $B_{\mu \nu },$ the
dilaton field $\Phi $ and possible other background fields, by introducing
supersymmetry, higher loop corrections and another generalizations. It
should be noted here that propagation of (super) strings may be considered
on arbitrary (super) manifolds. For instance, in Refs. \cite%
{vstring,vstr2,vmon1}, the corresponding background (super) spaces were
treated as (super) bundles provided with nonlinear connection
(N--connection) structure and, in result, there were constructed some types
of generalized (super) Finsler corrections to the usual Einstein and to
locally anisotropic (Finsler type, or theirs generalizations) gravity
theories.

The aim of this section is to demonstrate that anisotropic corrections and
extensions may be computed both in Einstein and string gravity [derived for
string propagation in usual (pseudo) Riemannian backgrounds] if the approach
is developed following a more rigorous geometrical formalism with
off--diagonal metrics and anholonomic frames. We note that (super) frames
[vielbeins] were used in general form, for example, in order to introduce
spinors and supersymmetry in sting theory but the anholonomic transforms
with mixed holonomic--anholonomic variables, resulting in diagonalization of
off--diagonal (super) metrics and effective anisotropic structures, were not
investigated in the previous literature on string/M--theory.

\subsection{Strings in general manifolds and bundles}

\subsubsection{Generalized nonlinear sigma models (some basics)}

The first quantized string theory was constructed in flat
Minkowski spacetime of dimension $k\geq 4.$ \ Then the analysis
was extended to more general manifolds with (pseudo) Riemannian
metric $\underline{g}_{\mu \nu },$ antisymmetric $B_{\mu \nu }$
and dilaton field $\Phi $ and possible other
background fields, including  tachyonic matter associated to a field $%
U $ in a tachyon state. The starting point in investigating the string
dynamics in the background of these fields is the generalized nonlinear
sigma model action for the maps $u:\Sigma \rightarrow M$ \ from a two
dimensional surface $\Sigma $ to a spacetime manifold $M$ of dimension $k,$%
\begin{equation}
S=S_{\underline{g},B}+S_{\Phi }+S_{U},  \label{act1}
\end{equation}%
with%
\begin{eqnarray*}
S_{\underline{g},B}[u,g] &=&\frac{1}{8\pi l^{2}}\int\limits_{\Sigma }d\mu
_{g}\partial _{A}u^{\mu }\partial _{B}u^{\nu }\left[ g_{[2]}^{AB}\underline{g%
}_{\mu \nu }(u)+\varepsilon ^{AB}B_{\mu \nu }(u)\right] , \\
S_{\Phi }[u,g] &=&\frac{1}{2\pi }\int\limits_{\Sigma }d\mu _{g}R_{g}\Phi
(u),~S_{U}[u,g]=\frac{1}{4\pi }\int\limits_{\Sigma }d\mu _{g}U(u),
\end{eqnarray*}%
where $B_{\mu \nu }$ is the pullback of a two--form $B=B_{\mu \nu }du^{\mu
}\wedge du^{\nu }$ under the map $u,\,$written out in local coordinates $%
u^{\mu };~g_{[2]AB}$ is the metric on the two dimensional surface $\Sigma $
(indices $A,B=0,1);~$ $\varepsilon ^{AB}=\overline{\varepsilon }^{AB}/\sqrt{%
\det |g_{AB}|},\overline{\varepsilon }^{01}=-\overline{\varepsilon }^{10}=1;$
the integration measure $d\mu _{g}$ is defined by the coefficients of the
metric $g_{AB},$ $R_{g}$ is the Gauss curvature of $\Sigma .$ The constants
in the action are related as%
\begin{equation*}
k=\frac{1}{4\pi \alpha ^{\prime }}=\frac{1}{8\pi \ell ^{2}},\alpha ^{\prime
}=2\ell ^{2}
\end{equation*}%
where $\alpha ^{\prime }$ is the Regge slope parameter $\alpha ^{\prime }$
and $\ell \sim 10^{-33}cm$ is the Planck length scale. The metric
coefficients $\underline{g}_{\mu \nu }(u)$ are defined by the quadratic
metric element given with respect to the coordinate co--basis $d^{\mu
}=du^{\mu }$ (being dual to the local coordinate basis $\partial _{\mu
}=\partial /\partial _{\mu }),$%
\begin{equation}
ds^{2}=\underline{g}_{\mu \nu }(u)du^{\mu }du^{\nu }.  \label{metric}
\end{equation}

The parameter $\ell $ is a very small length--scale, compared to
experimental scales $L_{\exp }\sim 10^{-17}$ accessible at present. This
defines the so--called low energy, or $\alpha ^{\prime }$--expansion. A
perturbation theory may be carried out as usual by letting $u=u_{0}+\ell
u_{[1]}$ for some reference configuration $u_{0}$ and considering expansions
of the fields $\underline{g},B$ and $\Phi ,$ for instance,
\begin{equation}
\underline{g}_{\mu \nu }(u)=\underline{g}_{\mu \nu }(u_{0})+\ell \partial
_{\alpha }\underline{g}_{\mu \nu }(u_{0})u_{[1]}^{\alpha }+\frac{1}{2}\ell
^{2}\partial _{\alpha }\partial _{\beta }\underline{g}_{\mu \nu
}(u_{0})u_{[1]}^{\alpha }u_{[1]}^{\beta }+...  \label{ser1}
\end{equation}%
This reveals that the quantum field theory defined by the action (\ref{act1}%
) is with an infinite number of couplings; the independent couplings of this
theory correspond to the successive derivatives of the fields $\underline{g}%
,B$ and $\Phi $ at the expansion point $u_{0}.$ Following an analysis of the
general structure of the Weyl dependence of Green functions in the quantum
field theory, standard regularizations schemes (see, for instance, Refs. %
\cite{deligne}) and conditions of vanishing of Weyl anomalies,
computing the $\beta $--functions, one derive the low energy
string effective actions and field equations.

\subsubsection{Anholonomic frame transforms of background metrics}

Extending the general relativity principle to the string theory, we should
consider that the string dynamics in the background of fields $\underline{g}%
,B$ and $\Phi $ and possible another ones, defined in the low energy limit
by certain effective actions and moving equations, does not depend on
changing of systems of coordinates, $u^{\alpha ^{\prime }}\rightarrow
u^{\alpha ^{\prime }}\left( u^{\alpha }\right) ,$ for a fixed local basis
(equivalently, system, frame, or vielbein) of reference, $e_{\alpha }\left(
u\right) ,$ on spacetime $M$ \ (for which, locally, $u=u^{\alpha }e_{\alpha
}=$ $u^{\alpha ^{\prime }}e_{\alpha ^{\prime }},$ $e_{\alpha ^{\prime
}}=\partial u^{\alpha }/\partial u^{\alpha ^{\prime }}e_{\alpha },$ usually
one considers local coordinate bases when $e_{\alpha }=\partial /\partial
u^{\alpha })$ as well the string dynamics should not depend on changings of
frames like $e_{\underline{\alpha }}\rightarrow e_{\underline{\alpha }%
}^{~\alpha }\left( u\right) e_{\alpha },$ parametrized by non--degenerated
matrices $e_{\underline{\alpha }}^{~\alpha }\left( u\right) .$

Let us remember some details connected with the geometry of moving frames in
(pseudo) Riemannian spaces \cite{cartan} and discuss its applications in
string theory, where the orthonormal frames were introduced with the aim to
eliminate non--trivial dependencies on the metric $\underline{g}_{\mu \nu }$
and on the background field $u_{0}^{\mu }$ which appears in elaboration of
the covariant background expansion method \ for the nonlinear sigma models %
\cite{deligne,friedan}. Such orthonormal frames, in the framework of a $%
SO\left( 1,k-1\right) $ like gauge theory are stated by the conditions%
\begin{eqnarray}
\underline{g}_{\mu \nu }\left( u\right) &=&e_{\mu }^{~\underline{\mu }%
}\left( u\right) e_{\nu }^{~\underline{\nu }}\left( u\right) \eta _{%
\underline{\mu }\underline{\nu }},  \label{tetrad} \\
e_{\mu }^{~\underline{\mu }}e_{\ \underline{\mu }}^{\nu } &=&\delta _{\mu
}^{\nu },\quad e_{\mu }^{~\underline{\mu }}e_{\ \underline{\nu }}^{\mu
}=\delta _{\underline{\nu }}^{\underline{\mu }},  \notag
\end{eqnarray}%
where $\eta _{\underline{\mu }\underline{\nu }}=diag\left(
-1,+1,...,+1\right) $ is the flat Minkowski metric \thinspace and $\delta
_{\mu }^{\nu },\delta _{\underline{\nu }}^{\underline{\mu }}$ are
Kronecker's delta symbols. One considers the covariant derivative $D_{\mu }$
with respect to an affine connection $\Gamma $ and a corresponding spin
connection $\omega _{\mu ~\underline{\beta }}^{~\underline{\alpha }}$ for
which the frame $e_{\mu }^{~\underline{\mu }}$ is covariantly constant,%
\begin{equation*}
D_{\mu }e_{\nu }^{~\underline{\alpha }}\equiv \partial _{\mu }e_{\nu }^{~%
\underline{\alpha }}-\Gamma _{~\mu \nu }^{\alpha }e_{\alpha }^{~\underline{%
\alpha }}+\omega _{\mu ~\underline{\beta }}^{~\underline{\alpha }}e_{\alpha
}^{~\underline{\beta }}=0.
\end{equation*}%
One also uses the covariant derivative
\begin{equation}
\mathcal{D}_{\mu }e_{\nu }^{~\underline{\alpha }}=D_{\mu }e_{\nu }^{~%
\underline{\alpha }}+\frac{1}{2}H_{\mu \nu }^{\quad \rho }e_{\rho }^{~%
\underline{\alpha }}  \label{cd}
\end{equation}%
including the torsion tensor $H_{\mu \nu \rho }$ which is the field strength
of the field $B_{\nu \rho },$ given by $H=dB,$ or, in component notation,
\begin{equation}
H_{\mu \nu \rho }\equiv \partial _{\mu }B_{\nu \rho }+\partial _{\nu
}B_{\rho \mu }+\partial _{\nu }B_{\rho \mu }.  \label{str1}
\end{equation}%
All tensors may be written with respect to an orthonormal frame basis, for
instance,
\begin{equation*}
H_{\mu \nu \rho }=e_{\mu }^{~\underline{\mu }}e_{\nu }^{~\underline{\nu }%
}e_{\rho }^{~\underline{\rho }}H_{\underline{\mu }\underline{\nu }\underline{%
\rho }}
\end{equation*}%
and
\begin{equation*}
\mathcal{R}_{\mu \nu \rho \sigma }=e_{\mu }^{~\underline{\mu }}e_{\nu }^{~%
\underline{\nu }}e_{\rho }^{~\underline{\rho }}e_{\sigma }^{~\underline{%
\sigma }}R_{\underline{\mu }\underline{\nu }\underline{\rho }\underline{%
\sigma }},
\end{equation*}%
where the curvature $\mathcal{R}_{\mu \nu \rho \sigma }$ of the connection $%
\mathcal{D}_{\mu },$ defined as
\begin{equation*}
(\mathcal{D}_{\mu }\mathcal{D}_{\nu }-\mathcal{D}_{\nu }\mathcal{D}_{\mu
})\xi ^{\rho }\doteqdot \lbrack \mathcal{D}_{\mu }\mathcal{D}_{\nu }]\xi
^{\rho }=H_{\ \mu \nu }^{\sigma }\mathcal{D}_{\sigma }\xi ^{\rho }+\mathcal{R%
}_{\ \sigma \mu \nu }^{\rho }\xi ^{\sigma },
\end{equation*}%
can be expressed in terms of the Riemannian tensor $R_{\mu \nu \rho \sigma }$
and the torsion tensor $H_{\ \mu \nu }^{\sigma },$%
\begin{equation*}
\mathcal{R}_{\mu \nu \rho \sigma }=R_{\mu \nu \rho \sigma }+\frac{1}{2}%
D_{\rho }H_{\sigma \mu \nu }-\frac{1}{2}D_{\sigma }H_{\rho \mu \nu }+\frac{1%
}{4}H_{\rho \mu \alpha }H_{\sigma \nu }^{\quad \alpha }-\frac{1}{4}H_{\sigma
\mu \alpha }H_{\rho \nu }^{\quad \alpha }.
\end{equation*}

Let us consider a generic off--diagonal metric, a non-degenerated matrix of
dimension $k\times k$ with the coefficients $\underline{g}_{\mu \nu }(u)$ \
defined with respect to a local coordinate frame like in (\ref{metric}).
This metric can transformed into a block $\left( n\times n\right) \oplus
\left( m\times m\right) $ form, for $k=n+m,\,$\
\begin{equation*}
\underline{g}_{\mu \nu }(u)\rightarrow \{ g_{ij}(u),h_{ab}\left(
u\right) \}
\end{equation*}%
if we perform a frame map with the vielbeins%
\begin{eqnarray}
e_{\mu }^{~\underline{\mu }}(u) &=&\left(
\begin{array}{cc}
e_{i}^{~\underline{i}}(x^{j},y^{a}) & N_{i}^{a}(x^{j},y^{a})e_{a}^{~%
\underline{a}}(x^{j},y^{a}) \\
0 & e_{a}^{~\underline{a}}(x^{j},y^{a})%
\end{array}%
\right)  \label{vielbtr} \\
e_{\ \underline{\nu }}^{\mu }(u) &=&\left(
\begin{array}{cc}
e_{\ \underline{i}}^{i}(x^{j},y^{a}) & -N_{k}^{a}(x^{j},y^{a})e_{\
\underline{i}}^{k}(x^{j},y^{a}) \\
0 & e_{\ \underline{a}}^{a}(x^{j},y^{a})%
\end{array}%
\right)  \notag
\end{eqnarray}%
which conventionally splits the spacetime into two subspaces: the first
subspace is parametrized by coordinates $x^{i}$ $\ $\ provided with indices
of type $i,j,k,...$ running values from $1$ to $n$ and the second subspace
is parametrized by coordinates $y^{a}$ provided with indices of type $%
a,b,c,...$ running values from $1$ to $m.$ This splitting is induced by the
coefficients $N_{i}^{a}(x^{j},y^{a}).$ For simplicity, we shall write the
local coordinates as $u^{\alpha }=\left( x^{i},y^{a}\right) ,$ or $u=\left(
x,y\right) .$

The coordinate bases $\partial _{\alpha }=\left( \partial _{i},\partial
_{a}\right) $ and theirs duals $d^{\alpha }=du^{\alpha }=\left(
d^{i}=dx^{i},d^{a}=dy^{a}\right) $ are transformed under maps (\ref{vielbtr}%
) as $\ $%
\begin{equation*}
\partial _{\alpha }\rightarrow e_{\underline{\alpha }}=e_{\ \underline{%
\alpha }}^{\alpha }(u)\partial _{\alpha },d^{\alpha }\rightarrow e^{%
\underline{\alpha }}=e_{\alpha }^{~\underline{\alpha }}(u)d^{\alpha },
\end{equation*}%
or, in 'N--distinguished' form,%
\begin{eqnarray}
e_{\underline{i}} &=&e_{\ \underline{i}}^{i}\partial _{i}-N_{k}^{a}e_{\
\underline{i}}^{k}\partial _{a},e_{\underline{a}}=e_{\ \underline{a}%
}^{a}\partial _{a},  \label{dder1a} \\
e^{\underline{i}} &=&e_{i}^{~\underline{i}}d^{i},~e^{\underline{a}%
}=N_{i}^{a}e_{a}^{~\underline{a}}d^{i}+e_{a}^{~\underline{a}}d^{a}.
\label{ddif1a}
\end{eqnarray}%
The quadratic line element (\ref{metric}) may be written equivalently in the
form
\begin{equation}
ds^{2}=g_{\underline{i}\underline{j}}(x,y)e^{\underline{i}}e^{\underline{j}%
}+h_{\underline{a}\underline{b}}(x,y)e^{\underline{a}}e^{\underline{b}}
\label{dmetric1a}
\end{equation}%
with the metric $\underline{g}_{\mu \nu }(u)$ parametrized in the form%
\begin{equation}
\underline{g}_{\alpha \beta }=\left[
\begin{array}{cc}
g_{ij}+N_{i}^{a}N_{j}^{b}h_{ab} & h_{ab}N_{i}^{a} \\
h_{ab}N_{j}^{b} & h_{ab}%
\end{array}%
\right] .  \label{ansatz}
\end{equation}

If we choose $e_{i}^{~\underline{i}}(x^{j},y^{a})=\delta _{i}^{~\underline{i}%
}$ and $e_{a}^{~\underline{a}}(x^{j},y^{a})=\delta _{a}^{~\underline{a}},$
we may not distinguish the 'underlined' and 'non--underlined' indices. The
operators (\ref{dder1a}) and (\ref{ddif1a}) transform respectively into the
operators of 'N--elongated' partial derivatives and differentials
\begin{eqnarray}
e_{i} &=&\delta _{i}=\partial _{i}-N_{i}^{a}\partial _{a},e_{a}=\partial
_{a},  \label{viel1} \\
e^{i} &=&d^{i},~e^{a}=\delta ^{a}=d^{a}+N_{i}^{a}d^{i}  \notag
\end{eqnarray}%
(which means that the anholonomic frames (\ref{dder1a}) and (\ref{ddif1a})
generated by vielbein transforms (\ref{vielbtr}) \ are, in general,
anholonomic; see the respective formulas (\ref{dder}), (\ref{ddif}) and (\ref%
{anhol}) in the Appendix) and the quadratic line element (\ref{dmetric1a})
trasforms in a d--metric element (see (\ref{dmetric}) in the Appendix).

The physical treatment of the vielbein transforms (\ref{vielbtr}) and
associated $N$--coefficients depends on the types of constraints
(equivalently, anholonomies) we impose on the string dynamics and/or on the
considered curved background. There were considered different possibilities:

\begin{itemize}
\item Ansatz of type (\ref{ansatz}) were used in Kaluza--Klein gravity \cite%
{salam}, as well in order to describe toroidal Kaluza--Klein reductions in
string theory (see, for instance, \cite{kir})). \ The coefficients $%
N_{i}^{a},$ usually written as $A_{i}^{a},$ are considered as the potentials
of some, in general, non--Abelian gauge fields, which in such theories are
generated by a corresponding compactification. In this case, the coordinates
$x^{i}$ can be used for the four dimensional spacetime and the coordinates $%
y^{a}$ are for extra dimensions.

\item Parametrizations of type (\ref{ansatz}) were considered in order to
elaborate an unified approach on vector/tangent bundles to
Finsler geometry
and its generalizations \cite%
{ma,miron,bejancu,vspinors,vmon1,vsuper,vstring,vstr2}. The coefficients $%
N_{i}^{a}$ were supposed to define a nonlinear connection \
(N--connection) structure in corresponding (super) bundles and
the metric coefficients $g_{ij}(u)$ and $g_{ab}\left( u\right) $
were taken for a
corresponding Finsler metric, or its generalizations (see formulas (\ref%
{fmetric}), (\ref{ncc}), (\ref{dmetricf}), (\ref{mfl}) and related
discussions in Appendix). The coordinates $x^{i}$ were defined on base
manifolds and the coordinates $y^{a}$ were used for fibers of bundles.

\item In a series of papers \cite%
{vexsol,vankin,vmethod,vbel,vsingl,vsingl1,vsolsp} the concept of
N--connection was introduced for (pseudo) Riemannian spaces
provided with off--diagonal metrics and/or anholonomic frames. In
a such approach the coefficients $N_{i}^{a}$ are associated to an
anholonomic frame structure describing a gravitational and matter
fields dynamics with mixed holonomic and anholonomic variables.
The coordinates $x^{i}$ are defined with respect to the subset of
holonomic frame vectors, but $y^{a}$ are given with respect to
the subset of anholonomic, N--ellongated, frame vectors. It was
proven that by using vielbein transforms of type (\ref{vielbtr})
the off--diagonal
metrics could be diagonalized and, for a very large class of ansatz of type (%
\ref{ansatz}), with the coefficients depending on 2,3 or 4 coordinate
variables, it was shown that the corresponding vacuum and non--vacuum
Einstein equations may be integrated in general form. This allowed an
explicit construction of new classes of exact solutions parametrized by
off--diagonal metrics with some anholonomically deformed symmetries. Two new
and very surprising conclusions were those that the Finsler like (and
another type) anisotropies may be modeled even in the framework of the
general relativity theory and its higher/lower dimension modifications, as
some exact solutions of the Einstein equations, and that the anholonomic
frame method is very efficient for constructing such solutions.
\end{itemize}

There is an important property of the off--diagonal metrics $\underline{g}%
_{\mu \nu }$ (\ref{ansatz}) which does not depend on the type of
space (a pseudo--Riemannian manifold, or a vector/tangent bundle)
this metric is given. With respect to the coordinate frames it is
defined a unique torsionless and metric compatible linear
connection derived as the usual Christoffel symbols (or the Levi
Civita connection). If anholonomic frames are introduced into
consideration, we can define an infinite number of metric
connections constructed from the coefficients of off--diagonal
metrics and induced by the anholonomy coefficients (see formulas
(\ref{lcsym}) and (\ref{lccon}) and the related discussion from
Appendix); this property is also mentioned in the monograph
\cite{mtw} (pages 216, 223, 261) for anholonomic frames but
without any particularities related to associated N--connection
structures. In this case there is an infinite number of metric
compatible linear connections, constructed from metric and
vielbein coefficients, all of them having non--trivial torsions
and transforming into the usual Christoffel symbols for $N_{i}^{a}%
\rightarrow 0$ and $m\rightarrow 0.$ For off--diagonal metrics
considered, and even diagonalized, with respect to anholonomic
frames and associated N--connections, we can not select a linear
connection being both torsionless and metric. The problem of
establishing of a physical linear connection structure
constructed from metric/frame coefficients is to be solved
together with that of fixing of a system of reference on a curved
spacetime which is not a pure dynamical task but depends on the
type of prescribed constraints, symmetries and boundary
conditions are imposed on interacting fields and/or string
dynamics.

In our further consideration we shall suppose that both a metric $\underline{%
g}_{\mu \nu }$ (equivalently, a set $\{ g_{ij},g_{ab},N_{i}^{a} \}
$) and metric linear connection $\Gamma _{~\beta \gamma }^{\alpha
},$ i.e. satisfying the conditions $D_{\alpha }g_{\alpha \beta
}=0,$ exist in the background spacetime. Such spaces will be
called locally anisotropic (equivalently, anolonomic) because the
anholonomic frames structure imposes locally a kind of anisotropy
with respective constraints on string and effective string
dynamics. For such configurations the torsion, induced as an
anholonomic frame effect, vanishes only with respect coordinate
frames. Here we note that in the string theory there are also
another type of torsion contributions to linear connections like
$H_{\ \mu \nu }^{\sigma },$ see formula (\ref{cd}).

\subsubsection{Anholonomic background field quantisation method}

We revise the perturbation theory around general field
configurations for background spaces provided with anholonomic
frame structures (\ref{dder1a}) and (\ref{ddif1a}), $\delta
_{\alpha }=(\delta _{i}=\partial _{i}-N_{i}^{a}\partial
_{a},\partial _{a})$ and $\delta ^{\alpha }=(d^{i},\delta
^{a}=d^{a}+N_{i}^{a}d^{i}),$ with associated N--connections,
$N_{i}^{a},$ and $\{ g_{ij},h_{ab}\} $ (\ref{dmetric1a}) adapted
to
such structures (distinguished metrics, or d--metrics, see formula (\ref%
{dmetric})). The linear connection in such locally anisotropic backgrounds
is considered to be compatible both to the metric and N--connection
structure (for simplicity, being a d--connection or an anholonomic variant
of Levi Civita connection, both with nonvanishing torsion, see formulas (\ref%
{dcon}), (\ref{lcsym}), (\ref{lccon}), and (\ref{dtors}), and related
discussions in the Appendix). The general rule for the tensorial calculus on
a space provided with N--connection structure is to split indices $\alpha
,\beta ,...$ into ''horozontal'', $i,j,...,$ and ''vertical'', $a,b,...,$
subsets and to apply for every type of indices the corresponding operators
of N--adapted partial and covariant derivations.

The anisotropic sigma model is to be formulated by anholonomic transforms of
the metric, $\underline{g}_{\mu \nu }\rightarrow \{g_{ij},h_{ab}\},$ partial
derivatives and differentials, $\partial _{\alpha }\rightarrow \delta
_{\alpha }$ and $d^{\alpha }\rightarrow \delta ^{a},$ volume elements, $d\mu
_{g}\rightarrow \delta \mu _{g}$ in the action (\ref{act1})

\begin{equation}
S=S_{g_{N},B}+S_{\Phi }+S_{U},  \label{act1a}
\end{equation}%
with%
\begin{eqnarray*}
S_{g_{N},B}[u,g] &=&\frac{1}{8\pi l^{2}}\int\limits_{\Sigma }\delta \mu
_{g}\{g^{AB}\left[ \partial _{A}x^{i}\partial _{B}x^{j}g_{ij}(x,y)+\partial
_{A}x^{a}\partial _{B}x^{b}h_{ab}(x,y)\right] \\
&&+\varepsilon ^{AB}\partial _{A}u^{\mu }\partial _{B}u^{\nu }B_{\mu \nu
}(u)\}, \\
S_{\Phi }[u,g] &=&\frac{1}{2\pi }\int\limits_{\Sigma }\delta \mu
_{g}R_{g}\Phi (u),~S_{U}[u,g]=\frac{1}{4\pi }\int\limits_{\Sigma }\delta \mu
_{g}U(u),
\end{eqnarray*}%
where the coefficients $B_{\mu \nu }$ are computed for a two--form $B=B_{\mu
\nu }\delta u^{\mu }\wedge \delta u^{\nu }.$

The perturbation theory has to be developed by changing the usual partial
derivatives into N--elongated ones, for instance, the decomposition (\ref%
{ser1}) is to be written
\begin{equation*}
\underline{g}_{\mu \nu }(u)=\underline{g}_{\mu \nu }(u_{0})+\ell \delta
_{\alpha }\underline{g}_{\mu \nu }(u_{0})u_{[1]}^{\alpha }+\ell ^{2}\delta
_{\alpha }\delta _{\beta }\underline{g}_{\mu \nu }(u_{0})u_{[1]}^{\alpha
}u_{[1]}^{\beta }+\ell ^{2}\delta _{\beta }\delta _{\alpha }\underline{g}%
_{\mu \nu }(u_{0})u_{[1]}^{\alpha }u_{[1]}^{\beta }+...,
\end{equation*}%
where we should take into account the fact that the operators $\delta
_{\alpha }$ do not commute but satisfy certain anholonomy relations (see (%
\ref{anhol})\ \ in Appendix).

The action (\ref{act1a}) is invariant under the group of diffeomorphisms on $%
\Sigma $ and $M$ (on spacetimes provided with N--connections the
diffeomorphisms may be adapted to such structures) and posses a $U(1)_{B}$
gauge invariance, acting by $B\rightarrow B+\delta \gamma $ for some $\gamma
\in \Omega ^{(1)}\left( M\right) ,$ where $\Omega ^{(1)}$ denotes the space
of 1--forms on $M.$ Wayl's conformal transformations of $\Sigma $ leave $%
S_{g_{N},B}$ invariant but result in anomalies under quantization. $S_{\Phi
} $ and $S_{U}$ fail to be conformal invariant even classically. We discuss
the renormalization of quantum filed theory defined by the action (\ref%
{act1a}) for general fields $g_{ij},h_{ab},N_{\mu }^{a},B_{\mu \nu }$ and $%
\Phi .$ We shall not discus in this work the effects of the tachyon field.

The string corrections to gravity (in both locally isotropic and locally
anisotropic cases) may be computed following some regularizatons schemes
preserving the classical symmetries and determining the general structure of
the Weyl dependence of Green functions specified by the action (\ref{act1a})
in terms of fixed background fields $g_{ij},h_{ab},N_{\mu }^{a},B_{\mu \nu }$
and $\Phi .$ One can consider unnormalized correlation functions of
operators $\phi _{1},...,\phi _{p},$ instead of points $\xi _{1},...,\xi
_{p}\in \Sigma $ \cite{deligne}.

By definition of the stress tensor $T_{AB},$ under conformal transforms on
the two dimensional hypersurface, $g_{[2]}\rightarrow \exp [2\delta \sigma
]g_{[2]}$ with support away from $\xi _{1},...,\xi _{p},$ we have%
\begin{equation*}
\triangle _{\sigma }<\phi _{1}...\phi _{p}>_{g_{[2]}}=\frac{1}{2\pi }%
\int_{\Sigma }\delta \mu _{g}\triangle \sigma <T_{A}^{~A}\phi _{1}...\phi
_{p}>_{g_{[2]}},
\end{equation*}%
when assuming throughout that correlation functions are covariant under the
diffeomorphisms on $\Sigma ,$ $\bigtriangledown ^{A}T_{AB}=0.$ The value $%
T_{A}^{~A}$ receives contributions from the explicit conformal
non--invariance of $S_{\Phi },$ from conformal (Weyl) anomalies which are
local functions of $u,$ i.e. dependent on $u$ and on finite order
derivatives on $u,$ and polynomial in the derivatives of $u.$ For spaces
provided with N--connection structures we should consider N--elongated
partial derivatives, choose a N--adapted linear connection structure with
some coefficients $\Gamma _{~\mu \nu }^{\alpha }$ (for instance the Levi
Civita connection (\ref{lcsym}), or d--connection (\ref{dcon})). The basic
properties of $T_{A}^{~A}$ are the same as for trivial values of $N_{i}^{a}$ %
\cite{deligne}, which allows us to write directly that
\begin{eqnarray*}
T_{A}^{~A} &=&g^{AB}[\partial _{A}x^{i}\partial _{B}x^{j}\beta
_{ij}^{g,N}(x,y)+\partial _{A}x^{i}\partial _{B}y^{b}\beta
_{ib}^{g,N}(x,y)+\partial _{A}y^{a}\partial _{B}x^{j}\beta _{aj}^{g,N}(x,y)
\\
&&+\partial _{A}y^{a}\partial _{B}y^{b}\beta _{ab}^{g,N}(x,y)+\varepsilon
^{AB}\partial _{A}u^{\alpha }\partial _{B}u^{\beta }\beta _{\alpha \beta
}^{B}(x,y)+\beta ^{\Phi }(x,y)R_{g},
\end{eqnarray*}%
where the functions $\beta _{\alpha \beta }^{g,N}=\{\beta _{ij}^{g,N},\beta
_{ab}^{g,N}\},\beta _{\alpha \beta }^{B}$ and $\beta ^{\Phi }(x,y)$ are
called beta functions. On general grounds, the expansions of $\beta $%
--functions are of type
\begin{equation*}
\beta (x,y)=\sum_{r=0}^{\infty }\ell ^{2r}\beta ^{\lbrack 2r]}(x,y).
\end{equation*}

One considers expandings up to and including terms with two derivatives on
the fields which consideres expansions up to order $r=0$ of $\beta _{\alpha
\beta }^{g,N}$ and $\beta _{\alpha \beta }^{B}$ and orders $s=0,2$ for $%
\beta ^{\Phi }.$ In this approximation, after cumbersome but simple
calculations (similar to those given in \cite{deligne}, in our case on
locally anisotropic backgrounds)%
\begin{eqnarray*}
\beta _{ij}^{g,N} &=&a_{1[1]}R_{ij}+a_{2[1]}g_{ij}+a_{3[1]}g_{ij}\widehat{R}%
+a_{4[1]}H_{i\rho \sigma }^{[N]}H_{j}^{[N]\rho \sigma
}+a_{5[1]}g_{ij}H_{\rho \sigma \tau }^{[N]}H^{[N]\rho \sigma \tau } \\
&&+a_{6[1]}D_{i}D_{j}\Phi +a_{7[1]}g_{ij}D^{2}\Phi +a_{8[1]}g_{ij}D^{\rho
}\Phi D_{\rho }\Phi ,
\end{eqnarray*}%
\begin{eqnarray}
\beta _{ib}^{g,N} &=&a_{1[2]}R_{ib}+a_{4[2]}H_{i\rho \sigma
}^{[N]}H_{b}^{[N]\rho \sigma }+a_{6[2]}D_{i}D_{b}\Phi ,  \label{beta1} \\
\beta _{aj}^{g,N} &=&a_{1[3]}R_{aj}+a_{4[3]}H_{a\rho \sigma
}^{[N]}H_{j}^{[N]\rho \sigma }+a_{6[3]}D_{a}D_{j}\Phi ,  \notag
\end{eqnarray}%
\begin{eqnarray*}
\beta _{ab}^{g,N}
&=&a_{1[4]}S_{ab}+a_{2[4]}h_{ab}+a_{3[4]}h_{ab}S+a_{4[4]}H_{a\rho \sigma
}^{[N]}H_{b}^{[N]\rho \sigma }+a_{5[4]}h_{ab}H_{\rho \sigma \tau
}^{[N]}H^{[N]\rho \sigma \tau } \\
&&+a_{6[4]}D_{a}D_{b}\Phi +a_{7[4]}h_{ab}D^{2}\Phi +a_{8[4]}h_{ab}D^{\rho
}\Phi D_{\rho }\Phi ,
\end{eqnarray*}%
\begin{eqnarray*}
\beta _{\alpha \beta }^{B} &=&b_{1}D^{\lambda }H_{\lambda \mu \nu
}^{[N]}+b_{2}(D^{\lambda }\Phi )H_{\lambda \mu \nu }^{[N]}, \\
&& \\
\beta ^{\Phi } &=&c_{0}+\ell ^{2}\left[ c_{1[1]}\widehat{R}%
+c_{1[2]}S+c_{2}D^{2}\Phi +c_{3}\left( D^{\lambda }\Phi \right) D_{\lambda
}\Phi +c_{4}H_{\rho \sigma \tau }^{[N]}H^{[N]\rho \sigma \tau }\right] ,
\end{eqnarray*}%
where $R_{\alpha \beta }=\{R_{ij},R_{ib},R_{aj},S_{ab}\}$ and $%
\overleftarrow{R}=\{\widehat{R},S\}$ are given respectively by the formulas (%
\ref{dricci}) and (\ref{dscalar}) and the $B$--strength $H_{\lambda \mu \nu
}^{[N]}$ is computed not by using partial derivatives, like in (\ref{str1}),
but with N--adapted partial derivatives,
\begin{equation}
H_{\mu \nu \rho }^{[N]}\equiv \delta _{\mu }B_{\nu \rho }+\delta _{\nu
}B_{\rho \mu }+\delta _{\nu }B_{\rho \mu }.  \label{htors}
\end{equation}%
The formulas for $\beta $--functions (\ref{beta1}) are adapted to the
N--connection structure being expressed via invariant decompositions for the
Ricci d--tensor and curvature scalar; every such invariant object was
provided with proper constants. In order to have physical compatibility with
the case $N\rightarrow 0$ we should take%
\begin{eqnarray*}
a_{z[1]} &=&a_{z[2]}=a_{z[3]}=a_{z[4]}=a_{z},~z=1,2,...,8; \\
c_{1[1]} &=&c_{1[2]}=c_{1,}
\end{eqnarray*}%
where $a_{z}$ and $c_{1}$ are the same as in the usual string theory,
computed from the 1-- and 2--loop $\ell $--dependence of graphs ($a_{2}=0,$ $%
a_{6}=1,$ $a_{7}=a_{8}=0$ and $b_{2}=1/2,$ $c_{3}=2)$ and by using the
background field method (in order to define the values $%
a_{1},a_{3},a_{4},a_{5},b_{1}$ and $c_{1},c_{2},c_{4}).$

\subsection{Low energy string anholonomic field equations}

\label{dcovrule}The effective action, as the generating functional for
1--particle irreducible Feynman diagrams in terms of a functional integral,
can be obtained following the background quantization method adapted, in our
constructions, to sigma models on spacetimes with N--connection structure.
On such spaces, we can also make use of the Riemannian coordinate expansion,
but taking into account that the coordinates are defined with respect to
N--adapted bases and that the covariant derivative $D$ is of type (\ref{dcon}%
), (\ref{lcsym}) or (\ref{lccon}), i. e. is d--covariant, defined by a
d--connection.

For two infinitesimally closed points $u_{0}^{\mu }=u^{\mu }(\tau _{0})$ and
$u^{\mu }(\tau ),$ with $\tau $ being a parameter on a curve connected the
points, we denote $\zeta ^{\alpha }=du^{\alpha }/d\tau _{\mid 0}$ and write $%
u^{\mu }=e^{\ell \zeta }u_{0}^{\mu }.$ We can consider diffeomorphism
invariant d--covariant expansions of d--tensors in powers of $\ell ,$ for
instance,
\begin{eqnarray*}
\Phi (u) &=&\Phi (u_{0})+\ell D_{\alpha }[\Phi (u)\zeta ^{\alpha }]_{\mid
u=u_{0}}+\frac{\ell ^{2}}{2}D_{\alpha }D_{\beta }[\Phi (u)\zeta ^{\alpha
}\zeta ^{\beta }]_{\mid u=u_{0}}+o\left( \ell ^{3}\right) , \\
A_{\alpha \beta }\left( u\right) &=&A_{\alpha \beta }\left( u_{0}\right)
+\ell D_{\alpha }[A_{\alpha \beta }(u)\zeta ^{\alpha }]_{\mid u=u_{0}}+\frac{%
\ell ^{2}}{2}\{D_{\alpha }D_{\beta }[A_{\mu \nu }(u)\zeta ^{\alpha }\zeta
^{\beta } \\
&&-\frac{1}{3}R_{~\alpha \mu \beta }^{[N]\rho }(u)A_{\rho \nu }(u)-\frac{1}{3%
}R_{~\alpha \nu \beta }^{[N]\rho }(u)A_{\rho \mu }(u)]\}_{\mid
u=u_{0}}+o\left( \ell ^{3}\right) ,
\end{eqnarray*}%
where the Riemannian curvature d--tensor $R_{~\alpha \mu \beta }^{[N]\rho
}=%
\{R_{h.jk}^{.i},R_{b.jk}^{.a},P_{j.ka}^{.i},P_{b.ka}^{.c},S_{j.bc}^{.i},S_{b.cd}^{.a}\}
$ has the invariant components given by the formulas
(\ref{dcurvatures}) from Appendix. \ Putting such expansions in
the action for the nonlinear
sigma model (\ref{act1a}), we obtain the decomposition%
\begin{equation*}
S_{g_{N},B}[u,g]=S_{g_{N},B}[u_{0},g]+\ell \int\limits_{\Sigma }\delta \mu
_{g}\zeta ^{\beta }S_{\beta }[u_{0},g]+\overline{S}[u,\zeta ,g],
\end{equation*}%
where $S_{\beta }$ is given by the variation
\begin{equation*}
S_{\beta }[u_{0},g]=(\det |g|)^{-1/2}\frac{\triangle S[e^{\chi }u_{0},g]}{%
\triangle \chi ^{\beta }}\mid _{\chi =0}
\end{equation*}%
and the last term $\overline{S}$ is an expansion on $\ell ,$%
\begin{equation*}
\overline{S}=\overline{S}_{[0]}+\ell \overline{S}_{[1]}+\ell ^{2}\overline{S}%
_{[2]}+o\left( \ell ^{3}\right) ,
\end{equation*}%
with%
\begin{eqnarray}
\overline{S}_{[0]} &=&\frac{1}{8\pi }\int\limits_{\Sigma }\delta \mu
_{g}\{g^{AB}[g_{ij}(u_{0})\mathcal{D}_{A}^{\ast }\zeta ^{i}\mathcal{D}%
_{B}^{\ast }\zeta ^{j}+h_{ab}(u_{0})\mathcal{D}_{A}^{\ast }\zeta ^{a}%
\mathcal{D}_{B}^{\ast }\zeta ^{b}]  \label{actser} \\
&&+\mathcal{R}_{\mu \nu \rho \sigma }^{[N]}(u_{0})[g^{AB}-\varepsilon
^{AB}]\partial _{A}u_{0}^{\mu }\partial _{B}u_{0}^{\rho }\zeta ^{\nu }\zeta
^{\sigma }\},  \notag \\
\overline{S}_{[1]} &=&\frac{1}{24\pi }\int\limits_{\Sigma }\delta \mu
_{g}H_{\mu \nu \rho }^{[N]}\varepsilon ^{AB}\zeta ^{\mu }\mathcal{D}%
_{A}^{\ast }\zeta ^{\nu }\mathcal{D}_{B}^{\ast }\zeta ^{\rho },  \notag \\
\overline{S}_{[2]} &=&\frac{1}{8\pi }\int\limits_{\Sigma }\delta \mu _{g}\{%
\frac{g^{AB}}{3}R_{\mu \nu \rho \sigma }^{[N]}\zeta ^{\nu }\zeta ^{\rho }%
\mathcal{D}_{A}^{\ast }\zeta ^{\mu }\mathcal{D}_{B}^{\ast }\zeta ^{\sigma }
\notag \\
&&-\frac{\varepsilon ^{AB}}{2}\mathcal{R}_{\mu \nu \rho \sigma }^{[N]}\zeta
^{\nu }\zeta ^{\rho }\mathcal{D}_{A}^{\ast }\zeta ^{\mu }\mathcal{D}%
_{B}^{\ast }\zeta ^{\sigma }+2D_{\alpha }D_{\beta }\Phi (u_{0})\zeta
^{\alpha }\zeta ^{\beta }R_{g}\}.  \notag
\end{eqnarray}%
The operator $\mathcal{D}_{A}^{\ast }\zeta ^{\nu }$ from (\ref{actser}) is
defined according the rule%
\begin{equation*}
\mathcal{D}_{A}^{\ast }\zeta ^{\nu }=D_{A}^{\ast }\zeta ^{\nu }+\frac{1}{2}%
H_{\quad \mu \rho }^{[N]\sigma }g_{AB}\varepsilon ^{BC}\partial _{C}u^{\mu
}\zeta ^{\rho },
\end{equation*}%
with $D_{A}^{\ast }$ being the covariant derivative on $T^{\ast }\Sigma
\otimes TM$ pulled back to $\Sigma $ by the map $u^{\alpha }$ and acting as
\begin{equation*}
D_{A}^{\ast }\partial _{B}u^{\nu }=\bigtriangledown _{A}\partial _{B}u^{\nu
}+\Gamma _{~\mu \nu }^{\alpha }\partial _{B}u^{\nu }\partial _{A}u^{\nu },
\end{equation*}%
with a h-- and v--invariant decomposition $\Gamma _{\ \beta \gamma }^{\alpha
}=\{L_{\ jk}^{i},L_{\ bk}^{a},C_{\ jc}^{i},C_{\ bc}^{a}\},$ see (\ref{dcon})
from Appendix, and the operator $\ \mathcal{R}_{\mu \nu \rho \sigma }^{[N]}$
is computed as
\begin{equation*}
\mathcal{R}_{\mu \nu \rho \sigma }^{[N]}=R_{\mu \nu \rho \sigma }^{[N]}+%
\frac{1}{2}D_{\rho }H_{\sigma \mu \nu }^{[N]}-\frac{1}{2}D_{\sigma }H_{\rho
\mu \nu }^{[N]}+\frac{1}{4}H_{\rho \mu \alpha }^{[N]}H_{\sigma \nu
}^{[N]\alpha }-\frac{1}{4}H_{\sigma \mu \alpha }^{[N]}H_{\rho \nu
}^{[N]\alpha }.
\end{equation*}

A comparative analysis of the expansion (\ref{actser}) with a similar one
for $N=0$ from the usual nonlinear sigma model (see, for instance, \cite%
{deligne}) define the 'geometric d--covariant rule': \ we may apply the same
formulas as in the usual covariant expansions but with that difference that
1) the usual spacetime partial derivatives and differentials are substituted
by N--elongated ones; 2) the Christoffell symbols of connection are changed
into certain d--connection ones, of type (\ref{dcon}), (\ref{lcsym}) or (\ref%
{lccon}); 3) the torsion $H_{\sigma \mu \nu }^{[N]}$ is computed via
N--elongated partial derivatives as in (\ref{htors}) and 4) the curvature $%
R_{\mu \nu \rho \sigma }^{[N]}$ is split into horizontal--vertical, in
brief, h--v--invariant, components according the the formulas (\ref%
{dcurvatures}). The geometric d--covariant rule allows us to transform
directly the formulas for spacetime backgrounds with metrics written with
respect to coordinate frames into the respective formulas with N--elongated
terms and splitting of indices into h-- and v-- subsets.

\subsubsection{Low energy string ansitropic field equations and effective
action}

Following the geometric d--covariant rule we may apply the results of the
holonomic sigma models in order to define the coefficients $%
a_{1},a_{3},a_{4},a_{5},b_{1}$ and $c_{1},c_{2},c_{4}$ of beta functions (%
\ref{beta1}) and to obtain the following equations of (in our case,
anholonomic) string dynamics,%
\begin{eqnarray*}
2\beta _{ij}^{g,N} &=&R_{ij}-\frac{1}{4}H_{i\rho \sigma
}^{[N]}H_{j}^{[N]\rho \sigma }+2D_{i}D_{j}\Phi =0, \\
2\beta _{ib}^{g,N} &=&R_{ib}-\frac{1}{4}H_{i\rho \sigma
}^{[N]}H_{b}^{[N]\rho \sigma }+2D_{i}D_{b}\Phi =0,
\end{eqnarray*}%
\begin{eqnarray*}
2\beta _{aj}^{g,N} &=&R_{aj}-\frac{1}{4}H_{a\rho \sigma
}^{[N]}H_{j}^{[N]\rho \sigma }+2D_{a}D_{j}\Phi =0, \\
2\beta _{ab}^{g,N} &=&S_{ab}-\frac{1}{4}H_{a\rho \sigma
}^{[N]}H_{b}^{[N]\rho \sigma }+2D_{a}D_{b}\Phi =0,
\end{eqnarray*}%
\begin{eqnarray}
2\beta _{\alpha \beta }^{B} &=&-\frac{1}{2}D^{\lambda }H_{\lambda \mu \nu
}^{[N]}+(D^{\lambda }\Phi )H_{\lambda \mu \nu }^{[N]}=0,  \label{streq} \\
&&  \notag \\
2\beta ^{\Phi } &=&\frac{n+m-26}{3}+\ell ^{2}\left[ \frac{1}{12}H_{\rho
\sigma \tau }^{[N]}H^{[N]\rho \sigma \tau }-\widehat{R}-S-4D^{2}\Phi
+4\left( D^{\lambda }\Phi \right) D_{\lambda }\Phi \right] =0,  \notag
\end{eqnarray}%
where $n+m$ denotes the total dimension of a spacetime with $n$ holonomic
and $m$ anholonomic variables. It should be noted that \ $\beta ^{g,N}=\beta
^{B}=0$ imply the condition that $\beta ^{\Phi }=const,$ which is similar to
the holonomic strings. The only way to satisfy $\beta ^{\Phi }=0$ with
integers $n$ and $m$ is to take $n+m=26.$

The equations (\ref{streq}) are similar to the Einstein equations for the
locally anisotropic gravity (see (\ref{einsteq2}) in Appendix) with the
matter energy--momentum d--tensor defined from the string theory. \ From
this viewpoint the fields $B_{\alpha \beta }$ and $\Phi $ can be viewed as
certain matter fields and the effective field equations (\ref{streq}) can be
derived from action%
\begin{equation}
S\left( g_{ij},h_{ab},N_{i}^{a},B_{\mu \nu },\Phi \right) =\frac{1}{2\kappa
^{2}}\int \delta ^{26}u\sqrt{|\det g_{\alpha \beta }|}e^{-2\Phi }\left[
\widehat{R}+S+4(D\Phi )^{2}-\frac{1}{12}H^{2}\right] ,  \label{act3}
\end{equation}%
where $\kappa $ is a constant and, for instance, $D\Phi =D_{\alpha }\Phi ,$ $%
H^{2}=H_{\mu }H^{\mu }$ and the critical dimension $n+m=26$ is taken. For $%
N\rightarrow 0$ and $m\rightarrow 0$ the metric $g_{\alpha \beta }$ is
called the string metric. We shall call $g_{\alpha \beta }$ the string
d--metric for nontrivial values of $N.$

Instead of action (\ref{act3}), a more standard action, for arbitrary
dimensions, \ can be obtained via a conformal transform of d--metrics of
type (\ref{dmetric}),
\begin{equation*}
g_{\alpha \beta }\rightarrow \widetilde{g}_{\alpha \beta }=e^{-4\Phi /\left(
n+m-2\right) }g_{\alpha \beta }.
\end{equation*}%
The action in d--metric $\widetilde{g}_{\alpha \beta }$ (by analogy with the
locally isotropic backgrounds we call it the Einstein d--metric) is written%
\begin{eqnarray*}
S\left( \widetilde{g}_{ij},\widetilde{h}_{ab},N_{i}^{a},B_{\mu \nu },\Phi
\right) &=&\frac{1}{2\kappa ^{2}}\int \delta ^{26}u\sqrt{|\det \widetilde{g}%
_{\alpha \beta }|}[\widetilde{\widehat{R}}+\widetilde{S} \\
&&+\frac{4}{n+m-2}(D\Phi )^{2}-\frac{1}{12}e^{-8\Phi /\left( n+m-2\right)
}H^{2}].
\end{eqnarray*}%
This action, for $N\rightarrow 0$ and $m\rightarrow 0,$ is known in
supergravity theory as a part of Chapline--Manton action, see Ref. \cite%
{deligne} and for the so--called locally anisotropic supergravity, \cite%
{vstr2,vmon1}. When we deal with superstirngs, the susperstring calculations
to the mentioned orders give the same results as the bosonic string except
the dimension. For anholonomic backrounds we have to take into account the
nontrivial contributions of $N_{i}^{a}$ and splittings into h-- and v--parts.

\subsubsection{Ahnolonomic Einstein and Finsler gravity from string theory}

\label{efs}It is already known that the $B$--field can be used for
generation of different types of noncommutative geometries from string
theories (see original results and reviews in Refs. \cite%
{strncg,dn,connes1,sw,vncf}). Under certain conditions such $B$--field
configurations may result in different variants of geometries with local
anistropy like anholonomic Riemannian geometry, Finsler like spaces and
their generalizations. There is also an alternative possibility when locally
anisotropic interactions are modeled by anholonomic frame fields with
arbitrary $B$--field contributions. In this subsection, we investigate both
type of anisotropic models contained certain low energy limits of string
theory.

\paragraph{B--fields and anholonomic Einstein--Finsler structures}

{\ }\newline
The simplest way to generate an anholonomic structure in a low energy limit
of string theory is to consider a background metric $g_{\mu \nu }=\left(
\begin{array}{cc}
g_{ij} & 0 \\
0 & h_{ab}%
\end{array}%
\right) $ with symmetric Christoffel symbols $\{_{\beta \gamma }^{\alpha }\}$
and such $B_{\mu \nu },$ with corresponding $H_{\mu \nu \rho }^{[N]}$ from (%
\ref{htors}), as there are the nonvanishing values $H_{\mu \nu }^{[N]\rho
}=\{H_{ij}^{[N]a},H_{bj}^{[N]a}=-H_{jb}^{[N]a}\}.$ The next step is to
consider a covariant operator $\mathcal{D}_{\mu }=D_{\mu }^{\{\}}+\frac{1}{2}%
H_{\mu \nu }^{[N]\rho }$ (\ref{cd}), where $\frac{1}{2}H_{\mu \nu }^{[N]\rho
}$ is identified with the torsion (\ref{torsion}). This way the torsion $%
H_{\mu \nu \rho }^{[N]}$ is associated to an aholonomic frame structure with
non--trivial $W_{ij}^{a}=\delta _{i}N_{j}^{a}-\delta
_{j}N_{i}^{a},~W_{ai}^{b}=-W_{ia}^{b}=-\partial _{a}N_{i}^{b}$ (\ref%
{anholncoef}), when $B_{\mu \nu }$ is parametrized in the form $B_{\mu \nu
}=\{B_{ij}=-B_{ji},B_{bj}=-B_{jb}\}$ by identifying
\begin{equation*}
g_{\mu \nu }W_{\gamma \beta }^{\nu }=\delta _{\mu }B_{\gamma \beta },
\end{equation*}%
i. .e.
\begin{equation}
h_{ca}W_{ij}^{a}=\partial _{c}B_{ij}\mbox{ and }h_{ca}W_{bj}^{a}=\partial
_{c}B_{bj}.  \label{aux01}
\end{equation}%
Introducing the formulas for the anholonomy coefficients (\ref{anholncoef})
into (\ref{aux01}), we find some formulas relating partial derivatives $%
\partial _{\alpha }N_{j}^{a}$ and the coefficients $N_{j}^{a}$ with partial
derivatives of $\{B_{ij},B_{bj}\},$%
\begin{eqnarray}
h_{ca}\left( \partial _{i}N_{j}^{a}-N_{i}^{b}\partial _{b}N_{j}^{a}-\partial
_{j}N_{i}^{a}+N_{j}^{b}\partial _{b}N_{i}^{a}\right) &=&\partial _{c}B_{ij},
\notag \\
-h_{ca}\partial _{b}N_{j}^{a} &=&\partial _{c}B_{bj}.  \label{aux02}
\end{eqnarray}%
So, given any data $\left( h_{ca},N_{i}^{a}\right) $ we can define from the
system of first order partial derivative equations (\ref{aux02}) \ the
coefficients $B_{ij}$ and $B_{bj},$ or, inversely, from the data $\left(
h_{ca},B_{ij},B_{bj}\right) $ we may construct some non--trivial values $%
N_{j}^{a}.$ We note that the metric coefficients $g_{ij}$ and the $B$--field
components $B_{ab}=-B_{ba}$ could be arbitrary ones, in the simplest
approach we may put $B_{ab}=0.$

The formulas (\ref{aux02}) define the conditions when a
$B$--field may be transformed into a N--connection structure, or
inversely, a N--connection
can be associated to a $B$--field for a prescribed d--metric structure $%
h_{ca},$ (\ref{dmetric}).

The next step is to decide what type of d--connection we consider on our
background spacetime. If the values $\{g_{ij},h_{ca}\}$ and $W_{\gamma \beta
}^{\nu }$ (defined by $N_{i}^{b}$ as in (\ref{anholncoef}), but also induced
from $\{B_{ij},B_{bj}\}$ following (\ref{aux02})) are introduced in formulas
(\ref{lcsym}) we construct a Levi Civita d--connection $\mathcal{D}_{\mu }$
with nontrivial torsion induced by anholonomic frames with associated
nonlinear connection structure. This spacetime is provided with a d--metric (%
\ref{dmetric}), $g_{\alpha \beta }=\{g_{ij},h_{ca}\},$ which is compatible
with $\mathcal{D}_{\mu },$ i. e. $\mathcal{D}_{\mu }g_{\alpha \beta }=0.$
The coefficients of $\mathcal{D}_{\mu }$ with respect to anholonomic frames (%
\ref{dder}) and (\ref{ddif}), $\Gamma _{\beta \gamma }^{\bigtriangledown
\tau },$ can be computed in explicit form by using formulas (\ref{lcsym}).
It is proven in the Appendix that on spacetimes provided with anholonomic
structures the Levi Civita connection is not a prioritary one being both
metric and torsion vanishing. We can construct an infinite number of metric
connections, for instance, the canonical d--connection with the coefficients
(\ref{dcon}), or, equivalently, following formulas (\ref{lccon}), to
substitute from the coefficients (\ref{lcsym}) the values $\frac{1}{2}%
g^{ik}\Omega _{jk}^{a}h_{ca},$ where the coefficients of N--connection
curvature are defined by $N_{i}^{a}$ as in (\ref{ncurv}). In general, all
such type of linear connections are with nontrivial torsion because of
anholonomy coefficients.

We may generate by $B$--fields an anholonomic (pseudo) Riemannian geometry
if (for given values of $g_{\alpha \beta }=\{g_{ij},h_{ca}\}$ and $N_{i}^{a},
$ satisfying the conditions (\ref{aux02})) the metric is considered in the
form (\ref{odm}) \ with respect to coordinate frames, or, equivalently, in
the form (\ref{dmetric}) with respected to N--adapted frame (\ref{ddif}).
The metric has to satisfy the gravitational field equations (\ref{einsteq2})
for the Einstein gravity of arbitrary dimensions with holonomic--anholonomic
variables, or the equations (\ref{streq}) if the gravity with anholonomic
constraints is induced in a low energy string dynamics. We emphasize that
the Ricci d--tensor coefficients from $\beta $--functions (\ref{streq})
should be computed by using the formulas (\ref{dricci}), derived from those
for d--curvatures (\ref{dcurvatures}) and for d--torsions (\ref{dtors}) for
a chosen variant of d--connection coefficients, for instance, (\ref{dcon})
or (\ref{lcsym}).

We note here that a number of particular ansatz of form (\ref{odm}) were
considered in Kaluza--Klein gravity \cite{salam} for different type of
compactifications. In Refs. \cite{vexsol,vbel,vsingl,vsingl1,vsolsp} there
were constructed and investigated a number exact solutions with
off--diagonal metrics and anholonomic frames with associated N--connection
structures in the Einstein gravity of different dimensions (see also the
Section \ref{exsol}).

Now, we discuss the possibility to generate a Finsler geometry from string
theory. We note that the standard definition of Finsler quadratic form $%
g_{ij}^{[F]}=(1/2)\partial ^{2}F/\partial y^{i}\partial y^{j}$ is considered
to be positively definite (see (\ref{fmetric}) in Appendix). There are
different possibilities to include Finsler like structures in string
theories. For instance, we can consider quadratic forms with non--constant
signatures and to generate (pseudo) Finsler geometries [similarly to
(pseudo) Eucliedean/Riemannian metrics], or, as a second approach, to
consider some embeddings of Finsler d--metrics (\ref{dmetricf}) of signature
$\left( ++...+\right) $ into a 26 dimensional pseudo-Riemannian anholonomic
background with signature $\left( -++...+\right) .$ In the last case, a
particular class of Finsler background d--metrics may be chosen in the form
\begin{equation}
G^{[F]}=-dx^{0}\otimes dx^{0}+dx^{1}\otimes dx^{1}+g_{i^{\prime }j^{\prime
}}^{[F]}(x,y)dx^{i^{\prime }}\otimes dx^{j^{\prime }}+g_{i^{\prime
}j^{\prime }}^{[F]}(x,y)\delta y^{i^{\prime }}\otimes \delta y^{j^{\prime }}
\label{dmfstr}
\end{equation}%
where $i^{\prime },j^{\prime },...$run values $1,2,..,n^{\prime }\leq 12$
for bosonic strings. The coefficients $g_{i^{\prime }j^{\prime }}^{[F]}$ are
of type (\ref{fmetric}) $\ $or may take the value $+\delta _{i^{\prime
}j^{\prime }}$ for some values of $i\neq i^{\prime },j\neq j^{\prime }.$ We
may consider some static Finsler backgrounds if $g_{i^{\prime }j^{\prime
}}^{[F]}$ do not depend on coordinates $\left( x^{0},x^{1}\right) ,$ but, in
general, we are not imposed to restrict ourselves only to such
constructions. The $N$--coefficients from $\delta y^{i^{\prime
}}=dy^{i^{\prime }}+N_{j^{\prime }}^{i^{\prime }}dx^{i^{\prime }}$ must be
of the form (\ref{ncc}) if we wont to generate in the low energy string
limit a Finsler structure with Cartan nonlinear connection (there are
possible different variants of nonlinear and distinguished nonlinear
connections, see details in Refs. \cite{finsler,ma,bejancu} and Appendix).

Let us consider in details how a Finsler metric can be included in a low
energy string dynamics. We take a Finsler metric $F$ which generate the
metric coefficients $g_{i^{\prime }j^{\prime }}^{[F]}$ and the N--connection
coefficients $N_{j^{\prime }}^{[F]i^{\prime }},$ respectively, via formulas (%
\ref{fmetric}) and (\ref{ncc}). The Cartan's N--connection structure $%
N_{j^{\prime }}^{[F]i^{\prime }}$ may be induced by a $B$--field if there
are some nontrivial values, let us denote them $\{B_{ij}^{[F]},B_{bj}^{[F]}%
\},$ which satisfy the conditions (\ref{aux02}). This way the $B$--field is
expressed via a Finsler metric $F\left( x,y\right) $ and induces a d--metric
(\ref{dmfstr}). This Finsler structure follows from a low energy string
dynamics if the Ricci d--tensor $R_{\alpha \beta
}=\{R_{ij},R_{ia},R_{ai},R_{ab}\}$ (\ref{dricci}) and the torsion $H_{\mu
\nu }^{[N]\rho }=\{H_{ij}^{[N]a},H_{bj}^{[N]a}=-H_{jb}^{[N]a}\}$ related
with $N_{j^{\prime }}^{[F]i^{\prime }}$as in (\ref{aux01}), all computed for
d--metric (\ref{dmfstr}) are solutions of the motion equations (\ref{streq})
for any value of the dilaton field $\Phi .$ In the Section \ \ref{exsol} we
shall consider an explicit example of a string--Finsler metric.

Here it should be noted that instead of a Finsler structure, in a similar
manner, we may select from a string locally anisotropic dynamics a Lagrange
structure if the metric coefficients $g_{i^{\prime }j^{\prime }}$ are
generated by a Lagrange function $L(x,y)$ (\ref{mfl}). The N--connection may
be an arbitrary one, or of a similar Cartan form. We omit such constructions
in this paper.

\paragraph{Anholonomic Einstein--Finsler structures for arbitrary B--fields}

{\ }\newline
Locally anisotropic metrics may be generated by anholonomic frames with
associated N--connections which are not induced by some $B$--field
configurations.

For an anholonomic (pseudo) Riemannian background we consider an ansatz of
form (\ref{odm}) which by anholonomic transform can be written as an
equivalent d--metric (\ref{dmetric}). The coefficients $N_{i}^{a}$ and $%
B_{\mu \nu }$ are related only via the string motion equations (\ref{streq})
which must be satisfied by the Ricci d--tensor (\ref{dricci}) computed, for
instance, for the cannonical d--connection (\ref{dcon}).

A Finsler like structure, not induced directly by $B$--fields, may be
emphasized if the d-metric is taken in the form (\ref{dmfstr}), but the $\ $%
\ values $\delta y^{i^{\prime }}=dy^{i^{\prime }}+N_{j^{\prime }}^{i^{\prime
}}dx^{i^{\prime }}$ being elongated by some $N_{j^{\prime }}^{i^{\prime }}$
are not obligatory constrained by the conditions (\ref{aux02}). Of course,
the Finsler metric $F$ and $B_{\mu \nu }$ are not completely independent;
these fields must be chosen as to generate a solution of string--Finsler
equations (\ref{streq}).

In a similar manner we can model as some alternative low energy limits of
the string theory, with corresponding nonlinear sigma models, different
variants of spacetime geometries with anholonomic and N--connection
structures, derived on manifold or vector bundles when the metric, linear
and N--connection structures are proper for a Lagrange, generalized Lagrange
or anholonomic Riemannian geometry \cite%
{ma,finsler,bejancu,kern,vexsol,vankin,vbel,vgauge,vsingl}.

\section{Superstrings and Anisotropic Supergravity}

The bosonic string theory, from which in the low energy limits we may
generate different models of anholonomic Riemannian--Finsler gravity,
suffers from at least four major problems: 1) there are tachyonic states
which violates the physical causality and divergence of transitions
amplitudes; 2) there are not included any fermionic states transforming
under a spinor representation of the spacetime Lorentz group; 3) it is not
clear why Yang--Mills gauge particles arise in both type of closed and open
string theories and to what type of strings should be given priority; 4)
experimentally there are 4 dimensions and not 26 as in the bosonic string
theory: it must be understood why the remaining dimensions are almost
invisible.

The first three problems may be resolved by introducing certain additional
dynamical degrees of freedom on the string worldsheet which results in
fermionic string states in the physical Hibert space and modifies the
critical dimension of spacetime. \ One tries to solve the forth problem by
developing different models of compactification.

There are distinguished five, consistent, tachyon free, spacetime
supersymmetric string theories in flat Minkowski spacetime (see, for
instance, \cite{deligne,kir} for basic results and references on types I,
IIA, IIB, Heterotic $Spin(32)/Z_{2}$ and Heterotic $E_{8}\times E_{8}$
string theories). The (super) string and (super) gravity theories in
geralized Finsler like, in general, supersymmetric backgrounds provided with
N--connection structure, and corresponding anisotropic superstring
perturbation theories, were investigated in \ Refs. \cite{vsuper,vstr2,vmon1}%
. The goal of this Section is to illustrate how anholonomic type structures
arise in the low energy limits of the mentioned string theories if the
backgrounds are considered with certain anholonomic frame and off--diagonal
metric structures. We shall consider the conditions when generalized Finsler
like geometries arise in (super) string theories.

We would like to start with the example of the two--dimensional $\mathcal{N}%
=1$ supergravity coupled to the dimension 1 superfields, containing a
bosonic coordinate $X^{\mu }$ and two fermionic coordinates, one
left--moving $\psi ^{\mu }$ and one right moving $\overline{\psi }^{\mu }$
(we use the symbol $\mathcal{N}$ for the supersimmetric dimension which must
be not confused with the symbol $N$ for a N--connection structure). We note
that the two dimensional $\mathcal{N}=1$ supergravity multiplet contains the
metric and a gravitino $\chi _{A}.$In order to develop models in backgrounds
distinguished by a N--connection structure, we have to consider splittings
into h-- and v--components, i. e. to write $X^{\mu }=\left(
X^{i},X^{a}\right) $ and $\psi ^{\mu }=\left( \psi ^{i},\psi ^{a}\right) ,%
\overline{\psi }^{\mu }=\left( \overline{\psi }^{i},\overline{\psi }%
^{a}\right) .$ The spinor differential geometry on anisotropic spacetimes
provided with N--connections (in brief, d--spinor geometry) was developed in
Refs. \cite{vspinors,vmon2}. Here we shall present only the basic formulas,
emphasizing the fact that the coefficients of d--spinors have the usual
spinor properties on separated h-- (v-) subspaces.

The simplest distinguished superstring model can be developed from an analog
of the bosonic Polyakov action,%
\begin{eqnarray}
S_{P} &=&\frac{1}{4\pi \alpha ^{\prime }}\int\limits_{\Sigma }\delta \mu
_{g}\{g^{AB}\left[ \partial _{A}X^{i}\partial _{B}X^{j}g_{ij}+\partial
_{A}X^{a}\partial _{B}X^{b}h_{ab}\right]  \label{actps} \\
&&+\frac{i}{2}[\psi ^{k}\gamma ^{A}\partial _{A}\psi ^{k}+\psi ^{a}\gamma
^{A}\partial _{A}\psi ^{a}]+\frac{i}{2}\left( \chi _{A}\gamma ^{B}\gamma
^{A}\psi ^{k}\right) \left( \partial _{B}X^{k}-\frac{i}{4}\chi _{B}\psi
^{k}\right)  \notag \\
&&+\frac{i}{2}\left( \chi _{A}\gamma ^{B}\gamma ^{A}\psi ^{a}\right) \left(
\partial _{B}X^{a}-\frac{i}{4}\chi _{B}\psi ^{a}\right) \}  \notag
\end{eqnarray}%
being invariant under transforms (i. e. being $\mathcal{N}=1$ left--moving $%
\left( 1,0\right) $ supersymmetric)%
\begin{eqnarray*}
\bigtriangleup g_{AB} &=&i\epsilon \left( \gamma _{A}\chi _{B}+\gamma
_{B}\chi _{A}\right) ,~\bigtriangleup \chi _{A}=2\bigtriangledown
_{A}\epsilon , \\
\bigtriangleup X^{i} &=&i\epsilon \psi ^{i},~\bigtriangleup \psi ^{k}=\gamma
^{A}\left( \partial _{A}X^{k}-\frac{i}{2}\chi _{A}\psi ^{k}\right) \epsilon
,~\bigtriangleup \overline{\psi }^{i}=0, \\
\bigtriangleup X^{a} &=&i\epsilon \psi ^{a},~\bigtriangleup \psi ^{a}=\gamma
^{A}\left( \partial _{A}X^{a}-\frac{i}{2}\chi _{A}\psi ^{a}\right) \epsilon
,~\bigtriangleup \overline{\psi }^{a}=0,
\end{eqnarray*}%
where the gamma matrices $\gamma _{A}$ and the covariant differential
operator $\bigtriangledown _{A}$ are defined on the two dimensional surface,
$\epsilon $ is a left--moving Majorana--Weyl spinor. There is also a similar
right--moving $\left( 0,1\right) $ supersymmetry involving a right moving
Majorana--Weyl spinor $\overline{\epsilon }$ and the fermions $\overline{%
\psi }^{\mu }$ which means that the model has a $\left( 1,1\right) $
supersymmetry. The superconformal gauge for the action (\ref{actps}) is
defined as%
\begin{equation*}
g_{AB}=e^{\Phi }\delta _{AB},~\chi _{A}=\gamma _{A}\zeta ,
\end{equation*}%
for a constant Majorana spinor $\zeta .$ This action has also certain matter
like supercurents $i\psi ^{\mu }\partial X^{\mu }$ and $i\overline{\psi }%
^{\mu }\overline{\partial }X^{\mu }.$

We remark that the so--called distinguished gamma matrices (d--matrices), $%
\gamma ^{\alpha }=\left( \gamma ^{i},\gamma ^{a}\right) $ and related spinor
calculus are derived from $\gamma $--decompositions of the h-- and v--
components of d--metrics \ $g^{\alpha \beta }=\{g^{ij},h^{ab}\}$ (\ref%
{dmetric})%
\begin{equation*}
\gamma ^{i}\gamma ^{j}+\gamma ^{j}\gamma ^{i}=-2g^{ij},~\gamma ^{a}\gamma
^{b}+\gamma ^{b}\gamma ^{a}=-2h^{ab},
\end{equation*}%
see details in Refs. \cite{vspinors,vmon2}.

In the next subsections we shall distinguish more realistic superstring
actions than (\ref{actps}) following the geometric d--covariant rule
introduced in subsection \ref{dcovrule}, when the curved spacetime geometric
objects like metrics, connections, tensors, spinors, ... as well the partial
and covariant derivatives and differentials are decomposed in invariant h--
and v--components, adapted to the N--connection structure. This will allow
us to extend directly the results for superstring low energy isotropic
actions to backgrounds with local anisotropy.

\subsection{Locally anisotropic supergravity theories}

We indicate that many papers on supergravity theories in various dimensions
are reprinted in a set of two volumes \cite{salamsezgin}. The bulk of
supergravity models contain locally anisotropic configurations which can be
emphasized by some vielbein transforms (\ref{vielbtr}) and metric anzatz (%
\ref{ansatz}) with associated N--connection. For corresponding
parametrizations of the d--metric coefficients, $g_{\alpha \beta
}(u)=\{g_{ij},h_{ab}\},$ N--connection, $N_{i}^{a}(x,y),$ and d--connection,
$\Gamma _{\ \beta \gamma }^{\alpha }=\left( L_{\ jk}^{i},L_{\ bk}^{a},C_{\
jc}^{i},C_{\ bc}^{a}\right) ,$ with possible superspace generalizations, we
can generate (pseudo) Riemannian off--diagonal metrics, Finsler or Lagrange
(super) geometries. \ In this subsection, we analyze the anholonomic frame
transforms of some supergravity actions which can be coupled to superstring
theory.

We note that the field components will be organized according to multiplets
of $Spin\left( 1,10\right) .$ We shall use 10 dimensional spacetime indices $%
\alpha ,\beta ...=0,1,2,...,9$ or 11 dimensional ones $\overline{\alpha },%
\overline{\beta }...=0,1,2,...,9,10.$ The coordinate $u^{10\text{ }}$could
be considered as a compactified one, or distinguished in a non--compactified
manner, by the N--connection structure. There is a general argument \cite%
{nahm} is that 11 is the largest possible dimension in which supersymmetric
multiplets can exist with spin less, or equal to 2, with \ a single local
supersymmetry. We write this as $n+m=11,$ which points to possible
splittings of indices like $\overline{\alpha }=\left( \overline{i},\overline{%
a}\right) $ where $\overline{i}$ and $\overline{a}$ run respectively $n$ and
$m$ values. A consistent superstring theory holds if $n+m=10.$ In this case,
indices are to be decomposed as $\alpha =\left( i,a\right) .$ For
simplicity, we shall consider that a metric tensor in $n+m=11$ dimensions
decomposes as $g_{\overline{\alpha }\overline{\beta }}\left( u^{\mu
},u^{10}\right) \rightarrow g_{\overline{\alpha }\overline{\beta }}\left(
u^{\mu }\right) $ and that in low energy approximation the fields are
locally anisotropically interacting and independent on $u^{10}.$ The
antisymmetric rank 3 tensor is taken to decompose as $A_{\overline{\alpha }%
\overline{\beta }\overline{\gamma }}\left( u^{\mu },u^{10}\right)
\rightarrow A_{\overline{\alpha }\overline{\beta }\overline{\gamma }}(u^{\mu
}).$ A fitting with superstring theory is to be obtained if $\left(
A_{\alpha \beta \gamma }^{[3]},B_{\mu \nu }\right) \rightarrow A_{\overline{%
\alpha }\overline{\beta }\overline{\gamma }}$ and consider for spinors
''dilatino'' fields $\left( \chi _{\mu }^{~\tau },\lambda _{\tau }\right)
\rightarrow \chi _{\overline{\mu }}^{~\tau },$ see, for instance, Refs. \cite%
{deligne} for details on couplings of supergravity and low energy
superstrings.

\subsubsection{$\mathcal{N}=1,n+m=11$ anisotropic supergravity}

The field content of $\mathcal{N}=1$ and 11 dimensional supergravity is
given by $g_{\overline{\alpha }\overline{\beta }}$ (graviton), $A_{\overline{%
\alpha }\overline{\beta }\overline{\gamma }}$ (U(1) gauge fields) and $\chi
_{\overline{\mu }}^{\alpha }$ (gravitino). The dimensional reduction is
stated by $g_{\alpha 10}=g_{10\alpha }=A_{\alpha }^{[1]}$ and $%
g_{10~10}=e^{-2\Phi },$ where the coefficients are given with respect to an
N--elongated basis. We suppose that an effective action
\begin{equation*}
S(g_{ij},h_{ab},N_{i}^{a},B_{\mu \nu },\Phi )=\frac{1}{2\kappa ^{2}}\int
\delta \mu _{\lbrack g,h]}e^{-2\Phi }\left[ -\widehat{R}-S+4(D\Phi )^{2}-%
\frac{1}{12}H^{2}\right] ,
\end{equation*}%
is to be obtained if the values $A_{\alpha }^{[1]},A_{\alpha \beta \gamma
}^{[3]},\chi _{\mu }^{~\tau },\lambda _{\tau }$ vanish. For $N\rightarrow
0,m\rightarrow 0$ this action results from the so--called NS sector of the
superstring theory, being related to the sigma model action (\ref{act3}). A
full $\mathcal{N}=1$ and 11 dimensional locally anisotropic supergravity can
be constructed similarly to the locally isotropic case \cite{cjs} but
considering that $H^{[N]}=\delta B$ and $F^{[N]}=\delta A$ are computed as
differential forms with respect to N--elongated differentials (\ref{ddif}),
\begin{eqnarray}
S\left( g_{ij},h_{ab},N_{i}^{a},A_{\alpha },\chi \right) &=&-\frac{1}{%
2\kappa ^{2}}\int \delta \mu _{\lbrack g,h]}[\widehat{R}+S-\frac{\kappa ^{2}%
}{12}F^{2}+\kappa ^{2}\overline{\chi }_{\overline{\mu }}\Gamma ^{\overline{%
\mu }\overline{\nu }\overline{\lambda }}D_{\overline{\nu }}\chi _{\overline{%
\lambda }}  \label{act4} \\
&&+\frac{\sqrt{2}\kappa ^{3}}{384}\left( \overline{\chi }_{\overline{\mu }%
}\Gamma ^{\overline{\mu }\overline{\nu }\overline{\rho }\overline{\sigma }%
\overline{\tau }\overline{\lambda }}\overline{\chi }_{\overline{\lambda }%
}+12\Gamma ^{\overline{\rho }\overline{\nu }\overline{\sigma }}\chi ^{%
\overline{\tau }}\right) \left( F+\widehat{F}\right) _{\overline{\nu }%
\overline{\rho }\overline{\sigma }\overline{\tau }}]  \notag \\
&&-\frac{\sqrt{2}\kappa }{81\times 56}\int A\wedge F\wedge F,  \notag
\end{eqnarray}%
where $\Gamma ^{\overline{\mu }\overline{\nu }\overline{\rho }\overline{%
\sigma }\overline{\tau }\overline{\lambda }}=\Gamma ^{\lbrack \overline{\mu }%
}\Gamma ^{\overline{\nu }}...\Gamma ^{\overline{\lambda }]}$ is the standard
notation for gamma matrices for 11 dimensional spacetimes, the field $%
\widehat{F}=F+\chi $--terms and $D_{\overline{\nu }}$ is the covariant
derivative with respect to $\frac{1}{2}\left( \omega +\widehat{\omega }%
\right) $ where
\begin{equation*}
\widehat{\omega }_{\overline{\mu }\overline{\alpha }\overline{\beta }%
}=\omega _{\overline{\mu }\overline{\alpha }\overline{\beta }}+\frac{1}{8}%
\chi ^{\overline{\nu }}\Gamma _{\overline{\nu }\overline{\mu }\overline{%
\alpha }\overline{\beta }\overline{\rho }}\chi ^{\overline{\rho }}
\end{equation*}%
with $\omega _{\overline{\mu }\overline{\alpha }\overline{\beta }}$ being
the spin connection determined by its equation of motion. We put the same
coefficients in the action (\ref{act4}) as in the locally isotropic case as
to have compatibility for such limits. Every object (tensors, connections,
connections) has a N--distinguished invariant character with indices split
into h-- and v--subsets. For simplicity we omit here further decompositions
of fields with splitting of \ indices.

\subsubsection{Type IIA anisotropic supergravity}

The action for a such model can be deduced from (\ref{act4}) if $A_{\alpha
\beta \gamma }=\kappa ^{1/4}A_{\alpha \beta \gamma }^{[3]}$ and $A_{\alpha
\beta 10}=\kappa ^{-1}B_{\alpha \beta }$ with further h-- and v--
decompositions of \ indices. The bosonic part of the type IIA locally
anisotropic supergravity is described by
\begin{eqnarray}
S\left( g_{ij},h_{ab},N_{i}^{a},\Phi ,A^{(1)},A^{(3)}\right) &=&-\frac{1}{%
2\kappa ^{2}}\int \delta \mu _{\lbrack g,h]}\{e^{-2\Phi }[\widehat{R}%
+S-4(D\Phi )^{2}+\frac{1}{12}H^{2}]  \notag \\
&&+\sqrt{\kappa }G_{[A]}+\frac{\sqrt{\kappa }}{12}F^{2}-\frac{\kappa ^{-3/2}%
}{288}\int B\wedge F\wedge F\},  \label{act4b}
\end{eqnarray}%
with $G_{[A]}=\delta A^{(1)},H=\delta B$ and $F=\delta A^{(3)}.$ This action
may be written directly from the locally isotropic analogous following the
d--covariant geometric rule.

\subsubsection{Type IIB, n+m=10, $\mathcal{N}=2$ anisotropic supergravity}

In a similar manner, geometrically, for d--objects, we may compute possible
anholnomic effects from an action describing a model of locally anisotropic
supergravity with a super Yang--Mills action (the bosonic part)%
\begin{equation}
S_{IIB}=-\frac{1}{2\kappa ^{2}}\int \delta \mu _{\lbrack g,h]}e^{-2\Phi }[%
\widehat{R}+S+4(D\Phi )^{2}-\frac{1}{12}\widetilde{H}^{2}-\frac{1}{4}F_{\mu
\nu }^{\widehat{a}}F^{\widehat{a}\mu \nu }],  \label{act5}
\end{equation}%
when the super--Yang--Mills multiplet is stated by the action
\begin{equation*}
S_{YM}=\frac{1}{\kappa }\int \delta \mu _{\lbrack g,h]}e^{-2\Phi }[-\frac{1}{%
4}F_{\mu \nu }^{\widehat{a}}F^{\widehat{a}\mu \nu }-\frac{1}{2}\overline{%
\psi }^{\widehat{a}}\Gamma ^{\mu }D_{\mu }\psi ^{\widehat{a}}].
\end{equation*}%
In these actions
\begin{equation*}
A=A_{\mu }^{\widehat{a}}t^{\widehat{a}}\delta u^{\mu }
\end{equation*}%
is the gauge d--field of $E_{8}\times E_{8}$ or $Spin\left( 32\right) /Z_{2}$
group (with generators $t^{\widehat{a}}$ labeled by the index $\widehat{a}),$
having the strength
\begin{equation*}
F=\delta A+g_{F}A\wedge A=\frac{1}{2}F_{\mu \nu }^{\widehat{a}}t^{\widehat{a}%
}\delta u^{\mu }\wedge \delta u^{\nu },
\end{equation*}%
$g_{F}$ being the coupling constant, and $\psi $ is the gaugino of $%
E_{8}\times E_{8}$ or $Spin\left( 32\right) /Z_{2}$ group (details on
constructions of locally anisotropic gauge and spinor theories can be found
in Refs. \cite{vgauge,vncf,vmon1,vmon2,vnonc,vspinors}). The action with $B$%
--field \ strength in (\ref{act5}) is defined as follows
\begin{equation*}
\widetilde{H}=\delta B-\frac{\kappa }{\sqrt{2}}\omega _{CS}\left( A\right) ,
\end{equation*}%
for
\begin{equation*}
\omega _{CS}\left( A\right) =tr\left( A\wedge \delta A+\frac{2}{3}%
g_{F}A\wedge A\wedge A\right) .
\end{equation*}%
Such constructions conclude in a theory with $S_{IIB}+S_{YM}+$ fermionic
terms with anholonomies and $\mathcal{N}=1$ supersymmetry.

Finally, we emphasize that the actions for supersymmetric anholonomic models
can considered in the framework of (super) geometric formulation of
supergravities in $n+m=10$ and $11$ dimensions on superbundles provided with
N--connection structure \cite{vsuper,vstr2,vmon1}.

\subsection{Superstring effective actions and anisotropic toroidal
compactifications}

The supergravity actions presented in the previous subsection can be
included in different supersymmetric string theories which emphasize
anisotropic effects if spacetimes provided with N--connection structure are
considered. In this subsection we analyze a model with toroidal
compactification when the background is locally anisotropic. In order to
obtain four--dimensional (4D) theories, the simplest way is to make use of
the Kaluza--Klein idea: to propose a model when some of the dimensions are
curled--up into a compact manifold, the rest of dimensions living only for
non--compact manifold. Our aim is to show that in result of toroidal
compactifications the resulting 4D theory could be locally anisotropic.

The action (\ref{act5}) can be obtained also as a 10 dimensional heterotic
string effective action (in the locally isotropic variant see, for instance,
Ref. \cite{kir})
\begin{equation}
\left( \alpha ^{\prime }\right) ^{8}S_{10-n^{\prime }-m^{\prime }}=\int
\delta ^{10}u\sqrt{|g_{\alpha \beta }|}e^{-\Phi ^{\prime }}[\widehat{R}%
+S+(D\Phi ^{\prime })^{2}-\frac{1}{12}\widetilde{H}^{2}-\frac{1}{4}F_{\mu
\nu }^{\widehat{a}}F^{\widehat{a}\mu \nu }]+o\left( \alpha ^{\prime }\right)
,  \label{act6}
\end{equation}%
where we redefined $2\Phi \rightarrow \Phi ^{\prime },$ use the string
constant $\alpha ^{\prime }$ and consider the $\left( n^{\prime },m^{\prime
}\right) $ as the (holonomic, anholonomic) dimensions of the compactified
spacetime (as a particular case we can consider $n^{\prime }+m^{\prime }=4,$
or $n^{\prime }+m^{\prime }<10$ for any brane configurations. Let us use
parametrizations of indices and of vierbeinds: Greek indices $\alpha ,\beta
,...\mu ...$ run values for a 10 dimensional spacetime and split as $\alpha
=\left( \alpha ^{\prime },\widehat{\alpha }\right) ,\beta =\left( \beta
^{\prime },\widehat{\beta }\right) ,...$ when primed indices $\alpha
^{\prime },\beta ^{\prime },...\mu ^{\prime }...$ run values for
compactified spacetime and split into h- and v--components like $\alpha
^{\prime }=\left( i^{\prime },a^{\prime }\right) ,$ $\beta ^{\prime }=\left(
j^{\prime },b^{\prime }\right) ,...;$ the frame coefficients are split as
\begin{equation*}
e_{\mu }^{~\underline{\mu }}(u)=\left(
\begin{array}{cc}
e_{\alpha ^{\prime }}^{~\underline{\alpha ^{\prime }}}(u^{\beta ^{\prime }})
& A_{\alpha ^{\prime }}^{\widehat{\alpha }}(u^{\beta ^{\prime }})e_{\widehat{%
\alpha }}^{~\underline{\widehat{\alpha }}}(u^{\beta ^{\prime }}) \\
0 & e_{\widehat{\alpha }}^{~\underline{\widehat{\alpha }}}(u^{\beta ^{\prime
}})%
\end{array}%
\right)
\end{equation*}%
where $e_{\alpha ^{\prime }}^{~\underline{\alpha ^{\prime }}}(u^{\beta
^{\prime }}),$ in their turn, are taken in the form (\ref{vielbtr}),
\begin{equation}
e_{\alpha ^{\prime }}^{~\underline{\alpha ^{\prime }}}(u^{\beta ^{\prime
}})=\left(
\begin{array}{cc}
e_{i^{\prime }}^{~\underline{i^{\prime }}}(x^{j^{\prime }},y^{a^{\prime }})
& N_{i^{\prime }}^{a^{\prime }}(x^{j^{\prime }},y^{a^{\prime }})e_{a^{\prime
}}^{~\underline{a^{\prime }}}(x^{j^{\prime }},y^{a^{\prime }}) \\
0 & e_{a^{\prime }}^{~\underline{a^{\prime }}}(x^{j^{\prime }},y^{a^{\prime
}})%
\end{array}%
\right) .  \label{frame8}
\end{equation}%
For the metric we have the recurrent ansatz%
\begin{equation*}
\underline{g}_{\alpha \beta }=\left[
\begin{array}{cc}
g_{\alpha ^{\prime }\beta ^{\prime }}(u^{\beta ^{\prime }})+N_{\alpha
^{\prime }}^{\widehat{\alpha }}(u^{\beta ^{\prime }})N_{\beta ^{\prime }}^{%
\widehat{\beta }}(u^{\beta ^{\prime }})h_{\widehat{\alpha }\widehat{\beta }%
}(u^{\beta ^{\prime }}) & h_{\widehat{\alpha }\widehat{\beta }}(u^{\beta
^{\prime }})N_{\alpha ^{\prime }}^{\widehat{\alpha }}(u^{\beta ^{\prime }})
\\
h_{\widehat{\alpha }\widehat{\beta }}(u^{\beta ^{\prime }})N_{\beta ^{\prime
}}^{\widehat{\beta }}(u^{\beta ^{\prime }}) & h_{\widehat{\alpha }\widehat{%
\beta }}(u^{\beta ^{\prime }})%
\end{array}%
\right] .
\end{equation*}%
where%
\begin{equation}
g_{\alpha ^{\prime }\beta ^{\prime }}=\left[
\begin{array}{cc}
g_{i^{\prime }j^{\prime }}(u^{\beta ^{\prime }})+N_{i^{\prime }}^{a^{\prime
}}(u^{\beta ^{\prime }})N_{j^{\prime }}^{b^{\prime }}(u^{\beta ^{\prime
}})h_{a^{\prime }b^{\prime }}(u^{\beta ^{\prime }}) & h_{a^{\prime
}b^{\prime }}(u^{\beta ^{\prime }})N_{i^{\prime }}^{a^{\prime }}(u^{\beta
^{\prime }}) \\
h_{a^{\prime }b^{\prime }}(u^{\beta ^{\prime }})N_{j^{\prime }}^{b^{\prime
}}(u^{\beta ^{\prime }}) & h_{a^{\prime }b^{\prime }}(u^{\beta ^{\prime }})%
\end{array}%
\right] .  \label{metr8}
\end{equation}%
The part of action (\ref{act6}) containing the gravity and dilaton terms
becomes%
\begin{eqnarray}
\left( \alpha ^{\prime }\right) ^{n^{\prime }+m^{\prime }}S_{n^{\prime
}+m^{\prime }}^{heterotic} &=&\int \delta ^{n^{\prime }+m^{\prime }}u\sqrt{%
|g_{\alpha \beta }|}e^{-\phi }[\widehat{R}^{\prime }+S^{\prime }+(\delta
_{\mu ^{\prime }}\phi )(\delta ^{\mu ^{\prime }}\phi )  \label{rterm} \\
&&+\frac{1}{4}(\delta _{\mu ^{\prime }}h_{\widehat{\alpha }\widehat{\beta }%
})(\delta ^{\mu ^{\prime }}h^{\widehat{\alpha }\widehat{\beta }})-\frac{1}{4}%
h_{\widehat{\alpha }\widehat{\beta }}F_{\mu ^{\prime }\nu ^{\prime }}^{[A]%
\widehat{\alpha }}F^{[A]\widehat{\beta }\mu ^{\prime }\nu ^{\prime }}],
\notag
\end{eqnarray}%
where $\phi =\Phi ^{\prime }-\frac{1}{2}\log \left( \det |h_{\widehat{\alpha
}\widehat{\beta }}|\right) $ and $F_{\mu ^{\prime }\nu ^{\prime }}^{[A]%
\widehat{\alpha }}=\delta _{\mu ^{\prime }}A_{\nu ^{\prime }}^{\widehat{%
\alpha }}-\delta _{\nu ^{\prime }}A_{\mu ^{\prime }}^{\widehat{\alpha }}$
and the h- and v--components of the induced scalar curvature, respectively, $%
\widehat{R}^{\prime }$ and $S^{\prime }$ (see formula (\ref{dscalar}) in
Appendix) are primed in order to point that these values are for the lower
dimensional space. The \ antisymmetric tensor part may be decomposed in the
form%
\begin{eqnarray}
-\frac{1}{12}\int \delta ^{10}u\sqrt{|g_{\alpha \beta }|}e^{-\Phi ^{\prime
}}H^{\mu \nu \rho }H_{\mu \nu \rho } &=&-\frac{1}{4}\int \delta ^{n^{\prime
}+m^{\prime }}u\sqrt{|g_{\alpha ^{\prime }\beta ^{\prime }}|}e^{-\phi }\times
\label{hterm} \\
&&[H^{\mu ^{\prime }\widehat{\alpha }\widehat{\beta }}H_{\mu ^{\prime }%
\widehat{\alpha }\widehat{\beta }}+H^{\mu ^{\prime }\nu ^{\prime }\widehat{%
\beta }}H_{\mu ^{\prime }\nu ^{\prime }\widehat{\beta }}+\frac{1}{3}H^{\mu
^{\prime }\nu ^{\prime }\rho ^{\prime }}H_{\mu ^{\prime }\nu ^{\prime }\rho
^{\prime }}],  \notag
\end{eqnarray}%
where, for instance,
\begin{equation*}
H_{\mu ^{\prime }\widehat{\alpha }\widehat{\beta }}=e_{\mu ^{\prime }}^{~%
\underline{\mu ^{\prime }}}e_{\underline{\mu ^{\prime }}}^{\mu }H_{\mu
\widehat{\alpha }\widehat{\beta }}
\end{equation*}%
and we have considered $H_{\widehat{\alpha }\widehat{\beta }\widehat{\gamma }%
}=0.$ In a similar manner we can decompose the action for gauge fields $%
\widehat{A}_{\mu }^{I}$ with index $I=1,...,32,$%
\begin{equation}
\int \delta ^{10}u\sqrt{|g_{\alpha \beta }|}e^{-\Phi ^{\prime
}}\sum\limits_{I=1}^{16}\widehat{F}^{I,\mu \nu }\widehat{F}_{\mu \nu
}^{I}=\int \delta ^{n^{\prime }+m^{\prime }}u\sqrt{|g_{\alpha ^{\prime
}\beta ^{\prime }}|}e^{-\phi }~\sum\limits_{I=1}^{16}[\widehat{F}^{I,\mu
^{\prime }\nu ^{\prime }}\widehat{F}_{\mu ^{\prime }\nu ^{\prime }}^{I}+2%
\widehat{F}^{I,\mu ^{\prime }\widehat{\nu }}\widehat{F}_{\mu ^{\prime }%
\widehat{\nu }}^{I}],  \label{fterm}
\end{equation}%
with
\begin{eqnarray*}
Y_{\widehat{\alpha }}^{I} &=&A_{\widehat{\alpha }}^{I},~A_{\alpha ^{\prime
}}^{I}=\widehat{A}_{\alpha ^{\prime }}^{I}-Y_{\widehat{\alpha }}^{I}A_{\mu
}^{\widehat{\alpha }},~ \\
\widehat{F}_{\mu ^{\prime }\nu ^{\prime }}^{I} &=&F_{\mu ^{\prime }\nu
^{\prime }}^{I}+Y_{\widehat{\alpha }}^{I}F_{\mu ^{\prime }\nu ^{\prime
}}^{[A]\widehat{\alpha }},~\widehat{F}_{\mu ^{\prime }\widehat{\nu }%
}^{I}=\delta _{\mu ^{\prime }}Y_{\widehat{\alpha }}^{I},~\widehat{F}_{\mu
^{\prime }\nu ^{\prime }}^{I}=\delta _{\mu ^{\prime }}A_{\nu ^{\prime
}}^{I}-\delta _{\nu ^{\prime }}A_{\mu ^{\prime }}^{I},
\end{eqnarray*}%
where the scalars $Y_{\widehat{\alpha }}^{I}$ coming from the
ten--dimensional vectors should be associated to a normal Higgs phenomenon
generating a mass matrix for the gauge fields. Thy are related to the fact
that a non--Abelian gauge field strength contains nonlinear terms not being
certain derivatives of potentials.

After a straighforword calculus of the actions' components (\ref{rterm}), (%
\ref{hterm}) and (\ref{fterm}) (for locally isotropic gauge theories and
strings, see a similar calculus, for instance, in Refs. \cite{kir}), putting
everything together, we can write the $n^{\prime }+m^{\prime }$ dimensional
action including anholonomic interactions in the form%
\begin{eqnarray}
S_{n^{\prime }+m^{\prime }}^{heterotic} &=&\int \delta ^{n^{\prime
}+m^{\prime }}u\sqrt{|g_{\alpha ^{\prime }\beta ^{\prime }}|}e^{-\phi }[%
\widehat{R}^{\prime }+S^{\prime }+(\delta _{\mu ^{\prime }}\phi )(\delta
^{\mu ^{\prime }}\phi )-\frac{1}{12}H^{\mu ^{\prime }\nu ^{\prime }\rho
^{\prime }}H_{\mu ^{\prime }\nu ^{\prime }\rho ^{\prime }}  \notag \\
&&-\frac{1}{4}(M^{-1})_{\overline{I}\overline{J}}F_{\mu ^{\prime }\nu
^{\prime }}^{\overline{I}}F^{\overline{J}\mu ^{\prime }\nu ^{\prime }}+\frac{%
1}{8}Tr\left( \delta _{\mu ^{\prime }}M\delta ^{\mu ^{\prime }}M^{-1}\right)
],  \label{act7}
\end{eqnarray}%
where ${\overleftarrow{R}}^{\prime }{=}\widehat{R}^{\prime }+S^{\prime }$ is
the d--scalar curvature of type (\ref{dscalar}) induced after toroidal
compactification, the $\left( 2p+16\right) \times \left( 2p+16\right) $
dimensional symmetric matrix $M$ has the structure%
\begin{equation*}
M=\left(
\begin{array}{ccc}
\underline{g}^{-1} & \underline{g}^{-1}C & \underline{g}^{-1}Y^{t} \\
C^{t}\underline{g}^{-1} & \underline{g}+C^{t}\underline{g}^{-1}C+Y^{t}Y &
C^{t}\underline{g}^{-1}Y^{t}+Y^{t} \\
Y\underline{g}^{-1} & Y\underline{g}^{-1}C+Y & I_{16}+Y\underline{g}%
^{-1}Y^{t}%
\end{array}%
\right)
\end{equation*}%
with the block sub--matrices
\begin{equation*}
\underline{g}=\left( \underline{g}_{\alpha \beta }\right) ,C=\left( C_{%
\widehat{\alpha }\widehat{\beta }}=B_{\widehat{\alpha }\widehat{\beta }}-%
\frac{1}{2}Y_{\widehat{\alpha }}^{I}Y_{\widehat{\beta }}^{I}\right)
,Y=\left( Y_{\widehat{\alpha }}^{I}\right) ,
\end{equation*}%
for which $I_{16}$ is the 16 dimensional unit matrix; for instance, $Y^{t}$
denotes the transposition of the matrix $Y.$ The dimension $p$ satisfies the
condition $n^{\prime }+m^{\prime }-p=16$ relevant to the heterotic string
describing $p$ left--moving bosons and $n^{\prime }+m^{\prime }$
right--moving ones with $m^{\prime }$ constrained degrees of freedom. To
have good modular properties $p-n^{\prime }-m^{\prime }$ should be a
multiple of eight. The indices $\overline{I},\overline{J}$ run values $%
1,2,...\left( 2p+16\right) .$ The action (\ref{act7}) describes a heterotic
string effective action with local anisotropies (contained in the values $%
\widehat{R}^{\prime },S^{\prime }$ and $\delta _{\mu ^{\prime
}})$ induced by the fact that the dynamics of the right--moving
bosons are subjected to certain constraints. The induced metric
$g_{\alpha ^{\prime }\beta ^{\prime }}$ is of type
(\ref{dmetric}) given with respect  to an N--elongated basis
(\ref{ddif}) (in this case,  primed), $\delta _{\mu ^{\prime
}}=\partial _{\mu ^{\prime }}+N_{\mu ^{\prime }}.$ For
$N_{i^{\prime }}^{a^{\prime }}\rightarrow 0$ and $m^{\prime },$
i. e. for a subclass of effective backgrounds with block
$n^{\prime }\times n^{\prime }\oplus m^{\prime }\times m^{\prime
}$ metrics $g_{\alpha ^{\prime }\beta ^{\prime }},$ the action
(\ref{act7}) transforms in the well known isotropic form (see,
for instance, formula (C22), from the Appendix C in Ref.
\cite{kir}, from which
following the 'geometric d--covariant rule' we could write down directly (%
\ref{act7}); this is a more formal approach which hides the physical meaning
and anholonomic character of the components (\ref{rterm}), (\ref{hterm}) and
(\ref{fterm})).

\subsection{4D NS--NS anholonomic field equations}

As a matter of principle, compactifications of all type in (super) string
theory can be performed in such ways as to include anholonomic frame effects
as in the previous subsection. The simplest way to define anisotropic
generalizations or such models is to apply the 'geometric d--covariant rule'
when the tensors, spinors and connections are changed into theirs
corresponding N--distinguished analogous. As an example, we write down here
the anholonomic variant of the toroidally compactified (from ten to four
dimensions) NS--NS action (we write in brief NS instead of Neveu--Schwarz) %
\cite{cosm},%
\begin{equation}
S=\int \delta ^{4}u\sqrt{|g_{\alpha ^{\prime }\beta ^{\prime }}|}e^{-\varphi
}[\widehat{R}^{\prime }+S^{\prime }+(\delta _{\mu ^{\prime }}\phi )(\delta
^{\mu ^{\prime }}\phi )-\frac{1}{2}(\delta _{\mu ^{\prime }}\beta )(\delta
^{\mu ^{\prime }}\beta )-\frac{1}{2}e^{2\varphi }(\delta _{\mu ^{\prime
}}\sigma )(\delta ^{\mu ^{\prime }}\sigma )],  \label{act9}
\end{equation}%
for a d--metric parametrized as
\begin{equation*}
\delta s^{2}=-\epsilon \delta (x^{0^{\prime }})^{2}+g_{\underline{\alpha
^{\prime }}\underline{\beta ^{\prime }}}\delta u^{\underline{\alpha ^{\prime
}}}\delta u^{\underline{\beta ^{\prime }}}+e^{\beta /\sqrt{3}}\delta _{%
\widehat{\alpha }\widehat{\gamma }}\delta u^{\widehat{\alpha }}\delta u^{%
\widehat{\beta }},
\end{equation*}%
where, for instance, $u^{\alpha ^{\prime }}=(x^{0^{\prime }},u^{\underline{%
\alpha ^{\prime }}}),\underline{\alpha ^{\prime }}=1,2,3$ and $\widehat{%
\alpha },\widehat{\beta },...=4,5,...9$ are indices of extra dimension
coordinates, $\epsilon =\pm 1$ depending on signature (in usual string
theory one takes $x^{0^{\prime }}=t$ and $\epsilon =-1),$ the modulus field $%
\beta $ is normalized in such a way that it becomes minimally coupled to
gravity in the Einstein d--frame, $\sigma $ is a pseudo--scalar axion
d--field, related with the anti--symmetric strength,%
\begin{equation*}
H^{\alpha ^{\prime }\beta ^{\prime }\gamma ^{\prime }}(u^{\alpha ^{\prime
}})=\varepsilon ^{\alpha ^{\prime }\beta ^{\prime }\gamma ^{\prime }\tau
^{\prime }}e^{\varphi (u^{\alpha ^{\prime }})}D_{\tau ^{\prime }}\sigma
(u^{\alpha ^{\prime }}),
\end{equation*}%
$\varepsilon ^{\alpha ^{\prime }\beta ^{\prime }\gamma ^{\prime }\tau
^{\prime }}$ being completely antisymmetric and $\varphi (u^{\alpha ^{\prime
}})=\Phi ^{\prime }(u^{\alpha ^{\prime }})-\sqrt{3}\beta (u^{\alpha ^{\prime
}}),$ with $\Phi ^{\prime }(u^{\alpha ^{\prime }})$ taken as in (\ref{act6}).

We can derive certain locally anisotropic field equations from the action (%
\ref{act9}) by varying with respect to N--adapted frames for massless
exitations of $g_{\alpha ^{\prime }\beta ^{\prime }},B_{\alpha ^{\prime
}\beta ^{\prime }},\beta $ and $\varphi ,$ which are given by%
\begin{eqnarray}
2\left[ R_{\mu ^{\prime }\nu ^{\prime }}-\frac{1}{2}\left( \widehat{R}%
^{\prime }+S^{\prime }\right) g_{\mu ^{\prime }\nu ^{\prime }}\right] =\frac{%
1}{2}H_{\mu ^{\prime }\lambda ^{\prime }\tau ^{\prime }}H_{\nu ^{\prime
}}^{~~\lambda ^{\prime }\tau ^{\prime }}-H^{2}g_{\mu ^{\prime }\nu ^{\prime
}}+ &&  \label{eqfstr} \\
\left( \delta _{\mu ^{\prime }}^{\lambda ^{\prime }}\delta _{\nu ^{\prime
}}^{\tau ^{\prime }}-\frac{1}{2}g_{\mu ^{\prime }\nu ^{\prime }}g^{\lambda
^{\prime }\tau ^{\prime }}\right) D_{\lambda ^{\prime }}\beta D_{\tau
^{\prime }}\beta -g_{\mu ^{\prime }\nu ^{\prime }}(D\varphi )^{2}+2\left(
g_{\mu ^{\prime }\nu ^{\prime }}g^{\lambda ^{\prime }\tau ^{\prime }}-\delta
_{\mu ^{\prime }}^{\lambda ^{\prime }}\delta _{\nu ^{\prime }}^{\tau
^{\prime }}\right) D_{\lambda ^{\prime }}D_{\tau ^{\prime }}\varphi &=&0,
\notag \\
D_{\mu ^{\prime }}\left( e^{-\varphi }H^{\mu ^{\prime }\nu ^{\prime }\lambda
^{\prime }}\right) &=&0,  \notag \\
D_{\mu ^{\prime }}\left( e^{-\varphi }D^{\mu ^{\prime }}\beta \right) &=&0,
\notag \\
2D_{\mu ^{\prime }}D^{\mu ^{\prime }}\varphi =-\widehat{R}^{\prime
}-S^{\prime }+(D\varphi )^{2}+\frac{1}{2}(D\beta )^{2}+\frac{1}{12}H^{2}
&=&0,  \notag
\end{eqnarray}%
where $H^{2}=H_{\mu ^{\prime }\lambda ^{\prime }\tau ^{\prime }}H^{~\mu
^{\prime }~\lambda ^{\prime }\tau ^{\prime }}$ and, for instance, $(D\varphi
)^{2}=D_{\mu ^{\prime }}\varphi D^{\mu ^{\prime }}\varphi .$ We may select a
consistent solution of these field equations when the internal space is
static with $D_{\mu ^{\prime }}\beta =0.$

The equations (\ref{eqfstr}) can be decomposed in invariant h--
and v--components like the Einstein d--equations (\ref{einsteq2})
(we omit a such trivial calculus). We recall \cite{deligne} that
the NS--NS sector is common to both the heterotic \ and type II
string theories and is comprised of the dilaton, graviton and
antisymmetric two--form potential. The obtained equations
(\ref{eqfstr}) \ define respective anisotropic string corrections
to the anholonomic Einstein gravity.

\subsection{Distinguishing anholonomic Riemannian--Finsler \newline
(super\-) gra\-vities}

There are two classes of general anisotropies contained in supergravity and
superstring effective actions:

\begin{itemize}
\item Generic local anisotropies contained in the higher dimension (11, for
supergravity models, or 10, for superstring models) which can be also
induced in lower dimension after compactification (like it was considered
for actions (\ref{act4}), (\ref{act4b}), (\ref{act5}) and (\ref{act6})).

\item Local anisotropies which are in induced on the lower dimensional
spacetime (for instance, actions (\ref{act7}) and (\ref{act9}) and
respective field equations).
\end{itemize}

All types of general supergravity/superstring anisotropies may be in their
turn to be distinguished to be of ''pure'' $B$--field origin, of ''pure''
anholonomic frame origin with arbitrary $B$--field, or of a mixed type when
local anisotropies are both induced in a nonlinear form by both anholonomic
(super) vielbeins and $B$--field (like we considered in subsection \ref{efs}
for bosonic strings). In explicit form, a model of locally anisotropic
superstring corrected gravity is to be constructed following the type of
parametrizations we establish for the N--coefficients, d--metrics and
d--connections.

For instance, if we choose the frame ansatz (\ref{frame8}) and corresponding
metric ansatz (\ref{metr8}) with general coefficients $g_{i^{\prime
}j^{\prime }}(x^{j^{\prime }},y^{c^{\prime }}),h_{a^{\prime }b^{\prime
}}(x^{j^{\prime }},y^{a^{\prime }})$ and $N_{i^{\prime }}^{a^{\prime
}}(x^{j^{\prime }},y^{a^{\prime }})$ satisfying the effective field
equations (\ref{eqfstr}) (containing also the fields $H_{\mu ^{\prime
}\lambda ^{\prime }\tau ^{\prime }},\varphi $ and $\beta )$ we define an
anholonomic gravity model corrected by toroidally compactified (from ten to
four dimensions) NS--NS \ superstring model. In four and five dimensional
Einstein/ Kaluza--Klein gravities, there were constructed a number of
anisotropic black hole, wormhole, solitonic, spinor waive and Taub/NUT
metrics \cite{vexsol,vmethod,vbel,vsingl,vsolsp}; in section \ref{exsol} we
shall consider some generalizations to string gravity.

Another possibility is to impose the condition that $g_{i^{\prime }j^{\prime
}},h_{a^{\prime }b^{\prime }}$ and $N_{i^{\prime }}^{a^{\prime }}$ are of
Finsler type, $g_{i^{\prime }j^{\prime }}^{[F]}=h_{i^{\prime }j^{\prime
}}^{[F]}=\partial ^{2}F^{2}/2\partial y^{i}\partial y^{j}$ (\ref{fmetric})
and $N_{j}^{[F]i}(x,y)=\partial \left[ c_{lk}^{\iota }(x,y)y^{l}y^{k}\right]
/$ $4\partial y^{j}$ (\ref{ncc}), with an effective d--metric (\ref{dmetricf}%
). If a such set of metric/N--connection coefficients can \ found as a
solution of some string gravity equations, we may construct a lower
dimensional Finsler gravity model induced from string theory (it depends of
what kind of effective action, (\ref{act7}) or (\ref{act9}), we consider).
Instead of a Finsler gravity we may search for a Lagrange model of string
gravity if the d--metric coefficients are taken in the form (\ref{mfl}).

We conclude this section by a remark that we may construct various type of
anholonomic Riemannian and generalized Finsler/Lagrange string gravity
models, with anisotropies in higher and/or lower dimensions by prescribing
corresponding parametrizations for $g_{ij},h_{ab}$ and $N_{i}^{a}$ (for
'higher' anisotropies) and $g_{i^{\prime }j^{\prime }},h_{a^{\prime
}b^{\prime }}$ and $N_{i^{\prime }}^{a^{\prime }}$ (for 'lower'
anisotropies). The anholonomic structures may be of mixed type, for
instance, in some dimensions being of Finsler configuration, in another ones
being with anholonomic Riemannian metric, in another one of Lagrange type
and different combinations and generalizations, see explicit examples in
Section \ref{exsol}.

\section{ Noncommutative Anisotropic Field Interactions}

We define the noncommutative field theory in a new form when spacetimes and
configuration spaces are provided with some anholonomic frame and associated
N--connection structures. The equations of motions are derived from
functional integrals in a usual manner but considering N--elongated partial
derivatives and differentials.

\subsection{Basic definitions and conventions}

The basic concepts on noncommutative geometry are outlined here in a
somewhat pedestrian way by emphasizing anholonomic structures. More rigorous
approaches on mathematical aspects of noncommutative geometry may be found
in Refs. \ \cite{nc,majid,qg,landi}, physical versions are given in Refs. %
\cite{dn,strncg,connes1,sw} \ (the review \cite{vncf} is a synthesis of
results on noncommutative geometry, N--connections and Finsler geometry,
Clifford structures and anholonomic gauge gravity \ based on monographs \ %
\cite{landi,vmon1,vmon2,ma}).

As a fundamental ingredient we use an associative, in general,
noncommutative algebra $\mathcal{A}$ with a product of some elements $a,b\in
\mathcal{A}$ denoted $ab=a\cdot b,$ or in the conotation to noncommutative
spaces, written as a ''star'' product $ab=a\star b.$ Every element $a\in
\mathcal{A}$ corresponds to a configuration of a classical complex scalar
field on a ''space'' $M,$ a topological manifold, which (in our approach)
can be enabled with a N--connection structure. This associated
noncommutative algebra generalize the algebra of complex valued functions $%
\mathcal{C}(M)$ on a manifold $M$ (for different theories we may consider
instead $M$ a tangent bundle $TM,$ or a vector bundle $E\left( M\right) ).$
We consider that all functions referring to the algebra $\mathcal{A},$ $\ $%
denoted as $\mathcal{A}\left( M\right) ,$ arising in resonable physical
considerations are of necessary class (continuous, smooth, subjected to
certain bounded conditions etc.).

\subsubsection{Matrix algebras and noncommutativity}

As the most elementary examples of noncommutative algebras, which are
largely applied in quantum field theory and noncommutative geometry, one
considers the algebra $Mat_{k}(\C)$ of complex $k\times k$ matrices and the
algebra $Mat_{k}\left( \mathcal{C}(M)\right) $ of $k\times k$ matrices whose
matrix elements are elements of $\mathcal{C}(M).$ The last algebra may be
also defined as a tensor product,
\begin{equation*}
Mat_{k}\left( \mathcal{C}(M)\right) =Mat_{k}(\C)\otimes \mathcal{C}(M).
\end{equation*}%
The last construction is easy to be generalized for arbitrary noncommutative
algebra $\mathcal{A}$ as
\begin{equation*}
Mat_{k}\left( \mathcal{A}\right) =Mat_{k}(\C)\otimes \mathcal{A},
\end{equation*}
which is just the algebra of $k\times k$ matrices with elements in $\mathcal{%
A}.$ The algebra $Mat_{k}\left( \mathcal{A}\right) $ admits an authomorphism
group $GL(k,\C)$ with the action defined as $a\rightarrow \varsigma
^{-1}a\varsigma ,$ for $a\in \mathcal{A},\varsigma \in GL(k,\C).$ One
considers the subgroup $U\left( k\right) \subset GL(k,\C)$ which is
preserved by hermitian conjugations, $a\rightarrow a^{+},$ and reality
conditions, $a=a^{+}.$ To define the hermitian conjugation, for which the
hermitian matrices $a=a^{+}$ have real eigenvalues, it is considered that $%
\left( a^{+}\right) ^{+}=a$ and $\left( ca\right) ^{+}=c^{\ast }a^{+},$ for $%
c\in \C $ and $c^{\ast }$ being the complex conjugated element of $c,$ i. e.
it defined an antiholomorphic involution.

\subsubsection{Noncommutative Euclidean space $\R_{\protect\theta }^{k}$}

Another simple example of a noncommutative space is the 'noncommutative
Euclidean space' $\R_{\theta }^{k}$ defined by all complex linear
combinations of products of variables $x=\{x^{j}\}$ satisfying%
\begin{equation}
\lbrack x^{j},x^{l}]=x^{j}x^{l}-x^{l}x^{j}=i\theta ^{jl},  \label{nceucl}
\end{equation}%
where $i$ is the complex 'imaginary' unity and $\theta ^{jl}$ are real
constants treated as some noncommutative parameters or a ''Poison tensor''
by analogy to the Poison bracket in quantum mechanics where the commutator $%
\left[ ...\right] $ of hermitian operators is antihermitian. A set of
partial derivatives $\partial _{j}=\partial /\partial x^{i}$ on $\R_{\theta
}^{k}$ can be defined by postulating the relations
\begin{eqnarray}
\partial _{j}x^{n} &=&\delta _{j}^{n},  \notag \\
\lbrack \partial _{j},\partial _{n}] &=&-i\Xi _{jn}  \label{nceucder}
\end{eqnarray}%
where $\Xi _{jn}$ may be zero, but in general is non--trivial if we wont to
incorporate some additional magentic fields or anholonomic relations. A
simplified noncommutative differential calculus can be constructed if $\Xi
_{jn}=-\left( \theta ^{-1}\right) _{jn}.$

The metric structure on $\R_{\theta }^{k}$ is stated by a constant symmetric
tensor $\eta _{nj}$ for which $\partial _{j}\eta _{nj}=0.$

Infinitesimal translations $x^{j}\rightarrow x^{j}+a^{j}$ on $\R_{\theta
}^{k}$ are defined as actions on functions $\varphi $ of type $%
\bigtriangleup \varphi =a^{j}\partial _{j}\varphi .$ Because the coordinates
are noncommuting there are formally defined inner derivations as%
\begin{equation}
\partial _{j}\varphi =\left[ -i\left( \theta ^{-1}\right) _{jn}x^{n},\varphi %
\right]  \label{inder}
\end{equation}%
which result in exponented global tanslations%
\begin{equation*}
\varphi \left( x^{j}+\epsilon ^{j}\right) =e^{-i\theta _{lj}\epsilon
^{l}x^{j}}\varphi \left( x^{j}\right) e^{i\theta _{lj}\epsilon ^{l}x^{j}}.
\end{equation*}

In order to understand the symmetries of the space $\R_{\theta }^{k}$ it is
better to write the metric and Poisson tensor in the forms%
\begin{eqnarray}
ds^{2} &=&\sum_{A=1}^{r}dz_{A}d\overline{z}_{A}+\sum_{B}dy_{B}^{2},
\label{strnce} \\
&=&dq_{A}^{2}+dp_{A}^{2}+dy_{B}^{2};  \notag \\
&&  \notag \\
\theta &=&\frac{1}{2}\sum_{A=1}\theta _{A}\partial _{z_{A}}\wedge \partial _{%
\overline{z}_{A}},~\theta _{A}>0,  \notag
\end{eqnarray}%
where $z_{A}=q_{A}+ip_{A}$ $\ $\ and $\ \overline{z}_{A}=q_{A}-ip_{A}$ are
some convenient complex coordinates for which there are satisfied the
commutation rules%
\begin{eqnarray*}
\left[ y_{A},y_{B}\right] &=&\left[ y_{B},q_{A}\right] =\left[ y_{B},p_{A}%
\right] =0, \\
\left[ q_{A},p_{B}\right] &=&i\theta _{A}\delta _{AB}.
\end{eqnarray*}%
Now, it is obvious that for fixed types of metric and Poisson structures (%
\ref{strnce}) there are two symmetry groups on $\R_{\theta }^{k},$ the group
of rotations, denoted $O\left( k\right) ,$ and the group of invariance of
the form $\theta ,$ denoted $Sp\left( 2r\right) .$

\subsubsection{The noncommutative derivative and integral}

In order to elaborate noncommutative field theories in terms of an
associative noncommutative algebra $\mathcal{A},$ additionally to the
derivatives $\partial _{j}$ we need an integral $\int Tr$ which following
the examples of noncommutative matrix spaces must contain also the ''trace''
operator. In this case we can not separate the notations of trace and
integral.

It should be noted here that the role of derivative $\partial _{j}$ can be
played by any sets of elements $d_{j}\in \mathcal{A}$ which some formal
derivatives as $\partial _{j}A=\left[ d_{j},A\right] ,$ for $A\in $ $%
\mathcal{A};$ derivations written in this form are called as inner
derivations while those which can not written in this form are referred to
as outer derivations.

The general derivation and integration operations are defined as some
general dual linear operators satisfying certain formal properties: 1) the
Leibnitz rule of the derivative, $\partial _{j}(AB)=\partial
_{j}(A)B+A(\partial _{j}B);$ 2) the integral of the trace of a total
derivative is zero, $\int Tr\partial _{j}A=0;$ 3) the integral of the trace
of a commutator is zero, $\int Tr[A,B]=0,$ for any $A,B\in $ $\mathcal{A}.$
For some particular classes of functions in some noncommutative models the
condition 2) and/or 3) may be violated, see details and discussion in Ref. %
\cite{dn}.

Given a noncommutative space induced by some relations (\ref{nceucl}), the
algebra of functions on $\R^{k}$ is deformed on $\R_{\theta }^{k}$ such that%
\begin{eqnarray}
f\left( x\right) \star \varphi \left( x\right) &=&e^{\frac{i}{2}\theta ^{jk}%
\frac{\partial }{\partial \xi ^{j}}\frac{\partial }{\partial \zeta ^{k}}%
}f\left( x+\xi \right) \varphi \left( x+\zeta \right) _{|\xi =\varsigma =0}
\notag \\
&=&f\varphi +\frac{i}{2}\theta ^{jk}\partial _{j}f\partial _{k}\varphi
+o\left( \theta ^{2}\right) ,  \label{moyal}
\end{eqnarray}%
which define the Moyal bracket (product), or star product ($\star $%
--product), of functions which is associative compatible with integration in
the sense that for matrix valued functions $f$ and $\varphi $ that vanish
rapidly enough at infinity we can integrate by parts in the integrals
\begin{equation*}
\int Tr~f\star \varphi =\int Tr~\varphi \star f.
\end{equation*}

In a more rigorous operator form the star multiplication is defined by
considering a space $M_{\theta },$ locally covered by coordinate carts with
noncommutative coordinates (\ref{nceucl}), and choosing a linear map $S$
from $M_{\theta }$ to $\mathcal{C}(M),$ called the ''symbol'' of the
operator, when $\widehat{f}$ $\rightarrow S\left[ \widehat{f}\right] .$ This
way, the original operator multiplication is expressed in terms of the star
product of symbols as
\begin{equation*}
\widehat{f}\widehat{\varphi }=S^{-1}\left[ S\left[ \widehat{f}\right] \star S%
\left[ \widehat{\varphi }\right] \right] .
\end{equation*}%
It should be noted that there could be many valid definitions of $S,$
corresponding to different choices of operator ordering prescription for $%
S^{-1}.$ One writes, for simplicity, $\int Tr~f\star \varphi =\int
Tr~f\varphi $ in some special cases.

\subsection{Anholonomic frames and noncommutative spacetimes}

One may consider that noncommutative relations for coordinates and
partial derivatives (\ref{nceucl}) and (\ref{nceucder}) are
introduced by specific form of anholonomic relations (\ref{anhol})
for some formal anholonomic frames of type (\ref{dder}) and/or
(\ref{ddif}) (see Appendix) when anholonomy coefficients are
complex and depend nonlinearly on frame coefficients. We shall
not consider in this work the method of complex nonlinear operator
anholonomic frames with associated nonlinear connection structure,
containing as particular cases various type of Finsler/Cartan and
Lagrange/Hamilton geometries in complexified form, which could
consist in a general complex geometric formalism for
noncommutative theories but we shall restrict our analysis to
noncommutative spaces for which the coordinates and partial
derivatives are distinguished by a N--connection structure into
certain holonomic and anholonomic subsets which generalize the
N--elongated \ commutative differential calculus (considered in
the previous Sections) to a variant of both $N$-- and $\theta
$--deformed one.

In order to emphasize the N--connection structure on respective spaces we
shall write $M_{\theta }^{N},TM_{\theta }^{N},E_{\theta }^{M}\left(
M_{\theta }\right) ,$ $\mathcal{C}(M^{N}),\mathcal{A}^{N}$ and $\mathcal{A}%
\left( M^{N}\right) .$ For a space $M^{N}$ provided with N--connection
structure, the matrix algebras considered in the previous subsection may be
denoted $Mat_{k}\left( \mathcal{C}(M^{N})\right) $ and $Mat_{k}\left(
\mathcal{A}^{N}\right) .$

\subsubsection{Noncommutative anholonomic derivatives}

\label{torus}We introduce splitting of indices, $\alpha =\left( i,a\right) ,$
$\beta =\left( j,b\right) ,...,$ and coordinates, $u^{\alpha }=\left(
x^{i},y^{a}\right) ,...,$ into 'horizontal' and 'vertical' components for a
space $M_{\theta }$ (being in general a manifold, tangeng/vector bundle, or
their duals, or higher order models \cite{miron,vspinors,vstr2,vmon1,vmon2}.
\ The derivatives $\partial _{i}$ satisfying the conditions (\ref{nceucder})
must be changed into some N--elongated ones if both anholnomy and
noncommutative structures are introduced into consideration.

In explicit form, the anholonomic analogous of (\ref{nceucl}) is stated by a
set of coordinates $u^{\alpha }=\left( x^{i},y^{a}\right) $ satisfying the
relations
\begin{equation}
\lbrack u^{\alpha },u^{\beta }]=i\Theta ^{\alpha \beta },  \label{nceucln}
\end{equation}%
with $\Theta ^{\alpha \beta }=\left( \Theta ^{ij},\Theta ^{ab}\right) $
parametrized as to have a noncommutative structure locally adapted to the
N--connection, and the analogouses of (\ref{nceucder}) redefined for
operators (\ref{dder}) as
\begin{eqnarray}
\delta _{\alpha }u^{\beta } &=&\delta _{\alpha }^{\beta },\mbox{
for }\delta _{\alpha }=\left( \delta _{i}=\partial
_{i}-N_{i}^{a}\partial _{a}, \partial _{b} \right) ,
\notag \\
\lbrack \delta _{\alpha },\delta _{\beta }] &=&-i\Xi _{\alpha \beta },
\label{nceucdern}
\end{eqnarray}%
where $\Xi _{\alpha \beta }=-\left( \Theta ^{-1}\right) _{\alpha \beta }$
for a simplified N--elongated noncommutative differential calculus. We
emphasize that if the vielbein transforms of type (\ref{vielbtr}) and frames
of type (\ref{dder1a}) and (\ref{ddif1a}) are considered, the values $\Theta
^{\alpha \beta }$ and $\Xi _{\alpha \beta }$ could be some complex functions
depending on variables $u^{\beta }$ including also the anholonomy
contributions of\ $N_{i}^{a}.$ In particular cases, they my constructed by
some anholonomic frame transforms from some constant real tensors.

An anholonomic noncommutative Euclidean space $\R_{N,\theta }^{n+m}$ is
defined as a usual one of dimension $k=n+m$ for which a N--connection
structure is prescribed by coefficients $N_{i}^{a}\left( x,y\right) $ which
states an N--elongated differential calculus. The d--metric $\eta _{\alpha
\beta }=\left( \eta _{ij},\eta _{ab}\right) $ and Poisson d--tensor $\Theta
^{\alpha \beta }=\left( \Theta ^{ij},\Theta ^{ab}\right) $ are introduced
via vielbein transforms (\ref{vielbtr}) depending on N--coefficients of the
corresponding constant values contained in (\ref{nceucl}) and (\ref{strnce}%
). As a matter of principle such noncommutative spaces are already curved.

The interior derivative (\ref{inder}) is to be extended on $\R_{N,\theta
}^{n+m}$ as%
\begin{equation*}
\delta _{\alpha }\varphi =\left[ -i\left( \Theta ^{-1}\right) _{\alpha \beta
}u^{\beta },\varphi \right] .
\end{equation*}%
In a similar form, by introducing operators $\delta _{\alpha }$ instead of $%
\partial _{\alpha },$ we can generalize the Moyal product (\ref{moyal}) for
anisotropic spaces:
\begin{eqnarray*}
f\left( x\right) \star \varphi \left( x\right) &=&e^{\frac{i}{2}\Theta
^{\alpha \beta }\frac{\delta }{\partial \xi ^{\alpha }}\frac{\delta }{%
\partial \zeta ^{\beta }}}f\left( x+\xi \right) \varphi \left( x+\zeta
\right) _{|\xi =\varsigma =0} \\
&=&f\varphi +\frac{i}{2}\Theta ^{\alpha \beta }\delta _{\alpha }f\delta
_{\beta }\varphi +o\left( \theta ^{2}\right) .
\end{eqnarray*}

\bigskip For elaborating of perturbation and scattering theory, the more
useful basis is the plane wave basis, which for anholonomic noncommutative
Euclidean spaces, consists of eigenfunctions of the derivatives%
\begin{equation*}
\delta _{\alpha }e^{ipu}=ip_{\alpha }e^{ipu},pu=p_{\alpha }u^{\alpha }.
\end{equation*}%
In this basis, the integral can be defined as
\begin{equation*}
^{N}\int Tr~e^{ipu}=\delta _{p,0}
\end{equation*}%
where the symbol $\int Tr$ is enabled with the left upper index $N$ in order
to emphasize that integration is to be performed on a N--deformed space (we
shall briefly call this as ''N--integration'') and the delta function may be
interpreted as usually (its value at zero represents the volume of physical
space, in our case, N--deformed). There is a specific multiplication low
with respect to the plane wave basis: for instance, by operator reordering,
\begin{equation*}
e^{ipu}\cdot e^{ip^{\prime }u}=e^{-\frac{1}{2}\Theta ^{\alpha \beta
}p_{\alpha }p_{\beta }^{\prime }}~e^{i\left( p+p^{\prime }\right) u},
\end{equation*}%
when $\Theta ^{\alpha \beta }p_{\alpha }p_{\beta }^{\prime }$ may be written
as $p\times p^{\prime }\equiv \Theta ^{\alpha \beta }p_{\alpha }p_{\beta
}^{\prime }=p\times _{\Theta }p^{\prime }.$ There is another example of
multiplication, when the N--elongated partial derivative is involved,
\begin{equation*}
e^{ipu}\cdot f\left( u\right) \cdot e^{-ipu}=e^{-\Theta ^{\alpha \beta
}p_{\alpha }\delta _{\beta }}f\left( u\right) =f\left( u^{\beta }-\Theta
^{\alpha \beta }p_{\alpha }\right) ,
\end{equation*}%
which shows that multiplication by a plane wave in anholonomic
noncommutative Euclidean space translates and N--deform a general function
by $u^{\beta }\rightarrow u^{\beta }-\Theta ^{\alpha \beta }p_{\alpha }.$
This exhibits both the nonlocality and anholonomy of the theory and
preserves the principles that large momenta lead to large nonlocality which
can be also locally anisotropic.

\subsubsection{Noncommutative anholonomic torus}

Let us define the concept of noncommutative anholonomic torus, $\mathbf{T}%
_{N,\theta }^{n+m},$ i. e. the algebra of functions on a noncommutative
torus with some splitting of coordinates into holonomic and anholonomic
ones. We note that a function $f$ on a anholonomic torus $\mathbf{T}%
_{N}^{n+m}$ with N--decomposition is a function on $\R$$_{N}^{n+m}$ which
satisfies a periodicity condition, $f\left( u^{\alpha }\right) =f\left(
u^{\alpha }+2\pi z^{\alpha }\right) $ for d--vectors $z^{\alpha }$ with
integer coordinates. Then the noncommutative extension is to define $\mathbf{%
T}_{N,\theta }^{n+m}$ as the algebra of all sums of products of arbitrary
integer powers of the set of distinguished $n+m$ variables $U_{\alpha
}=\left( U_{i},U_{j}\right) $ satisfying%
\begin{equation}
U_{\alpha }U_{\beta }=e^{-i\Theta ^{\alpha \beta }}U_{\beta }U_{\alpha }.
\label{nctorus}
\end{equation}%
The variables $U_{\alpha }$ are taken instead of $e^{iu^{\alpha }}$ for
plane waves and the derivation of a Weyl algebra from (\ref{nceucln}) is
possible if we take
\begin{eqnarray*}
\lbrack \delta _{\alpha }U_{\beta }] &=&i\delta _{\alpha \beta }U_{\beta },
\\
^{N}\int Tr~U_{1}^{z_{1}}...U_{n+m}^{z_{n+m}} &=&\delta _{\overrightarrow{z}%
,0}.
\end{eqnarray*}%
In addition to the usual topological aspects for nontrivial values of $N$%
--connection there is much more to say in dependence of the fact what type
of topology is induced by the $N$--connection curvature. We omit such
consideration in this paper. The introduced in this subsection formulas and
definitions transform into usual ones from noncommutative geometry if $%
N,m\rightarrow 0.$

\subsection{Anisotropic field theories and anholonomic symmetries}

In a formal sense, every field theory, commutative or noncommutative, can be
anholonomically transformed by changing partial derivatives into
N--elongated ones and redefining the integrating measure in corresponding
Lagrangians. We shall apply this rule to noncommutative scalar, gauge and
Dirac fields and make them to be locally anisotropic and to investigate
their anholonomic symmetries.

\subsubsection{Locally anisotropic matrix scalar field theory}

A generic matrix locally anisotropic matrix scalar field theory with a
hermitian matrix valued field $\phi \left( u\right) =\phi ^{+}\left(
u\right) $ and anholonomically N--deformed Euclidean action%
\begin{equation*}
S=~^{N}\int \delta ^{n+m}u\sqrt{|g_{\alpha \beta }|}\left[ \frac{1}{2}%
g^{\alpha \beta }Tr~\delta _{\alpha }\phi ~\delta _{\beta }\phi +V\left(
\phi \right) \right]
\end{equation*}%
where $V\left( \phi \right) $ is polynomial in variable $\phi ,g_{\alpha
\beta }$ is a d--metric of type (\ref{dmetric}) and $\delta _{\alpha }$ are
N--elongated partial derivatives (\ref{dder}). It is easy to check that if
we replace the matrix algebra by a general associative noncommutative
algebra $\mathcal{A},$ the standard procedure of derivation of motion
equations, classical symmetries from Noether's theorem and related physical
considerations go through but with N--elongated partial derivatives and
N--integration: The field equations are
\begin{equation*}
g^{\alpha \beta }\delta _{\alpha }~\delta _{\beta }\phi =\frac{\partial
V\left( \phi \right) }{\partial \phi }
\end{equation*}%
and the conservation laws
\begin{equation*}
\delta _{\alpha }J^{\alpha }=0
\end{equation*}%
for the curent $J^{\alpha }$ is associated to a symmetry $\bigtriangleup
\phi \left( \epsilon ,\phi \right) $ determined by the N--adapted
variational procedure, $\bigtriangleup S=~^{N}\int Tr~J^{\alpha }\delta
_{\alpha }\epsilon .$ We emphasize that these equations are obtained
according the prescription that we at the first stage perform a usual
variational calculus then we change the usual derivatives and differentials
into N--elongated ones. If we treat the N--connection as an object which
generates and associated linear connection with corresponding curvature we
have to introduce into the motion equations and conservation laws necessary
d--covariant objects curvature/torsion terms.

We may define the momentum operator
\begin{equation*}
P_{\alpha }=-i\left( \Theta ^{-1}\right) _{\alpha \beta }~^{N}\int
Tr~u^{\beta }T^{0},
\end{equation*}%
which follows from the anholonomic transform of the restricted \
stress--energy tensor' constructed from the Noether procedure with
symmetries $\bigtriangleup \phi =i[\phi ,\epsilon ]$ resulting in
\begin{equation*}
T^{\alpha }=ig^{\alpha \beta }[\phi ,\delta _{\beta }\phi ].
\end{equation*}%
We chosen the simplest possibility to define for noncommutative scalar
fields certain energy--momentum values and their anholonomic deformations.
In general, in noncommutatie field theory one introduced more conventional
stress--energy tensors \cite{abou}.

\subsubsection{Locally anisotropic noncommutative gauge fields}

\label{ncgt}Some models of locally anisotropic Yang--Mills and gauge gravity
noncommutative theories are analyzed in Refs. \cite{vnonc,vncf}. Here we say
only the basic facts about such theories with possible supersymmetry but not
concerning points of gauge gravity.

\paragraph{Anholonomic Yang--Mills actions and MSYM model}

{\qquad}

A gauge field is introduced as a one form $A_{\alpha }$ having each
component taking values in $\mathcal{A}$ and satisfying $A_{\alpha
}=A_{\alpha }^{+}$ and curvature (equivalently, field strength)
\begin{equation*}
F_{\alpha \beta }=\delta _{\alpha }A_{\beta }-\delta _{\beta }A_{\alpha }+i
\left[ A_{\alpha }A_{\beta }\right]
\end{equation*}%
with gauge locally anisotropic transformation laws,
\begin{equation}
\bigtriangleup F_{\alpha \beta }=i\left[ F_{\alpha \beta },\epsilon \right] %
\mbox{ for }\bigtriangleup A_{\alpha }=\delta _{\alpha }\epsilon +i\left[
A_{\alpha },\epsilon \right] .  \label{gauget}
\end{equation}%
Now we can introduce the noncommutative locally anisotropic Yang--Mills
action%
\begin{equation*}
S=-\frac{1}{4g_{YM}}^{N}\int Tr~F^{2}
\end{equation*}%
which describes the N--anholonomic dynamics of the gauge field $A_{\alpha }.$
Coupling to matter field can be introduced in a standard way by using
N--elongated partial derivatives $\delta _{\alpha },$%
\begin{equation*}
\bigtriangledown _{\alpha }\varphi =\delta _{\alpha }\varphi +i\left[
A_{\alpha },\varphi \right] .
\end{equation*}%
Here we note that by using $Mat_{Z}\left( \mathcal{A}\right) $ we can
construct both noncommutative and anisotropic analog of $U\left( Z\right) $
gauge theory, or, by introducing supervariables adapted to N--connections %
\cite{vstr2} and locally anisotropic spinors \cite{vspinors,vmon2}, we can
generate supersymmetric Yang--Mills theories. For instance, the maximally
supersymmetric Yang--Mills (MSYM) Lagrangian in ten dimensions with $%
\mathcal{N}=4$ can be deduced in anisotropic form, by corresponding
dimensional reductions and anholonomic constraints, as%
\begin{equation*}
S=~^{N}\int \delta ^{10}u~Tr~\left( F_{\alpha \beta }^{2}+i\overline{\chi }%
\overrightarrow{\bigtriangledown }\chi \right)
\end{equation*}%
where $\chi $ is a 16 component adjoint Majorana--Weyl fermion and the
spinor d--covariant derivating operator $\overrightarrow{\bigtriangledown }$
is writen by using N--anholonomic frames.

\paragraph{The emergence of locally anisotropic spacetime}

{\qquad}

It is well known that spacetime translations may arise from a gauge group
transforms in noncommutative gauge theory (see, for instance, Refs. \cite{dn}%
). If the same procedure is reconsidered for N--elongated partial
derivatives and distinguished noncommutative parameters, we can write
\begin{equation*}
\delta A_{\alpha }=v^{\beta }\delta _{\beta }A_{\alpha }
\end{equation*}
as a gauge transform (\ref{gauget}) when the parameter $\epsilon $ is
expressed as%
\begin{equation*}
\epsilon =v^{\alpha }\left( \Theta ^{-1}\right) _{\alpha \beta }u^{\beta
}=v^{i}\left( \Theta ^{-1}\right) _{jk}x^{k}+v^{a}\left( \Theta ^{-1}\right)
_{ab}x^{b},
\end{equation*}%
which generates
\begin{equation*}
\bigtriangleup A_{\alpha }=v^{\beta }\delta _{\beta }A_{\alpha }+v^{\beta
}\left( \Theta ^{-1}\right) _{\alpha \beta }.
\end{equation*}%
This way the spacetime anholonomy is induced by a noncommutative gauge
anisotropy. For another type of functions $\epsilon (u),$ we may generate
another spacetime locally anisot\-rop\-ic transforms. For instance, we can
generate a Poisson bracket $\{\varphi ,\epsilon \}$ with N--elongat\-ed
derivatives,%
\begin{equation*}
\bigtriangleup \varphi =i\left[ \varphi ,\epsilon \right] =\Theta ^{\alpha
\beta }\delta _{\alpha }\varphi \delta _{\beta }\epsilon +o\left( \delta
_{\alpha }^{2}\varphi \delta _{\beta }^{2}\epsilon \right) \rightarrow
\{\varphi ,\epsilon \}
\end{equation*}%
which proves that at leading order the locally anisotropic gauge transforms
preserve the locally anisotropic noncommutative structure of parameter $%
\Theta ^{\alpha \beta }.$

Now, we demonstrate that the Yang--Mills action may be rewritten as a
''matrix model'' action even the spacetime background is N--deformed. This
is another side of unification of noncommutative spacetime and gauge field
with anholonomically deformed symmetries. We can absorb a inner derivation
into a vector potential by associating the covariant operator $%
\bigtriangledown _{\alpha }=\delta _{\alpha }+iA_{\alpha }$ to connection
operators in $\R$$_{N,\theta }^{n+m},$%
\begin{equation*}
\bigtriangledown _{\alpha }\varphi \rightarrow \left[ C_{\alpha },\varphi %
\right]
\end{equation*}%
for%
\begin{equation}
C_{\alpha }=\left( -i\Theta ^{-1}\right) _{\alpha \beta }u^{\beta
}+iA_{\alpha }.  \label{coper}
\end{equation}%
As in usual noncommutative gauge theory we introduce the ''covariant
coordinates'' but distinguished by the N--connection,%
\begin{equation*}
Y^{\alpha }=u^{\alpha }+\Theta ^{\alpha \beta }A_{\beta }\left( u\right) .
\end{equation*}%
For invertible $\Theta ^{\alpha \beta },$ one considers another notation, $%
Y^{\alpha }=i\Theta ^{\alpha \beta }C_{\beta }.$ Such transforms allow to
express $F_{\alpha \beta }=i\left[ \bigtriangledown _{\alpha
},\bigtriangledown _{\beta }\right] $ as
\begin{equation*}
F_{\alpha \beta }=i\left[ C_{\alpha },C_{\beta }\right] -\left( \Theta
^{-1}\right) _{\alpha \beta }
\end{equation*}%
for which the Yang--Mills action transform into a matrix relation,%
\begin{eqnarray}
S &=&~^{N}Tr\sum_{\alpha ,\beta }\left( \acute{\imath}\left[ C_{\alpha
},C_{\beta }\right] -\left( \Theta ^{-1}\right) _{\alpha \beta }\right) ^{2}
\label{actymmatr} \\
&=&~^{N}Tr\{\left[ \acute{\imath}\left[ C_{k},C_{j}\right] -\left( \Theta
^{-1}\right) _{kj}\right] \left[ \acute{\imath}\left[ C^{k},C^{j}\right]
-\left( \Theta ^{-1}\right) ^{kj}\right]   \notag \\
&&+\left[ \acute{\imath}\left[ C_{a},C_{b}\right] -\left( \Theta
^{-1}\right) _{ab}\right] \left[ \acute{\imath}\left[ C^{a},C^{b}\right]
-\left( \Theta ^{-1}\right) ^{ab}\right] \}  \notag
\end{eqnarray}%
where we emphasize the N--distinguished components.

\paragraph{The noncommutative Dirac d--operator}

{\qquad}

If we consider multiplications $a\cdot \psi $ with $a\in \mathcal{A}$ on a
Dirac spinor $\psi ,$ we can have two different physics depending on the
orders of such multiplications we consider, $a\psi $ or $\psi a.$ In order
to avoid infinite spectral densities, in the locally isotropic
noncommutative gauge theory, one writes the Dirac operator as
\begin{equation*}
\overrightarrow{\bigtriangledown }\psi =\gamma ^{i}\left( \overrightarrow{%
\bigtriangledown }_{i}\psi -\psi \partial _{i}\right) =0.
\end{equation*}%
In the locally anisotropic case we have to introduce N--elongated partial
derivatives,%
\begin{eqnarray*}
\overrightarrow{\bigtriangledown }\psi &=&\gamma ^{\alpha }\left(
\overrightarrow{\bigtriangledown }_{\alpha }\psi -\psi \delta _{\alpha
}\right) \\
&=&\gamma ^{i}\left( \overrightarrow{\bigtriangledown }_{i}\psi -\psi \delta
_{i}\right) +\gamma ^{a}\left( \overrightarrow{\bigtriangledown }_{a}\psi
-\psi \delta _{a}\right) =0
\end{eqnarray*}%
and use a d--covariant spinor calculus \cite{vspinors,vmon2}.

\paragraph{The N--adapted stress--energy tensor}

{\qquad}

The action (\ref{actymmatr}) produces a stress--energy d--tensor%
\begin{equation*}
T_{\alpha \beta }\left( p\right) =\sum_{\gamma }\int_{0}^{1}ds~~^{N}\int
Tr~e^{isp_{\tau }Y^{\tau }}\left[ C_{\alpha },C_{\gamma }\right]
~e^{i(1-s)p_{\tau }Y^{\tau }}\left[ C_{\beta },C_{\gamma }\right]
\end{equation*}%
as a Noetther current derived by the variation $C_{\alpha }\rightarrow
C_{\alpha }+a_{\alpha }\left( p\right) e^{isp_{\tau }Y^{\tau }}.$ This
d--tensor has a property of conservation,
\begin{equation*}
p_{\tau }\Theta ^{\tau \lambda }T_{\lambda \beta }\left( p\right) =0
\end{equation*}%
for the solutions of field equations and seem to be a more natural object in
string theory, which admits an anholonomic generalizations by
''distinguishing of indices''.

\paragraph{The anholonomic Seiberg--Witten map}

{\qquad}

There are two different types of gauge theories: commutative and
noncommutative ones. They my be related by the so--called Seiberg--Witten
map \cite{sw} which explicitly transofrms a noncommutative vector potential
to a conventional Yang--Mills vector potential. This map can be generalized
in gauge gravity and for locally anisotropic gravity \cite{vnonc,vncf}. Here
we define the Seiberg--Witten map for locally anisotropic gauge fields with
N--elongated partial derivatives. \

The idea is that if there exists a standard, but locally anisotropic,
Yang--Mills potential $A_{\alpha }$ with gauge transformation laws
parametrized by the parameter $\epsilon $ like in (\ref{gauget}), a
noncommutative gauge potential $\widehat{A}_{\alpha }\left( A_{\alpha
}\right) $ with gauge transformation parameter $\widehat{\epsilon }\left(
A,\epsilon \right) ,$ when
\begin{equation*}
\widehat{\bigtriangleup }_{\widehat{\epsilon }}\widehat{A}_{\alpha }=\delta
_{\alpha }\widehat{\epsilon }+i\left( \widehat{A}_{\alpha }\star \widehat{%
\epsilon }-\widehat{\epsilon }\star \widehat{A}_{\alpha }\right) ,
\end{equation*}%
should satisfy the equation%
\begin{equation}
\widehat{A}\left( A\right) +\widehat{\bigtriangleup }_{\widehat{\epsilon }}%
\widehat{A}\left( A\right) =\widehat{A}\left( A+\bigtriangleup _{\epsilon
}A\right) ,  \label{swe}
\end{equation}%
where, for simplicity, the indices were omitted. This is the Seiberg--Witten
equation which, in our case, contains N--adapted operators $\delta _{\alpha
} $ (\ref{dder}) and d--vector gauge potentials, respectively, $\widehat{A}%
_{\alpha }=\left( \widehat{A}_{i},\widehat{A}_{a}\right) $ and $A_{\alpha
}=\left( A_{i},A_{a}\right) .$ To first order in $\Theta ^{\alpha \beta
}=\bigtriangleup \Theta ^{\alpha \beta },$ the equation (\ref{swe}) can be
solved in a usual way, by related respectively the potentials and
transformation parameters,%
\begin{eqnarray*}
\widehat{A}_{\alpha }\left( A_{\alpha }\right) -A_{\alpha } &=&-\frac{1}{4}%
\bigtriangleup \Theta ^{\beta \lambda }\left[ A_{\beta }\left( \delta
_{\lambda }A_{\alpha }+F_{\lambda \alpha }\right) +\left( \delta _{\lambda
}A_{\alpha }+F_{\lambda \alpha }\right) A_{\beta }\right] +o(\bigtriangleup
\Theta ^{2}), \\
\widehat{\epsilon }\left( A,\epsilon \right) -\epsilon &=&\frac{1}{4}%
\bigtriangleup \Theta ^{\beta \lambda }\left( \delta _{\beta }\epsilon
~A_{\lambda }+A_{\lambda }\delta _{\beta }\epsilon \right) +o(\bigtriangleup
\Theta ^{2}),
\end{eqnarray*}%
from which we can also find a first order relation for the field strength,%
\begin{eqnarray*}
\widehat{F}_{\lambda \alpha }-F_{\lambda \alpha } &=&\frac{1}{2}%
\bigtriangleup \Theta ^{\beta \tau }\left( F_{\lambda \beta }F_{\alpha \tau
}+F_{\alpha \tau }F_{\lambda \beta }\right) \\
&&-A_{\beta }\left( \bigtriangledown _{\tau }F_{\lambda \alpha }+\delta
_{\tau }F_{\lambda \alpha }\right) -\left( \bigtriangledown _{\tau
}F_{\lambda \alpha }+\delta _{\tau }F_{\lambda \alpha }\right) A_{\beta
}+o(\bigtriangleup \Theta ^{2}).
\end{eqnarray*}%
By a recurrent procedure the solution of (\ref{swe}) can be constructed in
all orders \ of $\bigtriangleup \Theta ^{\alpha \beta }$ as in the locally
isotropic case (see details on recent supersymmetric generalizations in
Refs. \cite{liu} which can be transformed at least in a formal form into
certain anisotropic analogs following the d--covariant geometric rule.

\section{Anholonomy and Noncommutativity: Relations to String/M--Theory}

The aim of this Section is to discuss how both noncommutative and locally
anisotropic field theories arise from string theory and M--theory. The first
use of noncommutative geometry in string theory was suggested by E. Witten
(see Refs. \cite{strncg,sw} for details and developments). Noncommutativity
is natural in open string theory: interactions of open strings with two ends
contains formal similarities to matrix multiplication which explicitly
results in noncommutative structures. In other turn, matix noncommutativity
is contained in off--diagonal metrics and anholonomic vielbeins with
associated N--connection and anholonomic relations (see (\ref{anhol}) and
related details in Appendix) which are used in order to develop locally
anisotropic geometries and field theories. We emphasize that the constructed
exact solutions with off--diagonal metrics in general relativity \ and extra
dimension gravity toghether with the existence of a string field framework
strongly suggest that noncommutative locally anisotropic structures have a
deep underlying significance in such theories \cite%
{vexsol,vbel,vsingl,vsingl1,vnonc,vncf,vstring,vstr2}.

\subsection{Noncommutativity and anholonomy in string theory}

In this subsection, we will analyze strings in curved spacetimes with
constant coefficients $\{g_{ij},h_{ab}\}$ of d--metric (\ref{dmetric}) (the
coefficients $N_{i}^{a}\left( x^{k},y^{a}\right) $ are not constant and the
off--diagonal metric (\ref{ansatz}) has a non--trivial curvature tensor).
With respect to N--adapted frames (\ref{dder}) and (\ref{ddif}) the string
propagation is like in constant Neveu--Schwarz constant $B$--field and with $%
Dp$--branes. We work under the conditions of string and brane
theory which results in noncommutative geometry \cite{strncg} but
the background under consideration here is an anholonomic one.
The $B$--field is a like constant magnetic field which is
polarized by the N--connection structure. The rank
of the matrix $B_{\alpha \beta }$ is denoted $k=n+m=11\leq p+1,$ where $%
p\geq 10$ is a constant. For a target space, defined with respect to
anholonomic frames, we will assume that $B_{0\beta }=0$ with $"0"$ the time
direction (for a Euclidean signature, this condition is not necessary). We
can similarly consider another dimensions than 11, or to suppose that some
dimenisons are compactified. We can pick some torus like coordinates, in
general anholonomic, by certain conditions, $u^{\alpha }\sim $ $u^{\alpha
}+2\pi k^{\alpha }.$ For simplicity, we parametrize $B_{\alpha \beta
}=const\neq 0$ for $\alpha ,\beta =1,...,k$ and $g_{\alpha \beta }=0$ for $%
\alpha =1,...,r,\beta \neq 1,...,k=n+m$ with a further distinguishing of
indices

There are two possibilities of writing out the worldsheet action,%
\begin{eqnarray}
S &=&\frac{1}{4\pi \alpha ^{\prime }}\int_{\Sigma }\delta \underline{\mu }%
_{g}\left( g_{\underline{\alpha }\underline{\beta }}\partial _{A}u^{%
\underline{\alpha }}\partial ^{A}u^{\underline{\beta }}-2\pi \alpha ^{\prime
}iB_{\underline{\alpha }\underline{\beta }}\varepsilon ^{AB}\partial _{A}u^{%
\underline{\alpha }}\partial _{B}u^{\underline{\beta }}\right)  \label{act10}
\\
&=&\frac{1}{4\pi \alpha ^{\prime }}\int_{\Sigma }\delta \underline{\mu }%
_{g}g_{\underline{\alpha }\underline{\beta }}\partial _{A}u^{\underline{%
\alpha }}\partial ^{A}u^{\underline{\beta }}-\frac{i}{2}\int_{\partial
\Sigma }\delta \underline{\mu }_{g}B_{\underline{\alpha }\underline{\beta }%
}~u^{\underline{\alpha }}\partial _{\tan }u^{\underline{\beta }};  \notag \\
&&  \notag \\
&=&\frac{1}{4\pi \alpha ^{\prime }}\int_{\Sigma }\delta \mu
_{g}(g_{ij}\partial _{A}x^{i}\partial ^{A}x^{j}+h_{ab}\partial
_{A}y^{a}\partial ^{A}y^{b}  \notag \\
&&-2\pi \alpha ^{\prime }iB_{ij}\varepsilon ^{AB}\partial _{A}x^{i}\partial
_{B}x^{j}-2\pi \alpha ^{\prime }iB_{ab}\varepsilon ^{AB}\partial
_{A}y^{a}\partial _{B}y^{b})  \notag \\
&=&\frac{1}{4\pi \alpha ^{\prime }}\int_{\Sigma }\delta \mu _{g}\left(
g_{ij}\partial _{A}x^{i}\partial ^{A}x^{j}+h_{ab}\partial _{A}y^{a}\partial
^{A}y^{b}\right)  \notag \\
&&-\frac{i}{2}\int_{\partial \Sigma }\delta \mu _{g}B_{ij}~x^{i}\partial
_{\tan }x^{j}-\frac{i}{2}\int_{\partial \Sigma }\delta \mu
_{g}B_{ab}~y^{a}\partial _{\tan }y^{b},  \notag
\end{eqnarray}%
where the first variant is written by using metric ansatz $g_{\underline{%
\alpha }\underline{\beta }}$ (\ref{ansatz}) but the second variant is just
the term $S_{g_{N},B}$ from action (\ref{act1a}) with d--metric (\ref%
{dmetric}) and different boundary conditions and $\partial _{\tan }$ is the
tangential derivative along the worldwheet boundary $\partial \Sigma .$ We
emphasize that the values $g_{ij},h_{ab}$ and $B_{ij},B_{ab},$ given with
respect to N--adapted frames are constant, but the off--diagonal $g_{%
\underline{\alpha }\underline{\beta }}$ and $B_{\underline{\alpha }%
\underline{\beta }},$ in coordinate base, are some functions on $\left(
x,y\right) .$ The worldsheet $\Sigma $ is taken to be with Euclidean
signature (for a Lorentzian wolrdsheet the complex $i$ should be omitted
multiplying $B).$

The equation of motion of string in anholonomic constant background define
respective anholonomic, N--adapted boundary conditions. For coordinated $%
\alpha $ along the $Dp$--branes they are
\begin{eqnarray}
g_{\alpha \beta }\partial _{norm}u^{\beta }+2\pi i\alpha ^{\prime }B_{\alpha
\beta }\partial _{\tan }u^{\beta } &=&  \label{boundc} \\
g_{ij}\partial _{norm}x^{j}+h_{ab}\partial _{norm}y^{b}+2\pi i\alpha
^{\prime }B_{ij}\partial _{B}x^{j}-2\pi \alpha ^{\prime }iB_{ab}\partial
_{\tan }y^{b}|_{\partial \Sigma } &=&0,  \notag
\end{eqnarray}%
where $\partial _{norm}$ is a normal derivative to $\partial \Sigma .$ By
transforms of type $g_{\underline{\alpha }\underline{\beta }}=e_{\
\underline{\alpha }}^{\alpha }(u)e_{\ \underline{\beta }}^{\beta
}(u)g_{\alpha \beta }$ and $B_{\underline{\alpha }\underline{\beta }}=e_{\
\underline{\alpha }}^{\alpha }(u)e_{\ \underline{\beta }}^{\beta
}(u)B_{\alpha \beta }$ we can remove these boundary conditions into a
holonomic off--diagonal form which is more difficult to investigate. With
respect to N--adapted frames (with non--underlined indices) the analysis is
very similar to the case constant values of the metric and $B$--field. For $%
B=0,$ the boundary conditions (\ref{boundc}) are Neumann ones. If $B$ has
the rank $r=p$ and $B\rightarrow \infty $ (equivalently, $g_{\alpha \beta
}\rightarrow 0$ along the spactial directions of the brane, the boundary
conditions become of Dirichlet type). The effect of all such type conditions
and their possible interpolations can be investigated as in the usual open
string theory with constant $B$--field but, in this subsection, with respect
to N--adapted frames.

For instance, we can suppose that $\Sigma $ is a disc,
conformally and anholonomically mapped to the upper half plane
with complex variables $z$ and $\overline{z}$ and ${Im}\ z\geq 0.$
The propagator with such boundary conditions is the same as in
\cite{fradkin} with coordinates redefined to
anholonomic frames,%
\begin{eqnarray*}
<x^{i}(z)x^{j}(z^{\prime })> &=&-\alpha ^{\prime }[g^{ij}\log \frac{%
|z-z^{\prime }|}{|z-\overline{z}^{\prime }|}+H^{ij}\log |z-\overline{z}%
^{\prime }|^{2} \\
&&+\frac{1}{2\pi \alpha ^{\prime }}\Theta ^{ij}\log \frac{|z-\overline{z}%
^{\prime }|}{|\overline{z}-z^{\prime }|}+Q^{ij}], \\
<y^{a}(z)y^{b}(z^{\prime })> &=&-\alpha ^{\prime }[h^{ab}\log \frac{%
|z-z^{\prime }|}{|z-\overline{z}^{\prime }|}+H^{ab}\log |z-\overline{z}%
^{\prime }|^{2} \\
&&+\frac{1}{2\pi \alpha ^{\prime }}\Theta ^{ab}\log \frac{|z-\overline{z}%
^{\prime }|}{|\overline{z}-z^{\prime }|}+Q^{ab}],
\end{eqnarray*}%
where the coefficients are correspondingly computed,%
\begin{eqnarray}
H_{ij} &=&g_{ij}-(2\pi \alpha ^{\prime })^{2}\left( Bg^{-1}B\right)
_{ij},~H_{ab}=h_{ab}-(2\pi \alpha ^{\prime })^{2}\left( Bg^{-1}B\right)
_{ab},  \label{constants} \\
&&  \notag
\end{eqnarray}%
\begin{eqnarray*}
H^{ij} &=&\left( \frac{1}{g+2\pi \alpha ^{\prime }B}\right)
_{[sym]}^{ij}=\left( \frac{1}{g+2\pi \alpha ^{\prime }B}g\frac{1}{g-2\pi
\alpha ^{\prime }B}\right) ^{ij}, \\
H^{ab} &=&\left( \frac{1}{h+2\pi \alpha ^{\prime }B}\right)
_{[sym]}^{ij}=\left( \frac{1}{h+2\pi \alpha ^{\prime }B}h\frac{1}{h-2\pi
\alpha ^{\prime }B}\right) ^{ij},
\end{eqnarray*}%
\begin{eqnarray*}
\Theta ^{ij} &=&2\pi \alpha ^{\prime }\left( \frac{1}{g+2\pi \alpha ^{\prime
}B}\right) _{[antisym]}^{ij}=-(2\pi \alpha ^{\prime })^{2}\left( \frac{1}{%
g+2\pi \alpha ^{\prime }B}g\frac{1}{g-2\pi \alpha ^{\prime }B}\right) ^{ij},
\\
\Theta ^{ab} &=&2\pi \alpha ^{\prime }\left( \frac{1}{g+2\pi \alpha ^{\prime
}B}\right) _{[antisym]}^{ab}=-(2\pi \alpha ^{\prime })^{2}\left( \frac{1}{%
g+2\pi \alpha ^{\prime }B}g\frac{1}{g-2\pi \alpha ^{\prime }B}\right) ^{ab},
\end{eqnarray*}%
with $[sym]$ and $[antisym]$ prescribing, respectively, the symmetric and
antisymmetric parts of a matrix and constants $Q^{ij}$ and $Q^{ab}$ (in
general, depending on $B,$ but not on $z$ or $z^{\prime })$ do to not play
an essential role which allows to set them to a convenient value. The last
two terms are signed--valued (if the branch cut of the logarithm is taken in
lower half plane) and the rest ones are manifestly signe--valued.

Restricting our considerations to the open string vertex operators and
interactions with real $z=\tau $ and $z=\tau ^{\prime },$ evaluating at
boundary points of $\Sigma $ for a convenient value of $D^{\alpha \beta },$
the propagator (in non--distinguished form) becomes%
\begin{equation*}
<u^{\alpha }(\tau )u^{\beta }(\tau ^{\prime })>=-\alpha ^{\prime }H^{\alpha
\beta }\log \left( \tau -\tau ^{\prime }\right) ^{2}+\frac{i}{2}\Theta
^{\alpha \beta }\epsilon \left( \tau -\tau ^{\prime }\right)
\end{equation*}%
for $\epsilon \left( \tau -\tau ^{\prime }\right) $ being 1 for $\tau >\tau
^{\prime }$ and -1 for $\tau <\tau ^{\prime }.$ The d--tensor $H_{\alpha
\beta }$ $\ $defines the effective metric seen by the open string subjected
to some anholonomic constraints being constant with respect to N--adapted
frames. Working as in conformal field theory, one can compute commutators of
operators from the short distance behavior of operator products (by
interpreting time ordering as operator ordering with time $\tau )$ and find
that the coordinate commutator
\begin{equation*}
\left[ u^{\alpha }(\tau ),u^{\beta }(\tau )\right] =i\Theta ^{\alpha \beta }
\end{equation*}%
which is just the relation (\ref{nceucln}) for noncommutative coordinates
with constant noncommutativity parameter $\Theta ^{\alpha \beta }$
distinguished by a N--connection structure.

In a similar manner we can introduce gauge fields and consider worldsheet
supersymmetry together with noncommutative relations with respect to
N--adapted frames. This results in locally anisotropic modifications of the
results from \cite{strncg} via anholonomic frame transforms and
distinguished tensor and noncommutative calculus (we omit here the details
of such calculations).

We emphasize that even the values $H^{\alpha \beta }$ and $\Theta ^{\alpha
\beta }$ (\ref{constants}) are constant with respect to N--adapted frames
the anholonomic noncommutative string configurations are characterized by
locally anisotropic values $H^{\underline{\alpha }\underline{\beta }}$ and $%
\theta ^{\underline{\alpha }\underline{\beta }}$ which are defined with
respect to coordinate frames as
\begin{equation*}
H^{\underline{\alpha }\underline{\beta }}=e_{\alpha }^{~\underline{\alpha }%
}(u)e_{\beta }^{~\underline{\beta }}(u)H^{\alpha \beta }\mbox{ and }\theta ^{%
\underline{\alpha }\underline{\beta }}=e_{\alpha }^{~\underline{\alpha }%
}(u)e_{\beta }^{~\underline{\beta }}(u)\theta ^{\alpha \beta }
\end{equation*}%
with $e_{\alpha }^{~\underline{\alpha }}(u)$ (\ref{vielbtr}) defined by $%
N_{i}^{a}$ as in (\ref{viel1}), i. e.%
\begin{equation*}
e_{i}^{~\underline{i}}=\delta _{i}^{~\underline{i}},~e_{i}^{~\underline{a}%
}=-N_{i}^{\underline{a}}(u),~e_{a}^{~\underline{a}}=\delta _{a}^{\underline{a%
}},~e_{a}^{~\underline{i}}=0.
\end{equation*}

Now, we make use of he standard relation between world--sheet correlation
function of vertex operators, the S--matrix for string scattering and
effective actions which can reproduce this low energy string physics \cite%
{deligne} but generalizing them for anholonomic structures. We consider that
operators in the bulk of the world--sheet correspond to closed strings,
while operators on the boundary correspond to open strings and thus fields
which propagate on the world volume of a D--brane. The basic idea is that
each local world--sheet operator $V_{s}\left( z\right) $ corresponds to an
interaction with a spacetime field $\varphi _{s}\left( z\right) $ which
results in the effective Lagrangian
\begin{equation*}
\int \delta ^{p+1}u\sqrt{|\det g_{\alpha \beta }|}~^{N}Tr~\varphi
_{1}\varphi _{2}...\varphi _{s}
\end{equation*}%
which is computed by integrating on $z_{s}$ following the prescribed order
for the correlation function%
\begin{equation*}
\left\langle \int dz_{1}V_{1}\left( z_{1}\right) \int dz_{2}V_{2}\left(
z_{2}\right) ...\int dz_{s}V_{s}\left( z_{s}\right) \right\rangle
\end{equation*}%
on a world--sheet $\Sigma $ with disk topology, with operators $V_{s}$ as
successive points $z_{s}$ on the boundary $\partial \Sigma .$ The
integrating measure is constructed from N--elongated values and coefficients
of d--metric. In the leading limit of the S--matrix with vertex operators
only for the massless fields we reproduce a locally anisotropic variant of
the MSYM effective action which describes the physics of a D--brane with
arbitrarily large but anisotropically and slowly varying field strength,
\begin{equation}
S_{BNI}^{[anh]}=\frac{1}{g_{s}l_{s}\left( 2\pi l_{s}\right) ^{p}}\int \delta
^{p+1}u\sqrt{|\det (g_{\alpha \beta }+2\pi l_{s}^{2}(B+F))|}  \label{act12}
\end{equation}%
where $g_{s}$ is the string coupling, the constant $l_{s}$ is the usual one
from D--brane theory and $g_{\alpha \beta }$ is the induced d--metric on the
brane world--volume. The action (\ref{act12}) is just the
Nambu--Born--Infeld (NBI) action \cite{fradkin} but defined for d--metrics
and d--tensor fields with coefficients computed with respect to N--adapted
frames.

\subsection{Noncommutative anisotropic structures in M(atrix) theory}

For an introduction to M--theory, we refer to \cite{polch,deligne}.
Throughout this subsection we consider M--theory as to be not completely
defined but with a well--defined quantum gravity theory with the low energy
spectrum of the 11 dimensional supergravity theory \cite{cjs}, containing
solitonic ''branes'', the 2--brane, or supermemrane, and five--branes and
that from M--theory there exists connections to the superstring theories.
Our claim is that in the low energy limits the noncommutative structures
are, in general, locally anisotropic.

The simplest way to derive noncommutativity from M--theory is to start with
a matrix model action such in subsection \ref{ncgt} and by introducing
operators of type $C_{\alpha }$ $\ $(\ref{coper}) and actions (\ref%
{actymmatr}). For instance, we can consider the action for maximally
supersymmetric quantum mechanics, i. e. a trivial case with $p=0$ of MSYM,
when
\begin{equation}
S=\int \delta t~~^{N}Tr\sum_{\alpha =1}^{9}\left( D_{t}X\right)
^{2}-\sum_{\alpha <\beta }\left[ X^{\alpha },X^{\beta }\right] ^{2}+\chi
^{+}\left( D_{t}+\Gamma _{\alpha }X^{\alpha }\right) \chi ,  \label{act11}
\end{equation}%
where $D_{t}=\delta /\partial t+iA_{0}$ with d--derivative
(\ref{dder}) with varying $A_{0}$ which introduces constraints in
physical states because of restriction of unitary symmetry. This
action is written in anholonomic variables and generalizes the
approach of entering the M--theory as a regularized form of the
actions for the supermembranes \cite{wit}. In this interpretation
the the compact eleventh dimension does not disappear and the
M--theory is to be considered as to be anisotropically
compactified on a light--like circle.

In order to understand how anisotropic torus compactifications may be
performed (see subsection \ref{torus}) we use the general theory of
D--branes on quotient spaces \cite{taylor}. We consider $U_{\alpha }=\gamma
\left( \beta _{\alpha }\right) $ for a set of generators of $\Z$$^{n+m}$
with $\mathcal{A}=Mat_{n+m}\left( \C\right) $ which satisfy the equations%
\begin{equation*}
U_{\alpha }^{-1}X^{\beta }U_{\alpha }=X^{\beta }+\delta _{\alpha }^{\beta
}2\pi R_{\alpha }
\end{equation*}%
having solutions of type
\begin{equation*}
X_{\beta }=-i\delta /\partial \sigma ^{\beta }+A_{\beta }
\end{equation*}
for $A_{\beta }$ commuting with $U_{\alpha }$ and indices distinguished by a
N--connection structure as $\alpha =\left( i,a\right) ,\beta =\left(
j,b\right) .$ For such variables the action (\ref{act11}) leads to a locally
anisotropic MSYM on $T^{n+m}\times \R.$ Of course, this construction admits
a natural generalization for variables $U_{\alpha }$ satisfying relations (%
\ref{nceucln}) for noncommutative locally anisotropic tori which leads to
noncommutative anholonomic gauge theories \cite{vnonc,vncf}. In original
form this type of noncommutativity was introduced in M--theory (without
anisotropies) in Ref. \cite{connes1}.

The anisotropic noncommutativity in M--theory can related to string model
via nontrivial components $C_{\alpha \beta -}$ of a three--form potential
(''-'' denotes the compact light--like direction). This potential has as \ a
background value if the M(atrix) theory is treated as M--theory on a
light--like circle as in usual isotropic models. In the IIA string
interpretation of $C_{\alpha \beta -}$ as a Neveu--Schwarz $B$--field which
minimally coupled to the string world--sheet, we obtain the action (\ref%
{act10}) compactified on a $\R\times T^{n+m}$ spacetime where torus has
constant d--metric and $B$--field coefficients.

\section{Anisotropic Gravity on Noncommutative D--Bra\-nes}

We develop a model of locally anisotropic \ gravity on noncommutative
D--branes (see Refs. \cite{ardalan} for a locally isotropic variant). We
investigate what kind of deformations of the low energy effective action of
closed strings are induced in the presence of constant background
antisymmetric field (or it anholonomic transforms) and/or in the presence of
generic off--diagonal metric and associated nonlinear connection terms. It
should be noted that there were proposed and studied different models of
nocommutative deformations of gravity \cite{ncg}, which were not derived
from string theory but introduced ''ad hoc''. Anholonomic and/or gauge
transforms in noncommutative gravity were considered in Refs. \cite%
{vncf,vnonc}. In this Section, we illustrate how such gravity models with
generic anisotropy and noncommutativity can be embedded in D--brane physics.

We can compute the tree level bosonic string scattering amplitude of two
massless closed string off a noncommutative D--brane \ with locally
anisotropic contributions by considering boundary conditions and correlators
stated with respect to anholonomic frames. By using the 'geometric
d--covariant rule' of changing the tensors, spinors and connections into
theirs corresponding N--distinguished d--objects we derive the locally
anisotropic variant of effective actions in a straightforword manner.

For instance, the action which describes this amplitude to order of the
string constant $(\alpha ^{\prime })^{0}$ is just the so--called DBI and
Einstein--Hilbert action. With respect to the Einstein N--emphasized frame
the DBI action is
\begin{equation}
S_{D-brane}^{[0]}=-T_{p}\int \delta ^{p+1}u~e^{-\Phi }\sqrt{\left| \det
\left( e^{-\gamma \Phi }g_{\alpha \beta }+\mathcal{B}_{\alpha \beta
}+f_{\alpha \beta }\right) \right| }  \label{act13}
\end{equation}%
where $g_{\alpha \beta }$ is the induced metric on the D--brane, $\mathcal{B}%
_{\alpha \beta }=B_{\alpha \beta }-2\kappa b_{\alpha \beta }$ is the pull
back of the antisymmetric d--field $B$ being constant \ with respect to
N--adapted frames along D--brane, $f_{\alpha \beta }$ is the gauge d--field
strength and $\gamma =-4/\left( n+m-2\right) $ and the constant $T_{p}=%
\mathcal{C}(\alpha ^{\prime })^{2}/C\kappa ^{2}$ is taken as in Ref. \cite%
{ardalan} for usual D--brane theory (this allow to obtain in a limit the
Einstein--Hilber action in the bulk). There are used such parametrizations
of indices:
\begin{equation*}
\mu ^{\prime },\nu ^{\prime },...=0,...,25;\mu ^{\prime }=\left( \mu ,\hat{%
\mu}\right) ;~\hat{\mu},\hat{\nu}...=p+1,...,25;\hat{\mu}=\left( \hat{\imath}%
,\hat{a}\right)
\end{equation*}%
where $i$ takes $n$--dimensional 'horizontal' values and $a$ takes $m$%
--dimensional 'vertical' being used for a D--brane localized at $%
u^{p+1},...u^{25}$ with the boundary conditions given with respect to a
N--adapted frame,%
\begin{equation*}
g_{\alpha \beta }\left( \partial -\overline{\partial }\right) U^{\alpha
}+B_{\alpha \beta }\left( \partial +\overline{\partial }\right) U_{\mid z=%
\overline{z}}^{\alpha }=0,
\end{equation*}%
which should be distinguished in h- and v--components, and the two point
correlator of string anholonomic coordinates $U^{\alpha ^{\prime }}\left( z,%
\overline{z}\right) $ on the D--brane is%
\begin{eqnarray*}
&<&U_{\hat{\mu}}^{\alpha ^{\prime }}U_{\hat{\nu}}^{\beta ^{\prime }}>~=-%
\frac{\alpha ^{\prime }}{2}\{g^{\alpha ^{\prime }\beta ^{\prime }}\log \left[
\left( z_{\hat{\mu}}-z_{\hat{\nu}}\right) \left( \overline{z}_{\hat{\mu}}-%
\overline{z}_{\hat{\nu}}\right) \right] \\
&&+D^{\alpha ^{\prime }\beta ^{\prime }}\log \left( z_{\hat{\mu}}-\overline{z%
}_{\hat{\nu}}\right) +D^{\beta ^{\prime }\alpha ^{\prime }}\log \left(
\overline{z}_{\hat{\mu}}-z_{\hat{\nu}}\right) \}
\end{eqnarray*}%
where
\begin{equation*}
D^{\beta \alpha }=2\left( \frac{1}{\eta +B}\right) ^{\alpha \beta }-\eta
^{\alpha \beta },~D^{\hat{\mu}\hat{\nu}}=-\delta ^{\hat{\mu}\hat{\nu}%
},~D_{~\alpha ^{\prime }}^{\beta ^{\prime }}D^{\nu ^{\prime }\alpha ^{\prime
}}=\eta ^{\beta ^{\prime }\alpha ^{\prime }}
\end{equation*}%
for constant $\eta ^{\alpha ^{\prime }\beta ^{\prime }\text{ }}$ given with
respect to N--adapted frames.

The scattering amplitude of two closed strings off a D--brane is computed as
the integral
\begin{equation}
A=g_{c}^{2}~e^{-\lambda }\int d^{2}z_{\underline{1}}~d^{2}z_{\underline{2}%
}~<V\left( z_{\underline{1}},\overline{z}_{\underline{1}}\right) V\left( z_{%
\underline{2}},\overline{z}_{\underline{2}}\right) >,  \label{dbranea}
\end{equation}%
for $g_{c}$ being the closed string coupling constant, $\lambda $ being the
Euler number of the world sheet and the vertex operators for the massless
closed strings with the momenta $k_{\underline{i}\mu ^{\prime }}=\left( k_{%
\underline{i}i^{\prime }},k_{\underline{i}a^{\prime }}\right) $ $\ $\ and
polarizations $\epsilon _{\mu ^{\prime }\nu ^{\prime }}$ (satisfying the
conditions $\epsilon _{\mu ^{\prime }\nu ^{\prime }}k_{\underline{i}}^{\mu
^{\prime }}=\epsilon _{\mu ^{\prime }\nu ^{\prime }}k_{\underline{i}}^{\nu
^{\prime }}=0$ and $k_{\underline{i}\mu ^{\prime }}k_{\underline{i}}^{\mu
^{\prime }}=0$ taken as
\begin{equation*}
V\left( z_{i},\overline{z}_{i}\right) =\epsilon _{\mu ^{\prime }\nu ^{\prime
}}~~D_{~\alpha ^{\prime }}^{\nu ^{\prime }}:\partial X^{\mu ^{\prime
}}\left( z_{\underline{i}}\right) \exp \left[ ik_{\underline{i}}X\left( z_{%
\underline{i}}\right) \right] :~:\overline{\partial }X^{\alpha ^{\prime
}}\left( \overline{z}_{\underline{i}}\right) \exp \left[ ik_{\underline{i}%
\beta ^{\prime }}D_{~\tau ^{\prime }}^{\beta ^{\prime }}X^{\tau ^{\prime
}}\left( z_{\underline{i}}\right) \right] :.
\end{equation*}%
Calculation of such calculation functions can be performed as in usual
string theory with that difference that the tensors and derivatives are
distinguished by N--connections.

Decomposing the metric $g_{\alpha \beta }$ as
\begin{equation*}
g_{\alpha \beta }=\eta _{\alpha \beta }+2\kappa \chi _{\alpha \beta }
\end{equation*}%
where $\eta _{\alpha \beta }$ is constant (Minkowski metric but with respect
to N--adapted frames) and $\chi _{\alpha \beta }$ could be of (pseudo)
Riemannian or Finsler like type. Action (\ref{act13}) can be written to the
first order of $\chi ,$
\begin{equation}
S_{D-brane}^{[0]}=-\kappa T_{p}c\int \delta ^{p+1}u~\chi _{\alpha \beta
}Q^{\alpha \beta },  \label{act13a}
\end{equation}%
where
\begin{equation}
Q^{\alpha \beta }=\frac{1}{2}\left( \eta ^{\alpha \beta }+D^{\alpha \beta
}\right)  \label{qfield}
\end{equation}%
and $c=\sqrt{|\det \left( \eta _{\alpha \beta }+B_{\alpha \beta }\right) |},$
which exhibits a source for locally anisotropic gravity on D--brane,
\begin{equation*}
T_{\chi }^{\alpha \beta }=-\frac{1}{2}T_{p}\kappa C\left( \eta
^{\alpha \beta }+D_{(S)}^{\beta \alpha }\right) ,
\end{equation*}%
for $D_{(S)}^{\beta \alpha }$ denoting symmetrization of the matrix $%
D^{\beta \alpha }.$ This way we reproduce the same action as in superstring
theory \cite{gar} but in a manner when anholonomic effects and anisotropic
scattering can be included.

Next order terms on $\alpha ^{\prime }$ in the string amplitude are included
by the term%
\begin{eqnarray*}
S_{bulk}^{[1]} &=&\frac{\alpha ^{\prime }}{8\kappa ^{2}}\int
\delta
^{26}u^{\prime }~e^{\gamma \Phi }\sqrt{|g_{\mu ^{\prime }\nu ^{\prime }}|}%
[R_{h^{\prime }i^{\prime }j^{\prime }k^{\prime }}R^{h^{\prime }i^{\prime
}j^{\prime }k^{\prime }}+R_{a^{\prime }b^{\prime }j^{\prime }k^{\prime
}}R^{a^{\prime }b^{\prime }j^{\prime }k^{\prime }}+P_{j^{\prime }i^{\prime
}k^{\prime }a^{\prime }}P^{j^{\prime }i^{\prime }k^{\prime }a^{\prime }} \\
&&+P_{c^{\prime }d^{\prime }k^{\prime }a^{\prime }}P^{c^{\prime }d^{\prime
}k^{\prime }a^{\prime }}+S_{j^{\prime }i^{\prime }b^{\prime }c^{\prime
}}S^{j^{\prime }i^{\prime }b^{\prime }c^{\prime }}+S_{d^{\prime }e^{\prime
}b^{\prime }c^{\prime }}S^{d^{\prime }e^{\prime }b^{\prime }c^{\prime }} \\
&&-4\left( R_{i^{\prime }j^{\prime }}R^{\iota ^{\prime }j^{\prime
}}+R_{i^{\prime }a^{\prime }}R^{i^{\prime }a^{\prime }}+P_{a^{\prime
}i^{\prime }}P^{a^{\prime }i^{\prime }}+R_{a^{\prime }b^{\prime
}}R^{a^{\prime }b^{\prime }}\right) +(g^{i^{\prime }j^{\prime }}R_{i^{\prime
}j^{\prime }}+h^{a^{\prime }b^{\prime }}S_{a^{\prime }b^{\prime }})^{2}]
\end{eqnarray*}%
where the indices are split as $\mu ^{\prime }=\left( i^{\prime },a^{\prime
}\right) $ and we use respectively the d--curvatures (\ref{dcurvatures}),
Ricci d--tensors (\ref{dricci}) and d--scalars (\ref{dscalar}). Splitting of
''primed' indices reduces to splitting of D--brane values.

The DBI action on D--brane (\ref{dbranea}) is defined with a gauge field
strength
\begin{equation*}
f_{\alpha \beta }=\delta _{\alpha }a_{\beta }-\delta _{\beta }a_{\alpha }
\end{equation*}%
and with the induced metric
\begin{equation*}
g_{\alpha \beta }=\delta _{\alpha }X^{\mu ^{\prime }}\delta _{\beta }X_{\mu
^{\prime }}
\end{equation*}%
expanded around the flat space in the static gauge $U^{\mu }=u^{\mu },$%
\begin{equation*}
g_{\mu \nu }=\eta _{\mu \nu }+2\kappa \chi _{\mu \nu }+2\kappa \left( \chi _{%
\hat{\mu}\mu }\delta _{\nu }U^{\hat{\mu}}+\chi _{\hat{\mu}\nu }\delta _{\mu
}U^{\hat{\mu}}\right) +\delta _{\mu }U^{\hat{\mu}}\delta _{\nu }U_{\hat{\mu}%
}+2\kappa \chi _{\hat{\mu}\hat{\nu}}\delta _{\mu }U^{\hat{\mu}}\delta _{\nu
}U^{\hat{\nu}}.
\end{equation*}%
In order to describe D--brane locally anisotropic processes in the first
order in $\alpha ^{\prime }$ we need to add a new term to the DBI\ as follow,%
\begin{equation}
S^{1}=-\frac{\alpha ^{\prime }T_{p}}{2}\int \delta ^{p+1}u\sqrt{|\det q_{\mu
\nu }|}\{~R_{\alpha \beta \gamma \tau }q^{\alpha \tau }-\left( \Psi _{\alpha
\gamma }^{\hat{\mu}}\Psi _{\hat{\mu}\beta \tau }-\Psi _{\alpha \tau }^{\hat{%
\mu}}\Psi _{\hat{\mu}\beta \gamma }\right) \tilde{q}^{\alpha \tau }\}\tilde{q%
}^{\beta \gamma }  \label{act14}
\end{equation}%
where $q_{\mu \nu }=\eta _{\mu \nu }+B_{\mu \nu }+f_{\mu \nu },$ $q^{\mu \nu
}$ is the inverse of $q_{\mu \nu },\tilde{q}_{\mu \nu }=g_{\mu \nu }+B_{\mu
\nu }+f_{\mu \nu },$ $\tilde{q}^{\mu \nu }$ is the inverse of $\tilde{q}%
_{\mu \nu },$ the curvature d--tensor $~R_{\alpha \beta \gamma \tau }$ is
constructed from the induced d--metric by using the cannonical d--connection
(see (\ref{dcurvatures}) and (\ref{dcon})) and
\begin{equation*}
\Psi _{\alpha \beta }^{\hat{\mu}}=\kappa \left( -\delta ^{\hat{\mu}}\chi
_{\alpha \beta }+\delta _{\alpha }\chi _{~\beta }^{\hat{\mu}}+\delta _{\beta
}\chi _{~\alpha }^{\hat{\mu}}\right) +\delta _{\alpha }\delta _{\beta }U^{%
\hat{\mu}}.
\end{equation*}

The action (\ref{act14}) can be related to the Einstein--Hilbert action on
the D--brane if the the $B$--field is turned off. To see this we consider
the field $Q^{\alpha \beta }=$ $\eta ^{\alpha \beta }$ (\ref{qfield}) which
reduces (up to some total d--derivatives, which by the momentum conservation
relation have no effects in scattering amplitudes, and ignoring gauge fields
because they do not any contraction with gravitons because of antisymmetry
of $f_{\alpha \beta }$) to
\begin{equation}
S_{D-brane}^{[1]}=-\frac{\alpha ^{\prime }T_{p}}{2}\int \delta ^{p+1}u\sqrt{%
|\det g_{\mu \nu }|}\left( \widehat{R}+S+\Psi _{~\alpha }^{\hat{\mu}~\alpha
}\Psi _{\hat{\mu}\beta }^{\quad \beta }-\Psi _{\alpha \beta }^{\hat{\mu}%
}\Psi _{\hat{\mu}}^{\quad \alpha \beta }\right)  \label{act15}
\end{equation}%
were $\widehat{R}$ and $S$ are computed as d--scalar objects (\ref{dscalar})
and by following the relation at $\mathcal{O}(\chi ^{2}),$%
\begin{equation*}
\sqrt{|\det \eta _{\mu \nu }|}R_{\alpha \beta \gamma \tau }\eta ^{\alpha
\tau }g^{\beta \gamma }=\sqrt{|\det g_{\mu \nu }|}R_{\alpha \beta \gamma
\tau }g^{\alpha \gamma }g^{\beta \tau }+\mbox{total d--derivatives}.
\end{equation*}%
The action (\ref{act15}) transformes into the Einstein--Hilbert action as it
was proven for the locally isotropic D--brane theory \cite{corley} for
vanishing N--connections and trivial vertical (anisotropic) dimensions.

In conclusion, it has been shown in this Section that the D--brane dynamics
can be transformed into a locally anisotropic one, which in low energy
limits contains different models of generalized Lagrange/\ Finsler or
anholonomic Riemannian spacetimes, by introducing corresponding anholonomic
frames with associated N--connection structures and d--metric fields (like (%
\ref{ncc}) and (\ref{mfl}) and (\ref{dmetricf})).

\section{Exact Solutions: Noncommutative and/ or Locally Anisotropic
Structures}

\label{exsol}

In the previous sections we demonstrated that locally anisotropic
noncommutative geometric structures are hidden in string/ M--theory. Our aim
here is to construct and analyze four classes of exact solutions in string
gravity with effective metrics posessing generic off--diagonal terms which
for associated anholonomic frames and N--connections can be extended to
commutative or noncommutative string configurations.

\subsection{Black ellipsoids from string gravity}

A simple string gravity model with antisymmetric two form potential field $%
H^{\alpha ^{\prime }\beta ^{\prime }\gamma ^{\prime }},$ for constant
dilaton $\phi ,$ and static internal space, $\beta ,$ is to be found for the
NS--NS sector which is common to both the heterotic and type II string
theories \cite{lidsey}. The equations (\ref{eqfstr}) reduce to%
\begin{eqnarray}
R_{\mu ^{\prime }\nu ^{\prime }} &=&\frac{1}{4}H_{\mu ^{\prime }\lambda
^{\prime }\tau ^{\prime }}H_{\nu ^{\prime }}^{~~\lambda ^{\prime }\tau
^{\prime }},  \label{eq16} \\
D_{\mu ^{\prime }}H^{\mu ^{\prime }~\lambda ^{\prime }\tau ^{\prime }} &=&0,
\notag
\end{eqnarray}%
for
\begin{equation*}
H_{\mu ^{\prime }\nu ^{\prime }\lambda ^{\prime }}=\delta _{\mu ^{\prime
}}B_{\nu ^{\prime }\lambda ^{\prime }}+\delta _{\lambda ^{\prime }}B_{\mu
^{\prime }\nu ^{\prime }}+\delta _{\nu ^{\prime }}B_{\lambda ^{\prime }\mu
^{\prime }}.
\end{equation*}%
If $H_{\mu ^{\prime }\nu ^{\prime }\lambda ^{\prime }}=\sqrt{|g_{\mu
^{\prime }\nu ^{\prime }}|}\epsilon _{\mu ^{\prime }\nu ^{\prime }\lambda
^{\prime }},$ we obtain the vacuum equations for the gravity with
cosmological constant $\lambda $,%
\begin{equation}
R_{\mu ^{\prime }\nu ^{\prime }}=\lambda g_{\mu ^{\prime }\nu ^{\prime }},
\label{eq17}
\end{equation}%
for $\lambda =1/4$ where $R_{\mu ^{\prime }\nu ^{\prime }}$ is
the Ricci d--tensor (\ref{dricci}), with ''primed'' indices
emphasizing that the geometry is induced after a topological
compactification.

For an ansatz of type
\begin{eqnarray}
\delta s^{2} &=&g_{1}(dx^{1})^{2}+g_{2}(dx^{2})^{2}+h_{3}\left( x^{i^{\prime
}},y^{3}\right) (\delta y^{3})^{2}+h_{4}\left( x^{i^{\prime }},y^{3}\right)
(\delta y^{4})^{2},  \label{ansatz18} \\
\delta y^{3} &=&dy^{3}+w_{i^{\prime }}\left( x^{k^{\prime }},y^{3}\right)
dx^{i^{\prime }},\quad \delta y^{4}=dy^{4}+n_{i^{\prime }}\left(
x^{k^{\prime }},y^{3}\right) dx^{i^{\prime }},  \notag
\end{eqnarray}%
for the d--metric (\ref{dmetric}) the Einstein equations (\ref{eq17}) are
written (see \cite{vmethod,vexsol} for details on computation)
\begin{eqnarray}
R_{1}^{1}=R_{2}^{2}=-\frac{1}{2g_{1}g_{2}}[g_{2}^{\bullet \bullet }-\frac{%
g_{1}^{\bullet }g_{2}^{\bullet }}{2g_{1}}-\frac{(g_{2}^{\bullet })^{2}}{%
2g_{2}}+g_{1}^{^{\prime \prime }}-\frac{g_{1}^{^{\prime }}g_{2}^{^{\prime }}%
}{2g_{2}}-\frac{(g_{1}^{^{\prime }})^{2}}{2g_{1}}] &=&\lambda ,
\label{ricci1a} \\
R_{3}^{3}=R_{4}^{4}=-\frac{\beta }{2h_{3}h_{4}} &=&\lambda ,  \label{ricci2a}
\\
R_{3i^{\prime }}=-w_{i^{\prime }}\frac{\beta }{2h_{4}}-\frac{\alpha
_{i^{\prime }}}{2h_{4}} &=&0,  \label{ricci3a} \\
R_{4i^{\prime }}=-\frac{h_{4}}{2h_{3}}\left[ n_{i^{\prime }}^{\ast \ast
}+\gamma n_{i^{\prime }}^{\ast }\right] &=&0,  \label{ricci4a}
\end{eqnarray}%
where the indices take values $i^{\prime },k^{\prime }=1,2$ and $a^{\prime
},b^{\prime }=3,4.$ The coefficients of equations (\ref{ricci1a}) - (\ref%
{ricci4a}) are given by
\begin{equation}
\alpha _{i}=\partial _{i}{h_{4}^{\ast }}-h_{4}^{\ast }\partial _{i}\ln \sqrt{%
|h_{3}h_{4}|},\qquad \beta =h_{4}^{\ast \ast }-h_{4}^{\ast }[\ln \sqrt{%
|h_{3}h_{4}|}]^{\ast },\qquad \gamma =\frac{3h_{4}^{\ast }}{2h_{4}}-\frac{%
h_{3}^{\ast }}{h_{3}}.  \label{abc}
\end{equation}%
The various partial derivatives are denoted as $a^{\bullet }=\partial
a/\partial x^{1},a^{^{\prime }}=\partial a/\partial x^{2},a^{\ast }=\partial
a/\partial y^{3}.$ This system of equations (\ref{ricci1a})--(\ref{ricci4a})
can be solved by choosing one of the ansatz functions (\textit{e.g.} $%
g_{1}\left( x^{i}\right) $ or $g_{2}\left( x^{i}\right) )$ and one of the
ansatz functions (\textit{e.g.} $h_{3}\left( x^{i},y^{3}\right) $ or $%
h_{4}\left( x^{i},y^{3}\right) )$ to take some arbitrary, but
physically interesting form. Then the other ansatz functions can
be analytically determined up to an integration in terms of this
choice. In this way we can generate a lost of different
solutions, but we impose the condition that the initial,
arbitrary choice of the ansatz functions is ``physically
interesting'' which means that one wants to make this original
choice so that the generated final solution yield a well behaved
metric.

In references \cite{vbel} it is proved that for
\begin{eqnarray}
g_{1} &=&-1,\quad g_{2}=r^{2}\left( \xi \right) q\left( \xi \right) ,
\label{data10} \\
h_{3} &=&-\eta _{3}\left( \xi ,\varphi \right) r^{2}\left( \xi \right) \sin
^{2}\theta ,\quad  \notag \\
h_{4} &=&\eta _{4}\left( \xi ,\varphi \right) h_{4[0]}\left( \xi \right) =1-%
\frac{2\mu }{r}+\varepsilon \frac{\Phi _{4}\left( \xi ,\varphi \right) }{%
2\mu ^{2}},  \notag
\end{eqnarray}%
with coordinates $x^{1}=\xi =\int dr\sqrt{1-2m/r+\varepsilon /r^{2}}%
,x^{2}=\theta ,y^{3}=\varphi ,y^{4}=t$ (the $(r,\theta ,\varphi )$ being
usual radial coordinates), the ansatz (\ref{ansatz18}) is a vacuum solution
with $\lambda =0$ of the equations (\ref{eq17}) which defines a black
ellipsoid with mass $\mu ,$ eccentricity $\varepsilon $ and gravitational
polarizations $q\left( \xi \right) ,\eta _{3}\left( \xi ,\varphi \right) $
and $\Phi _{4}\left( \xi ,\varphi \right) .$ Such black holes are certain
deformations of the Schawarzschild metrics to static configurations with
ellipsoidal horizons which is possible if generic off--diagonal metrics and
anholonomic frames are considered. In this subsection we show that the data (%
\ref{data10}) can be extended as to generate exact black
ellipsoid solutions with nontrivial cosmological constant $\lambda
=1/4$ which can be imbedded in string theory.

At the first \ step, we find a class of solutions with $g_{1}=-1$ and $\quad
g_{2}=g_{2}\left( \xi \right) $ solving the equation (\ref{ricci1a}), which
under such parametrizations transforms to
\begin{equation*}
g_{2}^{\bullet \bullet }-\frac{(g_{2}^{\bullet })^{2}}{2g_{2}}=2g_{2}\lambda
.
\end{equation*}%
With respect to the variable $Z=(g_{2})^{2}$ this equation is written as
\begin{equation*}
Z^{\bullet \bullet }+2\lambda Z=0
\end{equation*}%
which can be integrated in explicit form, $Z=Z_{[0]}\sin \left( \sqrt{%
2\lambda }\xi +\xi _{\lbrack 0]}\right) ,$ for some constants $Z_{[0]}$ and $%
\xi _{\lbrack 0]}$ which means that
\begin{equation}
g_{2}=-Z_{[0]}^{2}\sin ^{2}\left( \sqrt{2\lambda }\xi +\xi _{\lbrack
0]}\right)  \label{aux2}
\end{equation}%
parametrize a class of solution of (\ref{ricci1a}) for the signature $\left(
-,-,-,+\right) .$ For $\lambda \rightarrow 0$ we can approximate $%
g_{2}=r^{2}\left( \xi \right) q\left( \xi \right) =-\xi ^{2}$ and $%
Z_{[0]}^{2}=1$ which has compatibility with the data
(\ref{data10}). The solution (\ref{aux2}) with cosmological
constant (of string or non--string origin) induces oscillations
in the ''horozontal'' part of the d--metric.

The next step is to solve the equation (\ref{ricci2a}),%
\begin{equation*}
h_{4}^{\ast \ast }-h_{4}^{\ast }[\ln \sqrt{|h_{3}h_{4}|}]^{\ast }=-2\lambda
h_{3}h_{4}.
\end{equation*}%
For $\lambda =0$ a class of solution is given by any $\widehat{h}_{3}$ and $%
\widehat{h}_{4}$ related as
\begin{equation*}
\widehat{h}_{3}=\eta _{0}\left[ \left( \sqrt{|\hat{h}_{4}|}\right) ^{\ast }%
\right]^2
\end{equation*}%
for a constant $\eta _{0}$ chosen to be negative in order to generate the
signature $\left( -,-,-,+\right) .$ For non--trivial $\lambda ,$ we may
search the solution as
\begin{equation}
h_{3}=\widehat{h}_{3}\left( \xi ,\varphi \right) ~q_{3}\left( \xi ,\varphi
\right) \mbox{ and }h_{4}=\widehat{h}_{4}\left( \xi ,\varphi \right) ,
\label{sol15}
\end{equation}%
which solves (\ref{ricci2a}) if $q_{3}=1$ for $\lambda =0$ and
\begin{equation*}
q_{3}=\frac{1}{4\lambda }\left[ \int \frac{\hat{h}_{3}\hat{h}_{4}}{\hat{h}%
_{4}^{\ast }}d\varphi \right] ^{-1}\mbox{ for }\lambda \neq 0.
\end{equation*}

Now it is easy to write down the solutions of equations (\ref{ricci3a})
(being a linear equation for $w_{i^{\prime }})$ and (\ref{ricci4a}) (after
two integrations of $n_{i^{\prime }}$ on $\varphi ),$%
\begin{equation}
w_{i^{\prime }}=\varepsilon \widehat{w}_{i^{\prime }}=-\alpha _{i^{\prime
}}/\beta ,  \label{aux3}
\end{equation}%
were $\alpha _{i^{\prime }}$ and $\beta $ are computed by putting (\ref%
{sol15}) $\ $into corresponding values from (\ref{abc}) (we chose the
initial conditions as $w_{i^{\prime }}\rightarrow 0$ for $\varepsilon
\rightarrow 0)$ and
\begin{equation*}
n_{1}=\varepsilon \widehat{n}_{1}\left( \xi ,\varphi \right)
\end{equation*}%
where
\begin{eqnarray}
\widehat{n}_{1}\left( \xi ,\varphi \right) &=&n_{1[1]}\left( \xi \right)
+n_{1[2]}\left( \xi \right) \int d\varphi \ \eta _{3}\left( \xi ,\varphi
\right) /\left( \sqrt{|\eta _{4}\left( \xi ,\varphi \right) |}\right)
^{3},\eta _{4}^{\ast }\neq 0;  \label{auxf4} \\
&=&n_{1[1]}\left( \xi \right) +n_{1[2]}\left( \xi \right) \int d\varphi \
\eta _{3}\left( \xi ,\varphi \right) ,\eta _{4}^{\ast }=0;  \notag \\
&=&n_{1[1]}\left( \xi \right) +n_{1[2]}\left( \xi \right) \int d\varphi
/\left( \sqrt{|\eta _{4}\left( \xi ,\varphi \right) |}\right) ^{3},\eta
_{3}^{\ast }=0;  \notag
\end{eqnarray}%
with the functions $n_{k[1,2]}\left( \xi \right) $ to be stated by boundary
conditions.

We conclude that the set of data $g_{1}=-1,$ with non--tivial $g_{2}\left(
\xi \right) ,h_{3},h_{4},w_{i^{\prime }},n_{1}$ stated respectively by (\ref%
{aux2}), (\ref{sol15}), (\ref{aux3}), (\ref{auxf4}) define a black ellipsoid
solution with explicit dependence on cosmological constant $\lambda ,$ i. e.
a d--metric (\ref{ansatz18}), which can be induced from string theory for $%
\lambda =1/4.$ The stability of such string static black ellipsoids can be
proven exactly as it was done in Refs. \cite{vbel} for the vanishing
cosmological constant.

\subsection{2D Finsler structures in string theory}

There are some constructions which prove that two dimensional (2D) Finsler
structures can be embedded into the Einstein's theory of gravity \cite{dv}.
Here we analyze the conditions when such Finsler configurations can be
generated from string theory. The aim is to include a 2D Finsler metric (\ref%
{fmetric}) into a d--metric (\ref{dmetric}) being an exact solution of the
string corrected Einstein's equations (\ref{eq17}).

If
\begin{equation*}
h_{a^{\prime }b^{\prime }}=\frac{1}{2}\frac{\partial ^{2}F^{2}(x^{i^{\prime
}},y^{c^{\prime }})}{\partial y^{a^{\prime }}\partial y^{b^{\prime }}}
\end{equation*}%
for $i^{\prime },j^{\prime },...=1,2$ and $a^{\prime },b^{\prime },...=3,4$
and following the homogeneity conditions for Finsler metric, we can write
\begin{equation*}
F\left( x^{i^{\prime }},y^{3},y^{4}\right) =y^{3}f\left( x^{i^{\prime
}},s\right)
\end{equation*}%
for $s=y^{4}/y^{3}$ with $f\left( x^{i^{\prime }},s\right) =F\left(
x^{i^{\prime }},1,s\right) ,$ that
\begin{eqnarray}
h_{33} &=&\frac{s^{2}}{2}(f^{2})^{\ast \ast }-s(f^{2})^{\ast }+f^{2},
\label{hcoeff} \\
h_{34} &=&-\frac{s^{2}}{2}(f^{2})^{\ast \ast }+\frac{1}{2}(f^{2})^{\ast },
\notag \\
h_{44} &=&\frac{1}{2}(f^{2})^{\ast \ast },  \notag
\end{eqnarray}%
in this subsection we denote $a^{\ast }=\partial a/\partial s.$ The
condition of vanishing of the off--diagonal term $h_{34}$ gives us the
trivial case, when $f^{2}\simeq s^{2}...+...s+...,$ i. e. Riemannian 2D
metrics, so we can not include some general Finsler coefficients (\ref%
{hcoeff}) directly into a diagonal d--metric ansatz (\ref{ansatz18}). There
is also another problem related with the Cartan's N--connection (\ref{ncc})
being computed directly from the coefficients (\ref{hcoeff}) generated by a
function $f^{2}:$ all such values substituted into the equations (\ref%
{ricci2a}) - (\ref{ricci4a}) result in systems of nonlinear equations
containing the 6th and higher derivatives of $f$ on $s$ which is very
difficult to deal with.

We can include 2D Finsler structures in the Einstein and string gravity via
additional 2D anholnomic frame transforms,%
\begin{equation*}
h_{ab}=e_{a}^{a^{\prime }}\left( x^{i^{\prime }},s\right) ~e_{b}^{b^{\prime
}}\left( x^{i^{\prime }},s\right) ~h_{a^{\prime }b^{\prime }}\left(
x^{i^{\prime }},s\right)
\end{equation*}%
where $h_{a^{\prime }b^{\prime }}$ are induced by a Finsler metric $f^{2}$
as in (\ref{hcoeff}) and $h_{ab}$ may be diagonal, $h_{ab}=diag[h_{a}].$ We
also should consider an embedding of the Cartan's N--connection into a more
general N--connection, $N_{b^{\prime }}^{a^{\prime }}\subset N_{i^{\prime
}}^{a^{\prime }},$ via transforms $N_{i^{\prime }}^{a^{\prime }}=\hat{e}%
_{i^{\prime }}^{b^{\prime }}\left( x^{i^{\prime }},s\right) N_{b^{\prime
}}^{a^{\prime }}$ where $\hat{e}_{i^{\prime }}^{b^{\prime }}\left(
x^{i^{\prime }},s\right) $ are some additionan frame transforms in the
off--diagonal sector of the ansatz (\ref{ansatz}). Such way generated
metrics,
\begin{eqnarray*}
\delta s^{2} &=&g_{i^{\prime }}(dx^{i^{\prime }})^{2}+e_{a}^{a^{\prime
}}~e_{a}^{b^{\prime }}h_{a^{\prime }b^{\prime }}(\delta y^{a})^{2}, \\
\delta y^{a} &=&dy^{a}+\hat{e}_{i^{\prime }}^{b^{\prime }}N_{b^{\prime
}}^{a^{\prime }}dx^{i^{\prime }}\quad
\end{eqnarray*}%
may be constrained by the condition to be an exact solution of the Einstein
equations with (or not) certain string corrections. As a matter of
principle, any string black ellipsoid \ configuration (of the type examined
in the previous subsection) can be related to a 2D Finsler configuration for
corresponding coefficients $e_{a}^{a^{\prime }}$ and $\hat{e}_{i^{\prime
}}^{b^{\prime }}.$ An explicit form of anisotropic configuration is to be
stated by corresponding boundary conditions and the type of anholonomic
transforms. Finally, we note that instead of a 2D Finsler metric (\ref%
{fmetric}) we can use a 2D Lagrange metric (\ref{mfl}).

\subsection{Moving soliton--black hole string configurations}

In this subsection, we consider that the primed coordinates are 5D ones
obtained after a string compactification background for the NS--NS sector
being common to both the heterotic and type II string theories. The $%
u^{\alpha ^{\prime }}=(x^{i^{\prime }},y^{a^{\prime }})$ are split into
coordinates $x^{i},$ with indices $i^{\prime },j^{\prime },k^{\prime
}...=1,2,3,$ and coordinates $y^{a^{\prime }},$ with indices $a^{\prime
},b^{\prime },c^{\prime },...=4,5.$ Explicitly the coordinates are of the
form
\begin{equation*}
x^{i^{\prime }}=(x^{1}=\chi ,\quad x^{2}=\phi =\ln \widehat{\rho },\quad
x^{3}=\theta )\quad \mbox{and }\quad y^{a^{\prime }}=\left( y^{4}=v,\qquad
y^{5}=p\right) ,
\end{equation*}%
where $\chi $ is the 5th extra--dimensional coordinate and $\widehat{\rho }$
will be related with the 4D Schwarzschild coordinate. We analyze a metric
interval written as
\begin{equation}
ds^{2}=\Omega ^{2}(x^{i^{\prime }},v)\hat{{g}}_{\alpha ^{\prime }\beta
^{\prime }}\left( x^{i^{\prime }},v\right) du^{\alpha ^{\prime }}du^{\beta
^{\prime }},  \label{cmetric}
\end{equation}%
were the coefficients $\hat{{g}}_{\alpha ^{\prime }\beta ^{\prime }}$ are
parametrized by the ansatz {\scriptsize
\begin{equation}
\left[
\begin{array}{ccccc}
g_{1}+(w_{1}^{\ 2}+\zeta _{1}^{\ 2})h_{4}+n_{1}^{\ 2}h_{5} &
(w_{1}w_{2}+\zeta _{1}\zeta _{2})h_{4}+n_{1}n_{2}h_{5} & (w_{1}w_{3}+\zeta
_{1}\zeta _{3})h_{4}+n_{1}n_{3}h_{5} & (w_{1}+\zeta _{1})h_{4} & n_{1}h_{5}
\\
(w_{1}w_{2}+\zeta _{1}\zeta _{2})h_{4}+n_{1}n_{2}h_{5} & g_{2}+(w_{2}^{\
2}+\zeta _{2}^{\ 2})h_{4}+n_{2}^{\ 2}h_{5} & (w_{2}w_{3}++\zeta _{2}\zeta
_{3})h_{4}+n_{2}n_{3}h_{5} & (w_{2}+\zeta _{2})h_{4} & n_{2}h_{5} \\
(w_{1}w_{3}+\zeta _{1}\zeta _{3})h_{4}+n_{1}n_{3}h_{5} & (w_{2}w_{3}++\zeta
_{2}\zeta _{3})h_{4}+n_{2}n_{3}h_{5} & g_{3}+(w_{3}^{\ 2}+\zeta _{3}^{\
2})h_{4}+n_{3}^{\ 2}h_{5} & (w_{3}+\zeta _{3})h_{4} & n_{3}h_{5} \\
(w_{1}+\zeta _{1})h_{4} & (w_{2}+\zeta _{2})h_{4} & (w_{3}+\zeta _{3})h_{4}
& h_{4} & 0 \\
n_{1}h_{5} & n_{2}h_{5} & n_{3}h_{5} & 0 & h_{5}%
\end{array}%
\right]  \label{ansatzc}
\end{equation}%
} The metric coefficients are necessary class smooth functions of the form:
\begin{eqnarray}
g_{1} &=&\pm 1,\qquad g_{2,3}=g_{2,3}(x^{2},x^{3}),\qquad
h_{4,5}=h_{4,5}(x^{i^{\prime }},v)=\eta
_{4,5}(x^{i},v)h_{4,5[0]}(x^{k^{\prime }}),  \notag \\
w_{i^{\prime }} &=&w_{i^{\prime }}(x^{k^{\prime }},v),\qquad n_{i^{\prime
}}=n_{i^{\prime }}(x^{k^{\prime }},v),\qquad \zeta _{i^{\prime }}=\zeta
_{i^{\prime }}(x^{k^{\prime }},v),\qquad \Omega =\Omega (x^{i^{\prime }},v).
\label{par1}
\end{eqnarray}%
The quadratic line element (\ref{cmetric}) with metric coefficients (\ref%
{ansatzc}) can be diagonalized by anholonmic transforms,
\begin{equation}
\delta s^{2}=\Omega ^{2}(x^{i^{\prime
}},v)[g_{1}(dx^{1})^{2}+g_{2}(dx^{2})^{2}+g_{3}(dx^{3})^{2}+h_{4}(\hat{{%
\delta }}v)^{2}+h_{5}(\delta p)^{2}],  \label{cdmetric}
\end{equation}%
with respect to the anholonomic co--frame $\left( dx^{i^{\prime }},\hat{{%
\delta }}v,\delta p\right) ,$ where
\begin{equation}
\hat{\delta}v=dv+(w_{i^{\prime }}+\zeta _{i^{\prime }})dx^{i^{\prime
}}+\zeta _{5}\delta p\qquad \mbox{
and }\qquad \delta p=dp+n_{i^{\prime }}dx^{i^{\prime }}  \label{ddif2}
\end{equation}%
which is dual to the frame $\left( \hat{{\delta }}_{i^{\prime }},\partial
_{4},\hat{{\partial }}_{5}\right) ,$ where
\begin{equation}
\hat{{\delta }}_{i^{\prime }}=\partial _{i^{\prime }}-(w_{i^{\prime }}+\zeta
_{i^{\prime }})\partial _{4}+n_{i^{\prime }}\partial _{5},\qquad \hat{{%
\partial }}_{5}=\partial _{5}-\zeta _{5}\partial _{4}.  \label{dder2}
\end{equation}%
The simplest way to compute the nontrivial coefficients of the Ricci tensor
for the (\ref{cdmetric}) is to do this with respect to anholonomic bases ( %
\ref{ddif2}) and (\ref{dder2}) (see details in \cite{vmethod,vsingl}), which
reduces the 5D vacuum Einstein equations to the following system (in this
paper containing a non--trivial cosmological constant):
\begin{eqnarray}
\frac{1}{2}R_{1}^{1}=R_{2}^{2}=R_{3}^{3}=-\frac{1}{2g_{2}g_{3}}%
[g_{3}^{\bullet \bullet }-\frac{g_{2}^{\bullet }g_{3}^{\bullet }}{2g_{2}}-%
\frac{(g_{3}^{\bullet })^{2}}{2g_{3}}+g_{2}^{^{\prime \prime }}-\frac{%
g_{2}^{^{\prime }}g_{3}^{^{\prime }}}{2g_{3}}-\frac{(g_{2}^{^{\prime }})^{2}%
}{2g_{2}}] &=&\lambda ,  \label{ricci7a} \\
R_{4}^{4}=R_{5}^{5}=-\frac{\beta }{2h_{4}h_{5}} &=&\lambda ,  \label{ricci8a}
\\
R_{4i^{\prime }}=-w_{i^{\prime }}\frac{\beta }{2h_{5}}-\frac{\alpha
_{i^{\prime }}}{2h_{5}} &=&0,  \label{ricci9a} \\
R_{5i^{\prime }}=-\frac{h_{5}}{2h_{4}}\left[ n_{i^{\prime }}^{\ast \ast
}+\gamma n_{i^{\prime }}^{\ast }\right] &=&0,  \label{ricci10a}
\end{eqnarray}%
with the conditions that
\begin{equation}
\Omega ^{q_{1}/q_{2}}=h_{4}~(q_{1}\mbox{ and }q_{2}\mbox{ are
integers}),  \label{confq}
\end{equation}%
and $\zeta _{i}$ satisfies the equations
\begin{equation}
\partial _{i^{\prime }}\Omega -(w_{i^{\prime }}+\zeta _{i^{\prime }})\Omega
^{\ast }=0,  \label{confeq}
\end{equation}%
The coefficients of equations (\ref{ricci7a}) - (\ref{ricci10a}) are given
by
\begin{equation}
\alpha _{i^{\prime }}=\partial _{i}{h_{5}^{\ast }}-h_{5}^{\ast }\partial
_{i^{\prime }}\ln \sqrt{|h_{4}h_{5}|},\qquad \beta =h_{5}^{\ast \ast
}-h_{5}^{\ast }[\ln \sqrt{|h_{4}h_{5}|}]^{\ast },\qquad \gamma =\frac{%
3h_{5}^{\ast }}{2h_{5}}-\frac{h_{4}^{\ast }}{h_{4}}.  \label{abc1}
\end{equation}%
The various partial derivatives are denoted as $a^{\bullet }=\partial
a/\partial x^{2},a^{^{\prime }}=\partial a/\partial x^{3},a^{\ast }=\partial
a/\partial v.$

The system of equations (\ref{ricci7a})--(\ref{ricci10a}), (\ref{confq}) and
(\ref{confeq}) can be solved by choosing one of the ansatz functions (%
\textit{e.g.} $h_{4}\left( x^{i^{\prime }},v\right) $ or $h_{5}\left(
x^{i^{\prime }},v\right) )$ to take some arbitrary, but physically
interesting form. Then the other ansatz functions can be analytically
determined up to an integration in terms of this choice. In this way one can
generate many solutions, but the requirement that the initial, arbitrary
choice of the ansatz functions be ``physically interesting'' means that one
wants to make this original choice so that the final solution generated in
this way yield a well behaved solution. To satisfy this requirement we start
from well known solutions of Einstein's equations and then use the above
procedure to deform this solutions in a number of ways as to include it in a
string theory. In the simplest case we derive 5D locally anisotropic string
gravity solutions connected to the the Schwarzschild solution in \textit{%
isotropic spherical coordinates} \cite{ll} given by the quadratic line
interval%
\begin{equation}
ds^{2}=\left( \frac{\widehat{\rho }-1}{\widehat{\rho }+1}\right)
^{2}dt^{2}-\rho _{g}^{2}\left( \frac{\widehat{\rho }+1}{\widehat{\rho }}%
\right) ^{4}\left( d\widehat{\rho }^{2}+\widehat{\rho }^{2}d\theta ^{2}+%
\widehat{\rho }^{2}\sin ^{2}\theta d\varphi ^{2}\right) .  \label{schw}
\end{equation}%
We identify the coordinate $\widehat{\rho }$ with the re--scaled isotropic
radial coordinate, $\widehat{\rho }=\rho /\rho _{g},$ with $\rho
_{g}=r_{g}/4 $; $\rho $ is connected with the usual radial coordinate $r$ by
$r=\rho \left( 1+r_{g}/4\rho \right) ^{2}$; $r_{g}=2G_{[4]}m_{0}/c^{2}$ is
the 4D Schwarzschild radius of a point particle of mass $m_{0}$; $%
G_{[4]}=1/M_{P[4]}^{2}$ is the 4D Newton constant expressed via the Planck
mass $M_{P[4]}$ (in general, we may consider that $M_{P[4]}$ may be an
effective 4D mass scale which arises from a more fundamental scale of the
full, higher dimensional spacetime); we set $c=1.$

The metric (\ref{schw}) is a vacuum static solution of 4D Einstein equations
with spherical symmetry describing the gravitational field of a point
particle of mass $m_{0}.$ It has a singularity for $r=0$ and a spherical
horizon at $r=r_{g},$ or at $\widehat{\rho }=1$ in the re--scaled isotropic
coordinates. This solution is parametrized by a diagonal metric given with
respect to holonomic coordinate frames. This spherically symmetric solution
can be deformed in various interesting ways using the anholonomic frames
method.

Vacuum gravitational 2D solitons in 4D Einstein vacuum gravity were
originally investigated by Belinski and Zakharov \cite{belinski}. In Refs. %
\cite{vsolsp} 3D solitonic configurations were constructed on anisotropic
Taub-NUT backgrounds. Here we show that 3D solitonic/black hole
configurations can be embedded into the 5D locally anisotropic string
gravity.

\subsubsection{3D solitonic deformations in string gravity}

The simplest way to construct a solitonic deformation of the off--diagonal
metric in equation (\ref{ansatzc}) is to take one of the ``polarization''
factors $\eta _{4}$, $\eta _{5}$ from (\ref{par1}) or the ansatz function $%
n_{i^{\prime }}$ as a solitonic solution of some particular non-linear
equation. The rest of the ansatz functions can then be found by carrying out
the integrations of equations (\ref{ricci7a})-- (\ref{confeq}).

As an example of this procedure we suggest to take $\eta _{5}(r,\theta ,\chi
)$ as a soliton solution of the Kadomtsev--Petviashvili (KdP) equation or
(2+1) sine-Gordon (SG) equation (Refs. \cite{kad} contain the original
results, basic references and methods for handling such non-linear equations
with solitonic solutions). In the KdP case $\eta _{5}(v,\theta ,\chi )$
satisfies the following equation
\begin{equation}
\eta _{5}^{\ast \ast }+\epsilon \left( \dot{\eta}_{5}-6\eta _{5}\eta
_{5}^{\prime }+\eta _{5}^{\prime \prime \prime }\right) ^{\prime }=0,\qquad
\epsilon =\pm 1,  \label{kdp}
\end{equation}%
while in the most general SG case $\eta _{5}(v,\chi )$ satisfies
\begin{equation}
\pm \eta _{5}^{\ast \ast }\mp \ddot{\eta}_{5}=\sin (\eta _{5}).
\label{sineq}
\end{equation}%
For simplicity, we can also consider less general versions of the SG
equation where $\eta _{5}$ depends on only one (\textit{e.g.} $v$ and $x_{1}$%
) variable. We use the notation $\eta _{5}=\eta _{5}^{KP}$ or $\eta
_{5}=\eta _{5}^{SG}$ ($h_{5}=h_{5}^{KP}$ or $h_{5}=h_{5}^{SG}$) depending on
if ($\eta _{5}$ ) ($h_{5}$) satisfies equation (\ref{kdp}), or (\ref{sineq})
respectively.

For a stated solitonic form for $h_{5}=h_{5}^{KP,SG},$ $h_{4}$ can be found
from
\begin{equation}
h_{4}=h_{4}^{KP,SG}=h_{[0]}^{2}\left[ \left( \sqrt{|h_{5}^{KP,SG}(x^{i^{%
\prime }},v)|}\right) ^{\ast }\right] ^{2}  \label{p1b}
\end{equation}%
where $h_{[0]}$ is a constant. By direct substitution it can be shown that
equation (\ref{p1b}) solves equation (\ref{ricci8a}) with $\beta $ given by
( \ref{abc1}) when $h_{5}^{\ast }\neq 0$ but $\lambda =0.$ If $h_{5}^{\ast
}=0,$ then $\hat{h}_{4}$ is an arbitrary function $\hat{h}_{4}(x^{i^{\prime
}},v)$. In either case we will denote the ansatz function determined in this
way as $\hat{h}_{4}^{KP,SG}$ although it does not necessarily share the
solitonic character of $\hat{h}_{5}$. Substituting the values $\hat{h}%
_{4}^{KP,SG}$ and $\hat{h}_{5}^{KP,SG}$ into $\gamma $ from equation (\ref%
{abc}) gives, after two $v$ integrations of equation (\ref{ricci4a}), the
ansatz functions $n_{i^{\prime }}=n_{i^{\prime }}^{KP,SG}(v,\theta ,\chi ).$
Solutions with $\lambda \neq 0$ can be generated similarly as in (\ref{sol15}%
) by redefining
\begin{equation*}
h_{4}=\widehat{h}_{4}\left( x^{i^{\prime }},v\right) ~q_{4}\left(
x^{i^{\prime }},v\right) \mbox{ and }h_{5}=\widehat{h}_{5}\left(
x^{i^{\prime }},v\right) ,
\end{equation*}%
which solves (\ref{ricci8a}) if $q_{4}=1$ for $\lambda =0$ and
\begin{equation}
q_{4}=\frac{1}{4\lambda }\left[ \int \frac{\hat{h}_{5}\hat{h}_{4}}{\hat{h}%
_{5}^{\ast }}dv\right] ^{-1}\mbox{ for }\lambda \neq 0.  \label{quf}
\end{equation}

Here, for simplicity, we \ may set $g_{2}=-1$ but
\begin{equation}
g_{3}=-Z_{[0]}^{2}\sin ^{2}\left( \sqrt{2\lambda }x^{3}+\xi
_{\lbrack 0]}\right) ,\  Z_{[0]}, \xi _{[0]}=const, \label{aux5}
\end{equation}%
parametrize a class of solution of (\ref{ricci7a}) for the signature $\left(
-,-,-,-,+\right) $ like we constructed the solution (\ref{aux2}). In ref. %
\cite{vsolsp,vsingl} it was shown how to generate solutions using 2D
solitonic configurations for $g_{2}$ or $g_{3}.$

The main conclusion to be formulated here is that the ansatz (\ref{ansatzc}%
), when treated with anholonomic frames, has a degree of freedom that allows
one to pick one of the ansatz functions ($\eta _{4}$ , $\eta _{5}$ , or $%
n_{i^{\prime }}$) to satisfy some 3D solitonic equation. Then in terms of
this choice all the other ansatz functions can be generated up to carrying
out some explicit integrations and differentiations. In this way it is
possible to build exact solutions of the 5D string gravity equations with a
solitonic character.

\subsubsection{Solitonically propagating string black hole backgrounds}

\label{solitonical}

The Schwarzschild solution is given in terms of the parameterization in (\ref%
{ansatzc}) by
\begin{eqnarray}
g_{1} &=&\pm 1,\qquad g_{2}=g_{3}=-1,\qquad h_{4}=h_{4[0]}(x^{i^{\prime
}}),\qquad h_{5}=h_{5[0]}(x^{i^{\prime }}),  \notag \\
w_{i^{\prime }} &=&0,\qquad n_{i^{\prime }}=0,\qquad \zeta _{i^{\prime
}}=0,\qquad \Omega =\Omega _{\lbrack 0]}(x^{i^{\prime }}),  \notag
\end{eqnarray}%
with
\begin{equation}
h_{4[0]}(x^{i})=\frac{b(\phi )}{a(\phi )},\qquad h_{5[0]}(x^{i^{\prime
}})=-\sin ^{2}\theta ,\qquad \Omega _{\lbrack 0]}^{2}(x^{i^{\prime
}})=a(\phi )  \label{aux1}
\end{equation}%
or alternatively, for another class of solutions,
\begin{equation}
h_{4[0]}(x^{i^{\prime }})=-\sin ^{2}\theta ,\qquad h_{5[0]}(x^{i^{\prime }})=%
\frac{b(\phi )}{a(\phi )},  \label{aux2a}
\end{equation}%
were
\begin{equation}
a(\phi )=\rho _{g}^{2}\frac{\left( e^{\phi }+1\right) ^{4}}{e^{2\phi }}%
\qquad \mbox{ and }\qquad b(\phi )=\frac{\left( e^{\phi }-1\right) ^{2}}{%
\left( e^{\phi }+1\right) ^{2}},  \label{ab}
\end{equation}%
Putting this together gives
\begin{equation}
ds^{2}=\pm d\chi ^{2}-a(\phi )\left( d\lambda ^{2}+d\theta ^{2}+\sin
^{2}\theta d\varphi ^{2}\right) +b\left( \phi \right) dt^{2}  \label{schw5}
\end{equation}%
which represents a trivial embedding of the 4D Schwarzschild metric (\ref%
{schw}) into the 5D spacetime. We now want to deform anisotropically the
coefficients of (\ref{schw5}) in the following way
\begin{eqnarray*}
h_{4[0]}(x^{i^{\prime }}) &\rightarrow &h_{4}(x^{i^{\prime }},v)=\eta
_{4}\left( x^{i^{\prime }},v\right) h_{4[0]}(x^{i^{\prime }}), \\
~h_{5[0]}(x^{i^{\prime }}) &\rightarrow &h_{5}(x^{i^{\prime }},v)=\eta
_{5}\left( x^{i^{\prime }},v\right) h_{5[0]}(x^{i^{\prime }}), \\
\Omega _{\lbrack 0]}^{2}(x^{i^{\prime }}) &\rightarrow &\Omega
^{2}(x^{i^{\prime }},v)=\Omega _{\lbrack 0]}^{2}(x^{i^{\prime }})\Omega
_{\lbrack 1]}^{2}(x^{i^{\prime }},v).
\end{eqnarray*}%
The factors $\eta _{i^{\prime }}$ and $\Omega _{\lbrack 1]}^{2}$ can be
interpreted as re-scaling or ''renormalizing'' the original ansatz
functions. These gravitational ``polarization'' factors -- $\eta _{4,5}$ and
$\Omega _{\lbrack 1]}^{2}$ -- generate non--trivial values for $w_{i^{\prime
}}(x^{i^{\prime }},v),n_{i^{\prime }}(x^{i^{\prime }},v)$ and $\zeta
_{i^{\prime }}(x^{i^{\prime }},v),$ via the vacuum equations (\ref{ricci7a}%
)-- (\ref{confeq}). We shall also consider more general nonlinear
polarizations which can not be expresses as $h\sim $ $\eta h_{[0]}$ and show
how the coefficients $a(\phi )$ and $b(\phi )$ of the Schwarzschild metric
can be polarized by choosing the original, arbitrary ansatz function to be
some 3D soliton configuration.

The horizon is defined by the vanishing of the coefficient $b\left( \phi
\right) $ from equation (\ref{ab}). This occurs when $e^{\phi }=1$. In order
to create a solitonically propagating black hole we define the function $%
\tau =\phi -\tau _{0}\left( \chi ,v\right) $, and let $\tau _{0}\left( \chi
,v\right) $ be a soliton solution of either the 3D KdP equation (\ref{kdp}),
or the SG equation (\ref{sineq}). This redefines $b\left( \phi \right) $ as
\begin{equation*}
b\left( \phi \right) \rightarrow B\left( x^{i^{\prime }},v\right) =\frac{%
e^{\tau }-1}{e^{\phi }+1}.
\end{equation*}%
A class of 5D string gravity metrics can be constructed by parametrizing $%
h_{4}=\eta _{4}\left( x^{i^{\prime }},v\right) $ $h_{4[0]}(x^{i^{\prime }})$
and $h_{5}=B\left( x^{i^{\prime }},v\right) /a\left( \phi \right) $, or
inversely, $h_{4}=B\left( x^{i^{\prime }},v\right) /a\left( \phi \right) $
and $h_{5}=\eta _{5}\left( x^{i^{\prime }},v\right) h_{5[0]}(x^{i^{\prime
}}).$ The polarization $\eta _{4}\left( x^{i^{\prime }},v\right) $ \ (or \ $%
\eta _{5}\left( x^{i^{\prime }},v\right) )$ is determined from equation (\ref%
{p1b}) with the factor $q_{4}$ (\ref{quf}) \ included in $h^{2},$
\begin{equation*}
|\eta _{4}\left( x^{i^{\prime }},v\right) h_{4(0)}(x^{i^{\prime }})|=h^{2}
\left[ \left( \sqrt{\left| \frac{B\left( x^{i^{\prime }},v\right) }{a\left(
\phi \right) }\right| }\right) ^{\ast }\right] ^{2}
\end{equation*}%
or
\begin{equation*}
\left| \frac{B\left( x^{i},v\right) }{a\left( \phi \right) }\right|
=h^{2}h_{5(0)}(x^{i^{\prime }})\left[ \left( \sqrt{|\eta _{5}\left(
x^{i^{\prime }},v\right) |}\right) ^{\ast }\right] ^{2}.
\end{equation*}%
The last step in constructing of the form for these solitonically
propagating black hole solutions is to use $h_{4}$ and $h_{5}$ in equation (%
\ref{ricci4a}) to determine $n_{k^{\prime }}$
\begin{eqnarray}
n_{k^{\prime }} &=&n_{k^{\prime }[1]}(x^{i^{\prime }})+n_{k^{\prime
}[2]}(x^{i^{\prime }})\int \frac{h_{4}}{(\sqrt{|h_{5}|})^{3}}dv,\qquad
h_{5}^{\ast }\neq 0;  \label{nnn1} \\
&=&n_{k^{\prime }[1]}(x^{i^{\prime }})+n_{k^{\prime }[2]}(x^{i^{\prime
}})\int h_{4}dv,\qquad h_{5}^{\ast }=0;  \notag \\
&=&n_{k^{\prime }[1]}(x^{i^{\prime }})+n_{k^{\prime }[2]}(x^{i^{\prime
}})\int \frac{1}{(\sqrt{|h_{5}|})^{3}}dv,\qquad h_{4}^{\ast }=0,  \notag
\end{eqnarray}%
where $n_{k[1,2]}\left( x^{i^{\prime }}\right) $ are set by boundary
conditions.

The simplest version of the above class of solutions are the so--called $t$%
--solutions (depending on $t$--variable), defined by a pair of ansatz
functions, $\left[ B\left( x^{i^{\prime }},t\right) ,h_{5(0)}\right] ,$ with
$h_{5}^{\ast }=0$ and $B\left( x^{i^{\prime }},t\right) $ being a 3D
solitonic configuration. Such solutions have a spherical horizon when $%
h_{4}=0,$ \textit{i.e.} when $\tau =0$. This solution describes a
propagating black hole horizon. The propagation occurs via a 3D solitonic
wave form depending on the time coordinate, $t$, and on the 5$^{th}$
coordinate $\chi $. The form of the ansatz functions for this solution (both
with trivial and non-trivial conformal factors) is
\begin{eqnarray}
\mbox{$t$--solutions} &:&(x^{1}=\chi ,\qquad x^{2}=\phi ,\qquad x^{3}=\theta
,\qquad y^{4}=v=t,\qquad y^{5}=p=\varphi ),  \notag \\
g_{1} &=&\pm 1,g_{2}=-1,g_{3}=-Z_{[0]}^{2}\sin ^{2}\left( \sqrt{2\lambda }%
x^{3}+\xi _{\lbrack 0]}\right) ,\tau =\phi -\tau _{0}\left( \chi ,t\right) ,
\notag \\
h_{4} &=&B/a(\phi ),h_{5}=h_{5(0)}(x^{i^{\prime }})=-\sin ^{2}\theta ,\omega
=\eta _{5}=1,B\left( x^{i^{\prime }},t\right) =\frac{e^{\tau }-1}{e^{\phi }+1%
},  \notag \\
w_{i^{\prime }} &=&\zeta _{i^{\prime }}=0,\qquad n_{k^{\prime }}\left(
x^{i^{\prime }},t\right) =n_{k^{\prime }[1]}(x^{i^{\prime }})+n_{k^{\prime
}[2]}(x^{i^{\prime }})\int B\left( x^{i^{\prime }},t\right) dt,
\label{sol6t}
\end{eqnarray}%
where $q_{4}$ is chosen to preserve the condition $w_{i^{\prime }}=\zeta
_{i^{\prime }}=0.$

As a simple example of the above solutions we take $\tau _{0}$ to satisfy
the SG equation $\partial _{\chi \chi }\tau _{0}-\partial _{tt}\tau
_{0}=\sin (\tau _{0})$. This has the standard propagating kink solution
\begin{equation*}
\tau _{0}(\chi ,t)=4\tan ^{-1}\left[ \pm \gamma (\chi -Vt)\right]
\end{equation*}%
where $\gamma =(1-V^{2})^{-1/2}$ and $V$ is the velocity at which the kink
moves into the extra dimension $\chi $. To obtain the simplest form of this
solution we also take $n_{k^{\prime }[1]}(x^{i^{\prime }})=n_{k^{\prime
}[2]}(x^{i^{\prime }})=0.$ This example can be easily extended to solutions
with a non-trivial conformal factor $\Omega $ that gives an exponentially
suppressing factor, $\exp [-2k|\chi |],$ see details in Ref. \cite{vsingl}.
In this manner one has an effective 4D black hole which propagates from the
3D brane into the non-compact, but exponentially suppressed extra dimension,
$\chi $.

The solution constructed in this subsection describes propagating 4D
Schwarzschild black holes in a bulk 5D spacetime obtained from string
theory. The propagation arises from requiring that certain of the ansatz
functions take a 3D soliton form. In the simplest version of these
propagating solutions the parameters of the ansatz functions are constant,
and the horizons are spherical. It can be also shown that such propagating
solutions could be formed with a polarization of the parameters and/or
deformation of the horizons, see the non--string case in \cite{vsingl}.

\subsection{Noncommutative anisotropic wormholes and strings}

Let us construct and analyze an exact 5D solution of the string gravity
which can also considered as a noncommutative structure in string theory. \
The d--metric ansatz is taken in the form%
\begin{eqnarray}
\delta s^{2} &=&g_{1}(dx^{1})^{2}+g_{2}(dx^{2})^{2}+g_{3}(dx^{3})^{2}+h_{4}({%
\delta }y^{4})^{2}+h_{5}(\delta y^{5})^{2},  \notag \\
{\delta }y^{4} &=&{d}y^{4}+w_{k^{\prime }}\left( x^{i^{\prime }},v\right)
dx^{k^{\prime }},{\delta }y^{5}={d}y^{5}+n_{k^{\prime }}\left( x^{i^{\prime
}},v\right) dx^{k^{\prime }};i^{\prime },k^{\prime }=1,2,3,  \label{ans20}
\end{eqnarray}%
where
\begin{eqnarray}
g_{1} &=&1,\quad g_{2}=g_{2}(r),\quad g_{3}=-a(r),  \label{anz6a} \\
h_{4} &=&\hat{h}_{4}=\widehat{\eta }_{4}\left( r,\theta ,\varphi \right)
h_{4[0]}(r),\quad h_{5}=\hat{h}_{5}=\widehat{\eta }_{5}\left( r,\theta
,\varphi \right) h_{5[0]}(r,\theta )  \notag
\end{eqnarray}%
for the parametrization of coordinate of type
\begin{equation}
x^{1}=t,x^{2}=r,x^{3}=\theta ,y^{4}=v=\varphi ,y^{5}=p=\chi  \label{coord5}
\end{equation}%
where $t$ is the time coordinate, $\left( r,\theta ,\varphi \right) $ are
spherical coordinates, $\chi $ is the 5th coordinate; $\varphi $ is the
anholonomic coordinate; for this ansatz there is not considered the
dependence of d--metric coefficients on the second anholonomic coordinate $%
\chi .$ The data%
\begin{eqnarray}
g_{1} &=&1,~\hat{g}_{2}=-1,~g_{3}=-a(r),  \label{data6a} \\
~h_{4[0]}(r) &=&-r_{0}^{2}e^{2\psi (r)},~\eta _{4}=1/\kappa _{r}^{2}\left(
r,\theta ,\varphi \right) ,~h_{5[0]}=-a\left( r\right) \sin ^{2}\theta
,~\eta _{5}=1,  \notag \\
w_{1} &=&\widehat{w}_{1}=\omega \left( r\right) ,~w_{2}=\widehat{w}%
_{2}=0,w_{3}=~\widehat{w}_{3}=n\cos \theta /\kappa _{n}^{2}\left( r,\theta
,\varphi \right) ,  \notag \\
n_{1} &=&\widehat{n}_{1}=0,~n_{2,3}=\widehat{n}_{2,3}=n_{2,3[1]}\left(
r,\theta \right) \int \ln |\kappa _{r}^{2}\left( r,\theta ,\varphi \right)
|d\varphi  \notag
\end{eqnarray}%
for some constants $r_{0}$ $\ $\ and $n$ and arbitrary functions $a(r),\psi
(r)$ and arbitrary vacuum gravitational polarizations $\kappa _{r}\left(
r,\theta ,\varphi \right) $ and $\kappa _{n}\left( r,\theta ,\varphi \right)
$ define an exact vacuum 5D solution of Kaluza--Klein gravity \cite{vsingl1}
describing a locally anisotropic wormhole with elliptic gravitational vacuum
polarization of charges,
\begin{equation*}
\frac{q_{0}^{2}}{4a\left( 0\right) \kappa _{r}^{2}}+\frac{Q_{0}^{2}}{%
4a\left( 0\right) \kappa _{n}^{2}}=1,
\end{equation*}%
where $q_{0}=2\sqrt{a\left( 0\right) }\sin \alpha _{0}$ and $Q_{0}=2\sqrt{%
a\left( 0\right) }\cos \alpha _{0}$ are respectively the electric and
magnetic charges and $2\sqrt{a\left( 0\right) }\kappa _{r}$ and $2\sqrt{%
a\left( 0\right) }\kappa _{n}$ are ellipse's axes.

The first aim in this subsection is to prove that following the ansatz (\ref%
{ans20}) we can construct locally anisotropic wormhole metrics in string
gravity as solutions of the system of equations (\ref{ricci7a}) - (\ref%
{ricci10a}) with redefined coordinates as in (\ref{coord5}). Having the
vacuum data (\ref{data6a}) we may generalize the solution for a nontrivial
cosmological constant following the method presented in subsection \ref%
{solitonical}, when the new solutions are reprezented
\begin{equation}
h_{4}=\widehat{h}_{4}\left( x^{i^{\prime }},v\right) ~q_{4}\left(
x^{i^{\prime }},v\right) \mbox{ and }h_{5}=\widehat{h}_{5}\left(
x^{i^{\prime }},v\right) ,  \label{shift3}
\end{equation}%
with $\widehat{h}_{4,5}$ taken as in (\ref{anz6a}) which solves (\ref%
{ricci8a}) if $q_{4}=1$ for $\lambda =0$ and
\begin{equation*}
q_{4}=\frac{1}{4\lambda }\left[ \int \frac{\hat{h}_{5}\left( r,\theta
,\varphi \right) \hat{h}_{4}\left( r,\theta ,\varphi \right) }{\hat{h}%
_{5}^{\ast }\left( r,\theta ,\varphi \right) }d\varphi \right] ^{-1}%
\mbox{
for }\lambda \neq 0.
\end{equation*}%
This $q_{4}$ can be considered as an additional polarization to $\eta _{4}$
induced by the cosmological constant $\lambda .$ We state $g_{2}=-1$ but
\begin{equation*}
g_{3}=-\sin ^{2}\left( \sqrt{2\lambda }\theta +\xi _{\lbrack 0]}\right) ,
\end{equation*}%
which give of solution of (\ref{ricci7a}) with signature $\left(
+,-,-,-,-\right) $ which is different from the solution (\ref{aux2}). A
non--trivial $q_{4}$ results in modification of coefficients (\ref{abc1}),
\begin{eqnarray*}
\alpha _{i^{\prime }} &=&\hat{\alpha}_{i^{\prime }}+\alpha _{i^{\prime
}}^{[q]},~\beta =\hat{\beta}+\beta ^{\lbrack q]},~\gamma =\hat{\gamma}%
+\gamma ^{\lbrack q]}, \\
\hat{\alpha}_{i^{\prime }} &=&\partial _{i}{\hat{h}_{5}^{\ast }}-\hat{h}%
_{5}^{\ast }\partial _{i^{\prime }}\ln \sqrt{|\hat{h}_{4}\hat{h}_{5}|}%
,\qquad \hat{\beta}=\hat{h}_{5}^{\ast \ast }-\hat{h}_{5}^{\ast }[\ln \sqrt{|%
\hat{h}_{4}\hat{h}_{5}|}]^{\ast },\qquad \hat{\gamma}=\frac{3\hat{h}%
_{5}^{\ast }}{2\hat{h}_{5}}-\frac{\hat{h}_{4}^{\ast }}{\hat{h}_{4}} \\
\alpha _{i^{\prime }}^{[q]} &=&-h_{5}^{\ast }\partial _{i^{\prime }}\ln
\sqrt{|q_{4}|},\qquad \beta ^{\lbrack q]}=-h_{5}^{\ast }[\ln \sqrt{|q_{4}|}%
]^{\ast },\qquad \gamma ^{\lbrack q]}=-\frac{q_{4}^{\ast }}{q_{4}},
\end{eqnarray*}%
which following formulas (\ref{ricci9a}) and (\ref{ricci10a}) result in
additional terms to the N--connection coefficients, i. e.
\begin{equation}
w_{i^{\prime }}=\widehat{w}_{i^{\prime }}+w_{i^{\prime }}^{[q]}~\mbox{ and }%
n_{i^{\prime }}=\widehat{n}_{i^{\prime }}+n_{i^{\prime }}^{[q]},
\label{ncon05}
\end{equation}%
with $w_{i^{\prime }}^{[q]}$ and $n_{i^{\prime }}^{[q]}$ computed by using
respectively $\alpha _{i^{\prime }}^{[q]},\beta ^{\lbrack q]}$ and $\gamma
^{\lbrack q]}.$

The N--connection coefficients (\ref{ncon05}) can be transformed partially
into a $B$--field with $\{B_{i^{\prime }j^{\prime }},B_{b^{\prime }j^{\prime
}}\}$ defined by integrating the conditions (\ref{aux02}), i. e.%
\begin{equation}
B_{i^{\prime }j^{\prime }}=B_{i^{\prime }j^{\prime }[0]}\left( x^{k^{\prime
}}\right) +\int h_{4}\delta _{\lbrack i^{\prime }}w_{j^{\prime }]}d\varphi
,~B_{4j^{\prime }}=B_{4j^{\prime }[0]}\left( x^{k^{\prime }}\right) +\int
h_{4}w_{j^{\prime }}^{\ast }d\varphi ,  \label{last}
\end{equation}%
for some arbitrary functions $B_{i^{\prime }j^{\prime }[0]}\left(
x^{k^{\prime }}\right) $ and $B_{4j^{\prime }[0]}\left( x^{k^{\prime
}}\right) .$ The string background corrections are presented via nontrivial $%
w_{i^{\prime }}^{[q]}$ induced by $\lambda =1/4.$ The formulas (\ref{last})
consist the second aim of this subsection: to illustrate how a a $B$--field
inducing noncommutativity may be related with a N--connection inducing
locall anisotropy. This is an explicit example of locally anisotropic
noncommutative configuration contained in string theory. For the considered
class of wormhole solutions the coefficients $n_{i^{\prime }}$ do not
contribute into the noncommutative configuration, but, in general, following
(\ref{aux01}), they can be also related to noncommutativity.

\section{Comments and Questions}

In this paper, we have developed the method of anholonomic frames
 and associated nonlinear connections from a viewpoint of
application in noncommutative geometry and string theory. We note
in this retrospect that several futures connecting Finsler like
generalizations of gravity and gauge theories, which in the past
were considered ad hoc and sophisticated, actually have a very
natural physical and geometric interpretation in the
noncommutative and D--brane picture in string/M--theory. Such
locally anisotropic and/ or noncommutative configurations are
hidden even in general realtivity and its various Kaluza--Klein
like and supergravity extension. To emphazise them we have to
consider off--diagonal metrics which can be diagonalized in
result of certain anholonomic frame transforms which induce also
nonlinear connection structures in the curved spacetime, in
general, with noncompactified extra dimensions.

On general grounds, it could be said the the appearance of noncommutative
and Finsler like geometry when considering $B$--fields, off--diagonal
metrics and anholonomic frames (all parametrized, in general, by
noncommutative matrices) is a natural thing. Such implementations in the
presence of D--branes and matrix approaches to M--theory were proven here to
have explicit realisations and supported by six background constructions
elaborated in this paper:

First, both the local anisotropy and noncommutativity can be derived from
considering string propagation in general manifolds and bundles and in
various low energy string limits. This way the anholonomic Einstein and
Finsler generalized gravity models are generated from string theory.

Second, the anholonomic constructions with associated nonlinear connection
geometry can be explicitly modeled on superbundles which results in
superstring effective actions with anholonomic (super) field equations which
can be related to various superstring and supergravity theories.

Third, noncommutative geometries and associated differential calculi can be
distinguished in  anholonomic geometric form which allows  formulation of
locally anisotropic field theories with anholonomic symmetries.

Forth, anholonomy and noncommutativity can be related to string/M--theory \
following consequently the matrix algebra and geometry and/or associated to
nonlinear connections noncommutative covariant differential calculi.

Fifth, different models of locally anisotropic gravity with explicit limits
to string and Einstein gravity can be realized on noncommutative D--branes.

Sixth, the anholonomic frame method is a very powerful one in
constructing and investigating  new classes of exact solutions in
string and gravity theories; such solutions contain generic
noncommutativity and/or  local anisotropy and can be parametrized
as to describe  locally anisotropic black hole configurations,
Finsler like structures, anisoropic solitonic and moving string
black hole metrics, or noncommutative and anisotropic wormhole
structures which may be derived in Einstein gravity and/or its
Kaluza--Klein and (super) string generalizations.

The obtained in this paper results have a recent confirmation in
Ref. \cite{risi} where the spacetime noncommutativity is obtained
in string theory with a constant off--diagonal metric background
when an appropriate form is present and one of the spatial
direction has Dirichlet boundary conditions. We note that in Refs.
\cite{vmethod,vbel,vsolsp,vsingl,vsingl1} we constructed exact
solutions in the Einstein and extra dimension gravity with
off--diagonal metrics which were diagonalized by anholonomic
transforms to effective spacetimes with noncommutative structure
induced by anholonomic frames. Those results were extended to
noncommutative geometry and gauge gravity theories, in general,
containing local anisotropy, in Refs. \cite{vnonc,vncf}. The low
energy string spacetime with noncommutativity  constructed in
subsection 7.4 of this work is parametrized by an off--diagonal
metric which is a very general (non--constant) pseudo--Riemannian
one defining an exact solution in string gravity.

 \vskip6pt

Finally, our work raises a number of other interesting questions:

\begin{enumerate}
\item What kind of anholonomic quantum noncommutative structures are hidden
in string theory and gravity; how such constructions are to be
modeled by modern geometric methods.

\item How, in general, to relate the commutative and noncommutative gauge
models of (super) gravity with local anisotropy directly to string/M--theory.

\item What kind of quantum structure is more naturally associated to string
gravity and how to develop such anisotropic generalizations.

\item To formulate a nonlinear connection theory in quantum bundles and
relate it to various Finsler like quantum generalizations.

\item What kind of Clifford structures are more natural for developing a
unified geometric approach to anholonomic noncommutative and quantum
geometry following in various perturbative limits and non--perturbative
sectors of string/M--theory and when a such geometry is to be associated to
D--brane configurations.

\item To construct new classes of exact solutions with generic anisotropy
and noncommutativity and analyze theirs physical meaning and possible
applications.
\end{enumerate}

We hope to address some of these questions in future works.

\subsection*{Acknowledgements}

~~The author is grateful to S. Majid for collaboration and
discussions and  J.\ P. S. Lemos for hospitality and support.


\appendix

\section{Anholonomic Frames and N--Connections}

We outline the basic definitions and formulas on anholonomic frames and
associated nonlinear connection (N--connection) structures on vector bundles %
\cite{ma} and (pseudo) Riemannian manifolds \cite{vexsol,vmethod}.
  The
Einstein equations are written in mixed holonomic--anholonomic
variables. We state the conditions when locally anisotropic
structures (Finsler like and another type ones) can be modeled in
general relativity and its extra dimension generalizations. This
Abstract contains the necessary formulas in coordinate form taken
from a
 geometric paper under preparation together with a co-author.

\subsection{The N--connection geometry}

The concept of N--connection came from Finsler geometry (as a set of
coefficients it is present in the works of E. Cartan \cite{cartan}, then it
was elaborated in a more explicit fashion by A. Kawaguchi \cite{kaw}). The
global definition of N--connections in commutative spaces is due to W.
Barthel \ \cite{barthel}. The geometry of N--connections was developed in
details for vector, covector and higher order bundles \cite{ma,miron,bejancu}%
, spinor bundles \cite{vspinors,vmon2} and superspaces and superstrings \cite%
{vsuper,vmon1,vstring} with recent applications in modern anisotropic
kinetics and theormodynamics \cite{vankin} and elaboration of new methods of
constructing exact off--diagonal solutions of the Einstein equations \cite%
{vexsol,vmethod}. The concept of N--connection can be extended in a similar
manner from commutative to noncommutative spaces if a differential calculus
is fixed on a noncommutative vector (or covector) bundle or another type of
quantum manifolds \ \cite{vncf}.

\subsubsection{N--connections in vector bundles \ and (pseudo) Riemannian
spaces}

Let us consider a vector bundle $\xi =\left( E,\mu ,M\right) $ with typical
fibre $\R$$^{m}$ and the map%
\begin{equation*}
\mu ^{T}:TE\rightarrow TM
\end{equation*}%
being the differential of the map $\mu :E\rightarrow M.$ The map $\mu ^{T}$
is a fibre--preserving morphism of the tangent bundle $\left( TE,\tau
_{E},E\right) $ to $E$ and of tangent bundle $\left( TM,\tau ,M\right) $ to $%
M.$ The kernel of the morphism $\mu ^{T}$ is a vector subbundle of the
vector bundle $\left( TE,\tau _{E},E\right) .$ This kernel is denoted $%
\left( VE,\tau _{V},E\right) $ and called the vertical subbundle over $E.$
By
\begin{equation*}
i:VE\rightarrow TE
\end{equation*}%
it is denoted the inclusion mapping \ when the local coordinates of a point $%
u\in E$ are written $u^{\alpha }=\left( x^{i},y^{a}\right) ,$
where the values of indices are $i,j,k,...=1,2,...,n$ and
$a,b,c,...=1,2,...,m.$

A vector $X_{u}\in TE,$ tangent in the point $u\in E,$ is locally
represented
\begin{equation*}
\left( x,y,X,\widetilde{X}\right) =\left( x^{i},y^{a},X^{i},X^{a}\right) ,
\end{equation*}%
where $\left( X^{i}\right) \in $$\R$$^{n}$ and $\left( X^{a}\right) \in $$\R$%
$^{m}$ are defined by the equality
\begin{equation*}
X_{u}=X^{i}\partial _{i}+X^{a}\partial _{a}
\end{equation*}
[$\partial _{\alpha }=\left( \partial _{i},\partial _{a}\right) $ are usual
partial derivatives on respective coordinates $x^{i}$ and $y^{a}$]. For
instance, $\mu ^{T}\left( x,y,X,\widetilde{X}\right) =\left( x,X\right) $
and the submanifold $VE$ contains elements of type $\left( x,y,0,\widetilde{X%
}\right) $ and the local fibers of the vertical subbundle are isomorphic to $%
\R$$^{m}.$ Having $\mu ^{T}\left( \partial _{a}\right) =0,$ one comes out
that $\partial _{a}$ is a local basis of the vertical distribution $%
u\rightarrow V_{u}E$ on $E,$ which is an integrable distribution.

A nonlinear connection (in brief, N--connection) in the vector bundle $\xi
=\left( E,\mu ,M\right) $ is the splitting on the left of the exact sequence
\begin{equation*}
0\rightarrow VE\rightarrow TE/VE\rightarrow 0,
\end{equation*}%
i. e. a morphism of vector bundles $N:TE\rightarrow VE$ such that $C\circ i$
is the identity on $VE.$

The kernel of the morphism $N$ is a vector subbundle of $\left( TE,\tau
_{E},E\right) ,$ it is called the horizontal subbundle and denoted by $%
\left( HE,\tau _{H},E\right) .$ Every vector bundle $\left( TE,\tau
_{E},E\right) $ provided with a N--connection structure is Whitney sum of
the vertical and horizontal subbundles, i. e.
\begin{equation}
TE=HE\oplus VE.  \label{wihit}
\end{equation}
It is proven that for every vector bundle $\xi =\left( E,\mu ,M\right) $
over a compact manifold $M$ there exists a nonlinear connection \cite{ma}.

Locally a N--connection $N$ is parametrized by a set of coefficients\newline
$\left\{ N_{i}^{a}(u^{\alpha })=N_{i}^{a}(x^{j},y^{b})\right\} $ which
transform as
\begin{equation*}
N_{i^{\prime }}^{a^{\prime }}\frac{\partial x^{i^{\prime }}}{\partial x^{i}}%
=M_{a}^{a^{\prime }}N_{i}^{a}-\frac{\partial M_{a}^{a^{\prime }}}{\partial
x^{i}}y^{a}
\end{equation*}%
under coordinate transforms on the vector bundle $\xi =\left( E,\mu
,M\right) ,$%
\begin{equation*}
x^{i^{\prime }}=x^{i^{\prime }}\left( x^{i}\right) \mbox{ and
}y^{a^{\prime }}=M_{a}^{a^{\prime }}(x)y^{a}.
\end{equation*}

The well known class of linear connections consists a particular
parametization of the coefficients $N_{i}^{a}$ when
\begin{equation*}
N_{i}^{a}(x^{j},y^{b})=\Gamma _{bi}^{a}(x^{j})y^{b}
\end{equation*}%
are linear on variables $y^{b}.$

If a N--connection structure is associated to local frame (basis, vielbein)
on $\xi ,$ the operators of local partial derivatives $\partial _{\alpha
}=\left( \partial _{i},\partial _{a}\right) $ and differentials $d^{\alpha
}=du^{\alpha }=\left( d^{i}=dx^{i},d^{a}=dy^{a}\right) $ should be elongated
as to adapt the local basis (and dual basis) structure to the Whitney
decomposition of the vector bundle into vertical and horizontal subbundles, (%
\ref{wihit}):%
\begin{eqnarray}
\partial _{\alpha } &=&\left( \partial _{i},\partial _{a}\right) \rightarrow
\delta _{\alpha }=\left( \delta _{i}=\partial _{i}-N_{i}^{b}\partial
_{b},\partial _{a}\right) ,  \label{dder} \\
d^{\alpha } &=&\left( d^{i},d^{a}\right) \rightarrow \delta ^{\alpha
}=\left( d^{i},\delta ^{a}=d^{a}+N_{i}^{b}d^{i}\right) .  \label{ddif}
\end{eqnarray}%
The transforms can be considered as some particular case of frame transforms
of type
\begin{equation*}
\partial _{\alpha }\rightarrow \delta _{\alpha }=e_{\alpha }^{\beta
}\partial _{\beta }\mbox{ and }d^{\alpha }\rightarrow \delta ^{\alpha
}=(e^{-1})_{\beta }^{\alpha }\delta ^{\beta },
\end{equation*}%
$e_{\alpha }^{\beta }(e^{-1})_{\beta }^{\gamma }=\delta _{\alpha }^{\gamma
}, $ when the vielbein coefficients $e_{\alpha }^{\beta }$ are constructed
by using the Kronecker symbols $\delta _{a}^{b},\delta _{j}^{i}$ and $%
N_{i}^{b}. $

The bases $\delta _{\alpha }$ and $\delta ^{\alpha }$ satisfy, in general,
some anholonomy conditions, for instance,
\begin{equation}
\delta _{\alpha }\delta _{\beta }-\delta _{\beta }\delta _{\alpha
}=W_{\alpha \beta }^{\gamma }\delta _{\gamma },  \label{anhol}
\end{equation}%
where $W_{\alpha \beta }^{\gamma }$ are called the anholonomy coefficients.
\ An explicit calculus of commutators of operators (\ref{dder}) shows that
there are the non--trivial values:%
\begin{equation}
W_{ij}^{a}=R_{ij}^{a}=\delta _{i}N_{j}^{a}-\delta
_{j}N_{i}^{a},~W_{ai}^{b}=-W_{ia}^{b}=-\partial _{a}N_{i}^{b}.
\label{anholncoef}
\end{equation}

Tensor fields on a vector bundle $\xi =\left( E,\mu ,M\right) $ provided
with N--connection structure $N$ \ (we subject such spaces with the index $%
N, $ $\xi _{N})$ may be decomposed in N--adapted form with respect to the
bases $\delta _{\alpha }$ and $\delta ^{\alpha },$ and their tensor
products. For instance, for a tensor of rang (1,1) $T=\{T_{\alpha }^{~\beta
}=\left( T_{i}^{~j},T_{i}^{~a},T_{b}^{~j},T_{a}^{~b}\right) \}$ we have
\begin{equation}
T=T_{\alpha }^{~\beta }\delta ^{\alpha }\otimes \delta _{\beta
}=T_{i}^{~j}d^{i}\otimes \delta _{i}+T_{i}^{~a}d^{i}\otimes \partial
_{a}+T_{b}^{~j}\delta ^{b}\otimes \delta _{j}+T_{a}^{~b}\delta ^{a}\otimes
\partial _{b}.  \label{dten}
\end{equation}

Every N--connection with coefficients $N_{i}^{b}$ $\ $generates also a
linear connection on $\xi _{N}$ \ as $\Gamma _{\alpha \beta }^{(N)\gamma
}=\{N_{bi}^{a}=\partial N_{i}^{a}(x,y)/\partial y^{b}\}$ which defines a
covariant derivative
\begin{equation*}
D_{\alpha }^{(N)}A^{\beta }=\delta _{\alpha }A^{\beta }+\Gamma _{\alpha
\gamma }^{(N)\beta }A^{\gamma }.
\end{equation*}

Another important characteristic of a N--connection is its curvature $\Omega
=\{\Omega _{ij}^{a}\}$ with the coefficients
\begin{equation}
\Omega _{ij}^{a}=\delta _{j}N_{i}^{a}-\delta _{i}N_{j}^{a}=\partial
_{j}N_{i}^{a}-\partial _{i}N_{j}^{a}+N_{i}^{b}N_{bj}^{a}-N_{j}^{b}N_{bi}^{a}.
\label{ncurv}
\end{equation}

In general, on a vector bundle we may consider arbitrary linear connections
and metric structures adapted to the N--connection decomposition into
vertical and horizontal subbundles (one says that such objects are
distinguished by the N--connection, in brief, d--objects, like the d-tensor (%
\ref{dten}), d--connection, d--metric:

\begin{itemize}
\item The coefficients of linear d--connections $\Gamma =\{\Gamma _{\alpha
\gamma }^{\beta }=\left( L_{jk}^{i},L_{bk}^{a},C_{jc}^{i},C_{ac}^{b}\right)
\}$ are defined for an arbitrary covariant derivative $D$ on $\xi $ being
adapted to the $N$--connection structure as $D_{\delta _{\alpha }}(\delta
_{\beta })=\Gamma _{\beta \alpha }^{\gamma }\delta _{\gamma }$ with the
coefficients being invariant under horizontal and vertical decomposition
\begin{equation*}
\quad D_{\delta _{i}}(\delta _{j})=L_{ji}^{k}\delta _{k},~D_{\delta
_{i}}(\partial _{a})=L_{ai}^{b}\partial _{b},~D_{\partial _{c}}(\delta
_{j})=C_{jc}^{k}\delta _{k},~~D_{\partial _{c}}(\partial
_{a})=C_{ac}^{b}\partial _{b}.
\end{equation*}%
The operator of covariant differentiation $D$ splits into the horizontal
covariant derivative $D^{[h]},$ stated by the coefficients $\left(
L_{jk}^{i},L_{bk}^{a}\right) ,$ for instance, and the operator of vertical
covariant derivative $D^{[v]},$ stated by the coefficients $\left(
C_{jc}^{i},C_{ac}^{b}\right) .$ For instance, for $A=A^{i}\delta
_{i}+A^{a}\partial _{a}=A_{i}\partial ^{i}+A_{a}\delta ^{a}$ one holds the
d--covariant derivation rules,%
\begin{eqnarray*}
D_{i}^{[h]}A^{k} &=&\delta
_{i}A^{k}+L_{ij}^{k}A^{j},~D_{i}^{[h]}A^{b}=\delta _{i}A^{b}+L_{ic}^{b}A^{c},
\\
D_{i}^{[h]}A_{k} &=&\delta _{i}A_{k}-L_{ik}^{j}A_{j},D_{i}^{[h]}A_{b}=\delta
_{i}A_{b}-L_{ib}^{c}A_{c},~ \\
D_{a}^{[v]}A^{k} &=&\partial
_{a}A^{k}+C_{aj}^{k}A^{j},~D_{a}^{[v]}A^{b}=\partial
_{a}A^{b}+C_{ac}^{b}A^{c}, \\
D_{a}^{[v]}A_{k} &=&\partial
_{a}A_{k}-C_{ak}^{j}A_{j},D_{a}^{[v]}A_{b}=\partial
_{a}A_{b}-C_{ab}^{c}A_{c}.
\end{eqnarray*}

\item The d--metric structure $G=g_{\alpha \beta }\delta ^{a}\otimes \delta
^{b}$ which has the invariant decomposition as $g_{\alpha \beta }=\left(
g_{ij},g_{ab}\right) $ following from%
\begin{equation}
G=g_{ij}(x,y)d^{i}\otimes d^{j}+g_{ab}(x,y)\delta ^{a}\otimes \delta ^{b}.
\label{dmetric}
\end{equation}
\end{itemize}

We may impose the condition that a d--metric $g_{\alpha \beta }$ and a
d--connection $\Gamma _{\alpha \gamma }^{\beta }$ are compatible, i. e.
there are satisfied the conditions
\begin{equation}
D_{\gamma }g_{\alpha \beta }=0.  \label{metrcond}
\end{equation}

With respect to the anholonomic frames (\ref{dder}) and (\ref{ddif}), there
is a linear connection, called the canonical distinguished linear
connection, which is similar to the metric connection introduced by the
Christoffel symbols in the case of holonomic bases, i. e. being constructed
only from the metric components and satisfying the metricity conditions (\ref%
{metrcond}). It is parametrized by the coefficients,\ $\Gamma _{\ \beta
\gamma }^{\alpha }=\left( L_{\ jk}^{i},L_{\ bk}^{a},C_{\ jc}^{i},C_{\
bc}^{a}\right) $ where
\begin{eqnarray}
L_{\ jk}^{i} &=&\frac{1}{2}g^{in}\left( \delta _{k}g_{nj}+\delta
_{j}g_{nk}-\delta _{n}g_{jk}\right) ,  \label{dcon} \\
L_{\ bk}^{a} &=&\partial _{b}N_{k}^{a}+\frac{1}{2}h^{ac}\left( \delta
_{k}h_{bc}-h_{dc}\partial _{b}N_{k}^{d}-h_{db}\partial _{c}N_{k}^{d}\right) ,
\notag \\
C_{\ jc}^{i} &=&\frac{1}{2}g^{ik}\partial _{c}g_{jk},\ C_{\ bc}^{a}=\frac{1}{%
2}h^{ad}\left( \partial _{c}h_{db}+\partial _{b}h_{dc}-\partial
_{d}h_{bc}\right) .  \notag
\end{eqnarray}%
Instead of this connection one can consider on $\xi $ another types of
linear connections which are/or not adapted to the N--connection structure
(see examples in \cite{ma}).

\subsubsection{D--torsions and d--curvatures:}

The anholonomic coefficients $W_{\ \alpha \beta }^{\gamma }$ and
N--elongated derivatives give nontrivial coefficients for the torsion
tensor, $T(\delta _{\gamma },\delta _{\beta })=T_{\ \beta \gamma }^{\alpha
}\delta _{\alpha },$ where
\begin{equation}
T_{\ \beta \gamma }^{\alpha }=\Gamma _{\ \beta \gamma }^{\alpha }-\Gamma _{\
\gamma \beta }^{\alpha }+W_{\ \beta \gamma }^{\alpha },  \label{torsion}
\end{equation}%
and for the curvature tensor, $R(\delta _{\tau },\delta _{\gamma })\delta
_{\beta }=R_{\beta \ \gamma \tau }^{\ \alpha }\delta _{\alpha },$ where
\begin{equation}
R_{\beta \ \gamma \tau }^{\ \alpha }=\delta _{\tau }\Gamma _{\ \beta \gamma
}^{\alpha }-\delta _{\gamma }\Gamma _{\ \beta \tau }^{\alpha }+\Gamma _{\
\beta \gamma }^{\varphi }\Gamma _{\ \varphi \tau }^{\alpha }-\Gamma _{\
\beta \tau }^{\varphi }\Gamma _{\ \varphi \gamma }^{\alpha }+\Gamma _{\
\beta \varphi }^{\alpha }W_{\ \gamma \tau }^{\varphi }.  \label{curvature}
\end{equation}%
We emphasize that the torsion tensor on (pseudo) Riemannian spacetimes is
induced by anholonomic frames, whereas its components vanish with respect to
holonomic frames. All tensors are distinguished (d) by the N--connection
structure into irreducible (horizontal--vertical) h--v--components, and are
called d--tensors. For instance, the torsion, d--tensor has the following
irreducible, nonvanishing, h--v--components,\ $T_{\ \beta \gamma }^{\alpha
}=\{T_{\ jk}^{i},C_{\ ja}^{i},S_{\ bc}^{a},T_{\ ij}^{a},T_{\ bi}^{a}\},$
where
\begin{eqnarray}
T_{.jk}^{i} &=&T_{jk}^{i}=L_{jk}^{i}-L_{kj}^{i},\quad
T_{ja}^{i}=C_{.ja}^{i},\quad T_{aj}^{i}=-C_{ja}^{i},  \notag \\
T_{.ja}^{i} &=&0,\quad T_{.bc}^{a}=S_{.bc}^{a}=C_{bc}^{a}-C_{cb}^{a},
\label{dtors} \\
T_{.ij}^{a} &=&-\Omega _{ij}^{a},\quad T_{.bi}^{a}=\partial
_{b}N_{i}^{a}-L_{.bi}^{a},\quad T_{.ib}^{a}=-T_{.bi}^{a}  \notag
\end{eqnarray}%
(the d--torsion is computed by substituting the h--v--compo\-nents of the
canonical d--connection (\ref{dcon}) and anholonomy coefficients (\ref{anhol}%
) into the formula for the torsion coefficients (\ref{torsion})).

We emphasize that with respect to anholonomic frames the torsion is not zero
even for symmetric connections with $\Gamma _{\ \beta \gamma }^{\alpha
}=\Gamma _{\ \gamma \beta }^{\alpha }$ because the anholonomy coefficients $%
W_{\ \beta \gamma }^{\alpha }$ are contained in the formulas for the torsion
coefficients (\ref{torsion}). By straightforward computations we can prove
that for nontrivial N--connection curvatures, $\Omega _{ij}^{a}\neq 0,$ even
the Levi--Civita connection for the metric (\ref{dmetric}) contains
nonvanishing torsion coefficients. Of course, the torsion vanishes if the
Levi--Civita connection is defined as the usual Christoffel symbols with
respect to the coordinate frames, $\left( \partial _{i},\partial _{a}\right)
$ and $\left( d^{i},\partial ^{a}\right) ;$ in this case the d--metric (\ref%
{dmetric}) is redefined into, in general, off--diagonal metric containing
products of $N_{i}^{a}$ and $h_{ab}.$

The curvature d--tensor has the following irreducible, non-vanishing,
h--v--compon\-ents\ $R_{\beta \ \gamma \tau }^{\ \alpha
}=%
\{R_{h.jk}^{.i},R_{b.jk}^{.a},P_{j.ka}^{.i},P_{b.ka}^{.c},S_{j.bc}^{.i},S_{b.cd}^{.a}\},
$\ where
\begin{eqnarray}
R_{h.jk}^{.i} &=&\delta _{k}L_{.hj}^{i}-\delta
_{j}L_{.hk}^{i}+L_{.hj}^{m}L_{mk}^{i}-L_{.hk}^{m}L_{mj}^{i}-C_{.ha}^{i}%
\Omega _{.jk}^{a},  \label{dcurvatures} \\
R_{b.jk}^{.a} &=&\delta _{k}L_{.bj}^{a}-\delta
_{j}L_{.bk}^{a}+L_{.bj}^{c}L_{.ck}^{a}-L_{.bk}^{c}L_{.cj}^{a}-C_{.bc}^{a}%
\Omega _{.jk}^{c},  \notag \\
P_{j.ka}^{.i} &=&\partial _{a}L_{.jk}^{i}+C_{.jb}^{i}T_{.ka}^{b}-(\delta
_{k}C_{.ja}^{i}+L_{.lk}^{i}C_{.ja}^{l}-L_{.jk}^{l}C_{.la}^{i}-L_{.ak}^{c}C_{.jc}^{i}),
\notag \\
P_{b.ka}^{.c} &=&\partial _{a}L_{.bk}^{c}+C_{.bd}^{c}T_{.ka}^{d}-(\delta
_{k}C_{.ba}^{c}+L_{.dk}^{c\
}C_{.ba}^{d}-L_{.bk}^{d}C_{.da}^{c}-L_{.ak}^{d}C_{.bd}^{c}),  \notag \\
S_{j.bc}^{.i} &=&\partial _{c}C_{.jb}^{i}-\partial
_{b}C_{.jc}^{i}+C_{.jb}^{h}C_{.hc}^{i}-C_{.jc}^{h}C_{hb}^{i},  \notag \\
S_{b.cd}^{.a} &=&\partial _{d}C_{.bc}^{a}-\partial
_{c}C_{.bd}^{a}+C_{.bc}^{e}C_{.ed}^{a}-C_{.bd}^{e}C_{.ec}^{a}  \notag
\end{eqnarray}%
(the d--curvature components are computed in a similar fashion by using the
formula for curvature coefficients (\ref{curvature})).

\subsubsection{Einstein equations in d--variables}

In this subsection we write and analyze the Einstein equations on spaces
provided with anholonomic frame structures and associated N--connections.

The Ricci tensor $R_{\beta \gamma }=R_{\beta ~\gamma \alpha }^{~\alpha }$
has the d--components
\begin{equation}
R_{ij}=R_{i.jk}^{.k},\quad
R_{ia}=-^{2}P_{ia}=-P_{i.ka}^{.k},R_{ai}=^{1}P_{ai}=P_{a.ib}^{.b},\quad
R_{ab}=S_{a.bc}^{.c}.  \label{dricci}
\end{equation}%
In general, since $^{1}P_{ai}\neq ~^{2}P_{ia}$, the Ricci d-tensor is
non-symmetric (we emphasize that this could be with respect to anholonomic
frames of reference because the N--connection and its curvature
coefficients, $N_{i}^{a}$ and $\Omega _{.jk}^{a},$ as well the anholonomy
coefficients $W_{\ \beta \gamma }^{\alpha }$ and d--torsions $T_{\ \beta
\gamma }^{\alpha }$ are contained in the formulas for d--curvatures (\ref%
{curvature})). The scalar curvature of the metric d--connection, $%
\overleftarrow{R}=g^{\beta \gamma }R_{\beta \gamma },$ is computed
\begin{equation}
{\overleftarrow{R}}=G^{\alpha \beta }R_{\alpha \beta }=\widehat{R}+S,
\label{dscalar}
\end{equation}%
where $\widehat{R}=g^{ij}R_{ij}$ and $S=h^{ab}S_{ab}.$

By substituting (\ref{dricci}) and (\ref{dscalar}) into the Einstein
equations
\begin{equation}
R_{\alpha \beta }-\frac{1}{2}g_{\alpha \beta }R=\kappa \Upsilon _{\alpha
\beta },  \label{5einstein}
\end{equation}%
where $\kappa $ and $\Upsilon _{\alpha \beta }$ are respectively the
coupling constant and the energy--momentum tensor we obtain the
h-v-decomposition by N--connection of the Einstein equations
\begin{eqnarray}
R_{ij}-\frac{1}{2}\left( \widehat{R}+S\right) g_{ij} &=&\kappa \Upsilon
_{ij},  \label{einsteq2} \\
S_{ab}-\frac{1}{2}\left( \widehat{R}+S\right) h_{ab} &=&\kappa \Upsilon
_{ab},  \notag \\
^{1}P_{ai}=\kappa \Upsilon _{ai},\ ^{2}P_{ia} &=&\kappa \Upsilon _{ia}.
\notag
\end{eqnarray}%
The definition of matter sources with respect to anholonomic frames is
considered in Refs. \cite{vspinors,vmon1,ma}.

The vacuum locally anisotropic gravitational field equations, in invariant
h-- v--components, are written
\begin{eqnarray}
R_{ij} &=&0,S_{ab}=0,^{1}P_{ai}=0,\ ^{2}P_{ia}=0.  \label{einsteq3} \\
&&  \notag
\end{eqnarray}

We emphasize that general linear connections in vector bundles and even in
the (pseudo) Riemannian spacetimes have non--trivial torsion components if
off--diagonal metrics and anholomomic frames are introduced into
consideration. This is a ''pure'' anholonomic frame effect: the torsion
vanishes for the Levi--Civita connection stated with respect to a coordinate
frame, but even this metric connection contains some torsion coefficients if
it is defined with respect to anholonomic frames (this follows from the $w$%
--terms in (\ref{lcsym})). For the (pseudo) Riemannian spaces we conclude
that the Einstein theory transforms into an effective Einstein--Cartan
theory with anholonomically induced torsion if the general relativity is
formulated with respect to general frame bases (both holonomic and
anholonomic).

The N--connection geometry can be similarly formulated for a tangent bundle $%
TM$ of a manifold $M$ (which is used in Finsler and Lagrange geometry \cite%
{ma}), on cotangent bundle $T^{\ast }M$ and higher order bundles (higher
order Lagrange and Hamilton geometry \cite{miron}) as well in the geometry
of locally anisotropic superspaces \cite{vsuper}, superstrings \cite{vstr2},
anisotropic spinor \cite{vspinors} and gauge \ \cite{vgauge} theories or
even on (pseudo) Riemannian spaces provided with anholonomic frame
structures \cite{vmon2}.

\subsection{Anholonomic Frames in Commutative Gravity}

We introduce the concepts of generalized Lagrange and Finsler geometry and
outline the conditions when such structures can be modeled on a Riemannian
space by using anholnomic frames.

Different classes of commutative anisotropic spacetimes are
modeled by corresponding parametriztions of some compatible (or
even non--compatible) N--connection, d--connection and d--metric
structures on (pseudo) Riemannian spaces, tangent (or cotangent)
bundles, vector (or covector) bundles and their higher order
generalizations in their usual manifold,
supersymmetric, spinor, gauge like or another type approaches (see Refs. %
\cite{vexsol,miron,ma,bejancu,vspinors,vgauge,vmon1,vmon2}).

\subsubsection{Anholonomic structures on Riemannian spaces}

\bigskip We note that  the N--connection structure may be defined not
only in vector bundles but also on (pseudo) Riemannian spaces \cite{vexsol}.
\ In this case the N--connection is an object completely defined by
anholonomic frames, when the coefficients of frame transforms, $e_{\alpha
}^{\beta }\left( u^{\gamma }\right) ,$ are parametrized explicitly via
certain values $\left( N_{i}^{a},\delta _{i}^{j},\delta _{b}^{a}\right) ,$
where $\delta _{i}^{j}$ $\ $and $\delta _{b}^{a}$ are the Kronecker symbols.
By straightforward calculations we can compute that the coefficients of the
Levi--Civita metric connection
\begin{equation*}
\Gamma _{\alpha \beta \gamma }^{\bigtriangledown }=g\left( \delta _{\alpha
},\bigtriangledown _{\gamma }\delta _{\beta }\right) =g_{\alpha \tau }\Gamma
_{\beta \gamma }^{\bigtriangledown \tau },\,
\end{equation*}%
associated to a covariant derivative operator $\bigtriangledown ,$
satisfying the metricity condition $\bigtriangledown _{\gamma }g_{\alpha
\beta }=0$ for $g_{\alpha \beta }=\left( g_{ij},h_{ab}\right) $ and
\begin{equation}
\Gamma _{\alpha \beta \gamma }^{\bigtriangledown }=\frac{1}{2}\left[ \delta
_{\beta }g_{\alpha \gamma }+\delta _{\gamma }g_{\beta \alpha }-\delta
_{\alpha }g_{\gamma \beta }+g_{\alpha \tau }W_{\gamma \beta }^{\tau
}+g_{\beta \tau }W_{\alpha \gamma }^{\tau }-g_{\gamma \tau }W_{\beta \alpha
}^{\tau }\right] ,  \label{lcsym}
\end{equation}%
are given with respect to the anholonomic basis (\ref{ddif}) by the
coefficients
\begin{equation}
\Gamma _{\beta \gamma }^{\bigtriangledown \tau }=\left( L_{\ jk}^{i},L_{\
bk}^{a},C_{\ jc}^{i}+\frac{1}{2}g^{ik}\Omega _{jk}^{a}h_{ca},C_{\
bc}^{a}\right)  \label{lccon}
\end{equation}%
when $L_{\ jk}^{i},L_{\ bk}^{a},C_{\ jc}^{i},C_{\ bc}^{a}$ and $\Omega
_{jk}^{a}$ are respectively computed by the formulas (\ref{dcon}) and (\ref%
{ncurv}). A specific property of off--diagonal metrics is that they can
define different classes of linear connections which satisfy the metricity
conditions for a given metric, or inversely, there is a certain class of
metrics which satisfy the metricity conditions for a given linear
connection. \ This result was originally obtained by A. Kawaguchi \cite{kaw}
(Details can be found in Ref. \cite{ma}, see Theorems 5.4 and 5.5 in Chapter
III, formulated for vector bundles; here we note that similar proofs hold
also on manifolds enabled with anholonomic frames associated to a
N--connection structure).

With respect to anholonomic frames, we can not distinguish the Levi--Civita
connection as the unique one being both metric and torsionless. For
instance, both linear connections (\ref{dcon}) and (\ref{lccon}) contain
anholonomically induced torsion coefficients, are compatible with the same
metric and transform into the usual Levi--Civita coefficients for vanishing
N--connection and ''pure'' holonomic coordinates. This means that to an
off--diagonal metric in general relativity one may be associated different
covariant differential calculi, all being compatible with the same metric
structure (like in the non--commutative geometry, which is not a surprising
\ fact because the anolonomic frames satisfy by definition some
non--commutative relations (\ref{anhol})). In such cases we have to select a
particular type of connection following some physical or geometrical
arguments, or to impose some conditions when there is a single compatible
linear connection constructed only from the metric and N--coefficients. We
note that if $\Omega _{jk}^{a}=0$ the connections (\ref{dcon}) and (\ref%
{lccon}) coincide, i. e. $\Gamma _{\ \beta \gamma }^{\alpha }=\Gamma _{\beta
\gamma }^{\bigtriangledown \alpha }.$

If an anholonomic (equivalently, anisotropic) frame structure is defined on
a (pseu\-do) Riemannian space of dimension $(n+m)$ space, the space is
called to be an anholonom\-ic (pseudo) Riemannian one (denoted as $\
V^{(n+m)}).$ By fixing an anholonomic frame basis and co--basis with
associated N--connection $N_{i}^{a}(x,y),$ respectively, as (\ref{dder}) and
(\ref{ddif}), one splits the local coordinates $u^{\alpha }=(x^{i},y^{a})$
into two classes: the fist class consists from $n$ holonomic coorinates, $%
x^{i},$ and the second class consists from $m$ anholonomic coordinates, $%
y^{a}.$ The d--metric (\ref{dmetric}) on $V^{(n+m)}$,
\begin{equation}
G^{[R]}=g_{ij}(x,y)dx^{i}\otimes dx^{j}+h_{ab}(x,y)\delta y^{a}\otimes
\delta y^{b}  \label{dmetrr}
\end{equation}%
written with respect to a usual coordinate basis $du^{\alpha }=\left(
dx^{i},dy^{a}\right) ,$%
\begin{equation*}
ds^{2}=\underline{g}_{\alpha \beta }\left( x,y\right) du^{\alpha }du^{\beta }
\end{equation*}%
is a generic off--diagonal \ Riemannian metric parametrized as%
\begin{equation}
\underline{g}_{\alpha \beta }=\left[
\begin{array}{cc}
g_{ij}+N_{i}^{a}N_{j}^{b}g_{ab} & h_{ab}N_{i}^{a} \\
h_{ab}N_{j}^{b} & h_{ab}%
\end{array}%
\right] .  \label{odm}
\end{equation}%
Such type of metrics were largely investigated in the Kaluza--Klein gravity %
\cite{salam}, but also in the Einstein gravity \cite{vexsol}. An
off--diagonal metric (\ref{odm}) can be reduced to a block $\left( n\times
n\right) \oplus \left( m\times m\right) $ form $\left( g_{ij},g_{ab}\right)
, $ and even effectively diagonalized in result of a superposition of
ahnolonomic N--transforms. It can be defined as an exact solution of the
Einstein equations. With respect to anholonomic frames, in general, the
Levi--Civita connection obtains a torsion component (\ref{lcsym}). Every
class of off--diagonal metrics can be anholonomically equivalent to another
ones for which it is not possible to a select the Levi--Civita metric defied
as the unique torsionless and metric compatible linear connection. \ The
conclusion is that if anholonomic frames of reference, which authomatically
induce the torsion via anholonomy coefficients, are considered on a
Riemannian space we have to postulate explicitly what type of linear
connection (adapted both to the anholonomic frame and metric structure) is
chosen in order to construct a Riemannian geometry and corresponding
physical models. For instance, we may postulate the connection (\ref{lccon})
or the d--connection (\ref{dcon}). Both these connections are metric
compatible and transform into the usual Christoffel symbols if the
N--connection vanishes, i. e. the local frames became holonomic. But, in
general, anholonomic frames and off--diagonal Riemannian metrics are
connected with anisotropic configurations which allow, in principle, to
model even Finsler like structures in (pseudo) Riemannian spaces \cite%
{vankin,vexsol}.

\subsubsection{Finsler geometry and its almost Kahlerian model}

The modern approaches to Finsler geometry are outlined in Refs. \cite%
{finsler,ma,miron,bejancu,vmon1,vmon2}. Here we emphasize that a Finsler
metric can be defined on a tangent bundle $TM$ with local coordinates $%
u^{\alpha }=(x^{i},y^{a}\rightarrow y^{i})$ of dimension $2n,$ with a
d--metric (\ref{dmetric}) for which the Finsler metric, i. e. the quadratic
form
\begin{equation}
g_{ij}^{[F]}=h_{ab}=\frac{1}{2}\frac{\partial ^{2}F^{2}}{\partial
y^{i}\partial y^{j}}  \label{fmetric}
\end{equation}%
is positive definite, is defined in this way: \ 1) A Finsler metric on a
real manifold $M$ is a function $F:TM\rightarrow \R$ which on $\widetilde{TM}%
=TM\backslash \{0\}$ is of class $C^{\infty }$ and $F$ is only continuous on
the image of the null cross--sections in the tangent bundle to $M.$ 2) $%
F\left( x,\chi y\right) =\chi F\left( x,y\right) $ for every $\R_{+}^{\ast
}. $ 3) The restriction of $F$ to $\widetilde{TM}$ is a positive function.
4) $rank\left[ g_{ij}^{[F]}(x,y)\right] =n.$

The Finsler metric $F(x,y)$ and the quadratic form $g_{ij}^{[F]}$ can be
used to define the Christoffel symbols (not those from the usual Riemannian
geometry)%
\begin{equation*}
c_{jk}^{\iota }(x,y)=\frac{1}{2}g^{ih}\left( \partial
_{j}g_{hk}^{[F]}+\partial _{k}g_{jh}^{[F]}-\partial _{h}g_{jk}^{[F]}\right) ,
\end{equation*}%
where $\partial _{j}=\partial /\partial x^{j},$ $\ $which allows us to
define the Cartan nonlinear connection as
\begin{equation}
N_{j}^{[F]i}(x,y)=\frac{1}{4}\frac{\partial }{\partial y^{j}}\left[
c_{lk}^{\iota }(x,y)y^{l}y^{k}\right]  \label{ncc}
\end{equation}%
where we may not distinguish the v- and h- indices taking on $TM$ \ the same
values.

In Finsler geometry there were investigated different classes of remarkable
Finsler linear connections introduced by Cartan, Berwald, Matsumoto and
other ones (see details in Refs. \cite{finsler,ma,bejancu}). Here we note
that we can introduce $g_{ij}^{[F]}=g_{ab}$ and $N_{j}^{i}(x,y)$ in (\ref%
{dmetric}) and construct a d--connection via formulas (\ref{dcon}).

A usual Finsler space $F^{n}=\left( M,F\left( x,y\right) \right) $ is
completely defined by its fundamental tensor $g_{ij}^{[F]}(x,y)$ and Cartan
nonlinear connection $N_{j}^{i}(x,y)$ and its chosen d--connection
structure. But the N--connection allows us to define an almost complex
structure $I$ on $TM$ as follows%
\begin{equation*}
I\left( \delta _{i}\right) =-\partial /\partial y^{i}\mbox{ and
}I\left( \partial /\partial y^{i}\right) =\delta _{i}
\end{equation*}%
for which $I^{2}=-1.$

The pair $\left( G^{[F]},I\right) $ consisting from a Riemannian metric on $%
TM,$%
\begin{equation}
G^{[F]}=g_{ij}^{[F]}(x,y)dx^{i}\otimes dx^{j}+g_{ij}^{[F]}(x,y)\delta
y^{i}\otimes \delta y^{j}  \label{dmetricf}
\end{equation}%
and the almost complex structure $I$ defines an almost Hermitian structure
on $\widetilde{TM}$ associated to a 2--form%
\begin{equation*}
\theta =g_{ij}^{[F]}(x,y)\delta y^{i}\wedge dx^{j}.
\end{equation*}%
This model of Finsler geometry is called almost Hermitian and denoted $%
H^{2n} $ and it is proven \cite{ma} that is almost Kahlerian, i. e. the form
$\theta $ is closed. The almost Kahlerian space $K^{2n}=\left( \widetilde{TM}%
,G^{[F]},I\right) $ is also called the almost Kahlerian model of the Finsler
space $F^{n}.$

On Finsler (and their almost Kahlerian models) spaces one distinguishes the
almost Kahler linear connection of Finsler type, $D^{[I]}$ on $\widetilde{TM}
$ with the property that this covariant derivation preserves by parallelism
the vertical distribution and is compatible with the almost Kahler structure
$\left( G^{[F]},I\right) ,$ i.e.
\begin{equation*}
D_{X}^{[I]}G^{[F]}=0\mbox{ and }D_{X}^{[I]}I=0
\end{equation*}%
for \ every d--vector field on $\widetilde{TM}.$ This d--connection is
defined by the data
\begin{equation*}
\Gamma =\left( L_{jk}^{i},L_{bk}^{a}=0,C_{ja}^{i}=0,C_{bc}^{a}\rightarrow
C_{jk}^{i}\right)
\end{equation*}%
with $L_{jk}^{i}$ and $C_{jk}^{i}$ computed as in the formulas (\ref{dcon})
by using $g_{ij}^{[F]}$ and $N_{j}^{i}$ from (\ref{ncc}).

We emphasize that a Finsler space $F^{n}$ with a d--metric (\ref{dmetricf})
and Cartan's N--connection structure (\ref{ncc}), or the corresponding
almost Hermitian (Kahler) model $H^{2n},$ can be equivalently modeled on a
Riemannian space of dimension $2n$ provided with an off--diagonal Riemannian
metric (\ref{odm}). From this viewpoint a Finsler geometry is a
corresponding Riemannian geometry with a respective off--diagonal metric
(or, equivalently, with an anholonomic frame structure with associated
N--connection) and a corresponding prescription for the type of linear
connection chosen to be compatible with the metric and N--connection
structures.

\subsubsection{Lagrange and generalized Lagrange geometry}

Lagrange spaces were introduced in order to geometrize the fundamental
concepts in mechanics \cite{kern} and investigated in Refs. \cite{ma} (see %
\cite{vspinors,vgauge,vsuper,vstr2,vmon1,vmon2} for their spinor, gauge and
supersymmetric generalizations).

A Lagrange space $L^{n}=\left( M,L\left( x,y\right) \right) $ is defined as
a pair which consists of a real, smooth $n$--dimensional manifold $M$ and
regular Lagrangian $L:TM\rightarrow \R.$ Similarly as for Finsler spaces one
introduces the symmetric d--tensor field
\begin{equation}
g_{ij}^{[L]}=h_{ab}=\frac{1}{2}\frac{\partial ^{2}L}{\partial y^{i}\partial
y^{j}}.  \label{mfl}
\end{equation}%
So, the Lagrangian $L(x,y)$ is like the square of the fundamental Finsler
metric, $F^{2}(x,y),$ but not subjected to any homogeneity conditions.

In the rest me can introduce similar concepts of almost Hermitian
(Kahlerian) models of Lagrange spaces as for the Finsler spaces, by using
the similar definitions and formulas as in the previous subsection, but
changing $g_{ij}^{[F]}\rightarrow g_{ij}^{[L]}.$

R. Miron introduced the concept of generalized Lagrange space, GL--space
(see details in \cite{ma}) and a corresponding N--connection geometry on $TM$
when the fundamental metric function $g_{ij}=g_{ij}\left( x,y\right) $ is a
general one, not obligatory defined as a second derivative from a Lagrangian
as in (\ref{mfl}). The corresponding almost Hermitian (Kahlerian) models of
GL--spaces were investigated and applied in order to elaborate
generalizations of gravity and gauge theories \cite{ma,vgauge}.

Finally, a few remarks on definition of gravity models with generic local
anisotropy on anholonomic Riemannian, Finsler or (generalized) Lagrange
spaces and vector bundles. So, by choosing a d-metric (\ref{dmetric}) (in
particular cases (\ref{dmetrr}), or (\ref{dmetricf}) with $g_{ij}^{[F]},$ or
$g_{ij}^{[L]})$ we may compute the coefficients of, for instance,
d--connection (\ref{dcon}), d--torsion (\ref{dtors}) and (\ref{dcurvatures})
and even to write down the explicit form of Einstein equations (\ref%
{einsteq2}) which define such geometries. For instance, in a series of works %
\cite{vankin,vexsol,vmon2} we found explicit solutions when Finsler like and
another type anisotropic configurations are modeled in anisotropic kinetic
theory and irreversible thermodynamics and even in Einstein or
low/extra--dimension gravity as exact solutions of the vacuum (\ref{einsteq2}%
) and nonvacuum (\ref{einsteq3}) Einstein equations. From the viewpoint of
the geometry of anholonomic frames is not much difference between the usual
Riemannian geometry and its Finsler like generalizations. The explicit form
and parametrizations of coefficients of metric, linear connections,
torsions, curvatures and Einstein equations in all types of mentioned
geometric models depends on the type of anholomic frame relations and
compatibility metric conditions between the associated N--connection
structure and linear connections we fixed. Such structures can be
correspondingly picked up from a noncommutative functional model, for
instance, from some almost Hermitian structures over projective modules
and/or generalized to some noncommutative configurations \cite{vncf}.



\begin{thebibliography}{99}
\bibitem{abou} M. Abou--Zeid and H. Dorn, ''Comments on the energy--momentum
tensor in non--commutative field theories'', Phys. Lett. \textbf{B 514}
(2001)\ 183-189; A. Gerhold, J. Grimstrup, H. Grosse, L. Popp, M. Schweda
and R. Wulenhaar, ''The energy--momentum tensor on noncommutative spaces:
Some pedagogical comments'', hep--th/0012112.

\bibitem{ardalan} F. Ardalan, H. Arfaei, and M. M. Sheikh--Jabbari, \
''Mixed branes and M(atrix) theory on noncommutative torus, hep--th/9803067;
''Noncommutative geometry from string and branes'', JHEP \textbf{9902}
(1999) 016; ''Dirac quantization of open string and noncommutativity in
branes'', Nucl. Phys. \textbf{B 576} (2000)\ 578--; F. Ardalan, H. Arfaei,
M. R. Garousi, and A. Ghodsi, ''Gravity on noncommutative D--branes'',
hep--th/0204117.

\bibitem{barthel} W. Barthel, ''Nichtlineare Zusammenhange unde Deren
Holonomie Gruppen'', J. Reine Angew. Math. \textbf{212} (1963) 120--149.

\bibitem{bejancu} A. Bejancu, ''Finsler Geometry and Applications'', Ellis
Horwood, Chichester, England, (1990).

\bibitem{belinski} V.\ A. Belinski and V.\ E. Zakharov, \textsl{Soviet.
Phys. JETP}, \textbf{48} (1978) 985--994 [translated from:\ \textsl{Zh.
Eksper. Teoret. Fiz.} \textbf{75} (1978) 1955--1971, in Russian]; V.
Belinski and E. Verdaguer, \textit{Gravitational Solitons} (Cambridge
University Press, 2001)

\bibitem{cartan} E. Cartan, ''Les Espaces de Finsler'', Hermann, Paris
(1934);\ ''La Methode du Repere Mobile, la Theorie des Groupes Continus et
les Espaces Generalises'', Herman, Paris (1935) [Russian translation, Moscow
Univ. Press (1963);\ ''Les Systems Differentielles Exterieurs et Leuws
Application Geometricques'' (Herman and Cie Editeur, Paris, 1945);\ ''Expos$%
\acute{e}$s de G$\acute{e}$om$\acute{e}$trie'', in \emph{Series Actualit$%
\acute{e}$s Scientifiques et Industrielles}\ \textbf{79} (1936) [reprinted:
Herman, Paris (1971)]; ''Le\c{c}ons sur la th\'{e}orie des spineurs''. Tome
I: ''Les spineurs de l'espace \'{a} $n>3$ dimensions. Les Spineurs en g\`{e}%
ometrie reimannienne'', Hermann, Paris, (1938); ''The Theory of Spinors'',
Dover Publications, New York (1966);\ ''Sur les equations de la gravitation
d'Enstein'', Gautier Villars, Paris (1922);\ ''Riemannian geometry in an
orthogonal frame,'' \'{E}lie Cartan lectures at Sorbonne, 1926--27, World
Scientific Publishing Co., Inc., River Edge, NJ, (2001) [from Russian]; ''On
manifolds with an affine connection and the theory of general relativity.''
\ Monographs and Textbooks in Physical Science, 1. Bibliopolis, Naples
(1986) \ [from French];\ ''Prostranstva affinno\u{\i}, proektivno\u{\i} i
konformno\u{\i} svyaznosti'' [''Spaces of affine, projective and conformal
connectivity''], Monographs and Studies in Non-Euclidean Geometry, No. 3
Izdat. Kazan. Univ. (1962) [Russian, from French].

\bibitem{ncg} A. H. Chamseddine, G. Felder and J. Frohlich, ''Gravity in
Non--Commutative Geometry'', Commun. Math. Phys. \textbf{155} (1993)
205--217; A. H. Chamseddine, J. Frohlich and O. Grandjean, ''The
Gravitational Sector in the Connes--Lott Formulation of the Standard
Model'', J. Math. Phys. \textbf{36} (1995) 6255--6275; A.\ Connes,
''Noncommutative Geometry and Reality'', J. Math. Phys. 36 (1995)
6194--6231; A. Connes and J. Lott, ''Particle Models and Noncommutative
Geometry'', Nucl. Phys. B \ (Proc. Suppl) \textbf{18} (1990) 29; ''The
Metric Aspect on Noncommutative Geometry'', in: Proceedings of the 1991
Cargese summer school, Eds. J. Frolich et all, Plenum (1992); W. Kalau and
M. Walze, ''Gravity, Noncommutative Geometry and the Wodzicki Residue'', J.
Geom. Phys. \textbf{16} (1995) 327--344; G.\ Landi, Nguyen A. V. and K. C.
Wali, ''Gravity and Electromagnetism in Noncommutative Geometry, Phys. Lett.
\textbf{B326} (1994) 45--50; G. Landi and C.\ Rovelli, ''General Relativity
in Terms of Dirac Eigenvalues'', Phys. Rev. Lett. \textbf{78} (1997)
3051-3054; I. Vancea, ''Observables of the Euclidean Supergravity'', Phys.
Rev. Lett. \textbf{79} (1997) 3121--3124; Erratum--ibid, 80 (1998) 1355; A.
H. Chamseddine, ''Deforming Einstein's gravity'', Phys. Lett. \textbf{B 504}
(2001) 33--37; ''Complex gravity in noncommutative spaces'', Commun. Math.
Phys. \textbf{218} (2001) 283--292; J. W. Moffat, ''Noncommutative quantum
gravity'', Phys. Lett. \textbf{B 491} (2000) 345--352; ''Perturbative
noncommutative quantum gravity'', Phys. Lett. \textbf{B 493} (2000)
142--148; V. P. Nair, ''Gravitational fields on noncommutative space'', \
hep--th/0112114; S. Cacciatorii, A. H. Chamseddine, D. Klemm, L. Martucci,
W. A. Sabra, and D. Zanon, ''Noncommutative gravity in two dimensions''
Class.Quant.Grav. \textbf{19} (2002) 4029-4042; R. Jackiw and S. Y. Pi,
''Covariant coordinate transformations on noncommutative space'', Phys. Rev.
Lett. 88 (2002) 111603; S. Cacciatori, D. Klemm, L. Martucci and D. Zanon
''Noncommutative Einstein-AdS Gravity in three Dimensions'', Phys. Lett.
\textbf{B 536} (2002) 101-106.

\bibitem{nc} A. Connes, '' Noncommutative Geometry'' Academic Press (1994);\
J. Madore, ''An Introduction to Noncommutative Geometry and its Physical
Applications'', LMS lecture note Series 257, 2nd ed., Cambridge University
Press (1999);\ J. M. Gracia--Bondia, J. C. Varilly, and H. Figueroa,
''Elements of Noncommutative Geometry'', Birkh\"{a}user Advanced Texts:
Basler Lehrb\"{u}cher. [Birkh\"{a}user Advanced Texts: Basel Textbooks] Birkh%
\"{a}user Boston, Inc., Boston, MA (2001).

\bibitem{connes1} A. Connes, M.\ R. Douglas and A. Schwarz, ''Noncommutative
Geometry and Matrix Theory: Compactification on Tory'', JHEP \textbf{02}
(1998) 003.

\bibitem{corley} S.\ Corley, D. Lowe and S. Rangoolam, ''Einstein--Hilbert
action on the brane for the bulk graviton'', JHEP 0107 (2001) 030.

\bibitem{cjs} E. Cremmer, B. Julia and J. Scherk, ''Supergravity theory in
eleven--dimensions'', Phys. Lett. \textbf{B 76} (1978)\ 409-412.

\bibitem{dv} H. Dehnen and S. Vacaru, ''Nonlinear connections and nearly
autoparallel maps in general relativity'', gr--qc/0009038.

\bibitem{dn} M. R. Douglas and N. A. Nekrasov, ''Noncommutative Field
Theory'', \ Rev. Mod. Phys. \textbf{73} (2001) 977-1029; \ A. Konechny and
A. Schwarz, ''Introduction to M (matrix) Theory and Noncommutative
Geometry'', Phys. Rept. \textbf{360} (2002) 353-465.

\bibitem{haw} S. W. Hawking and G. F. R. Ellis, \textit{The Large Scale
Structure of Space--Time} (Cambridge University Press, 1973).

\bibitem{deligne} P. Deligne, P. Etingof, D. S. Freed et all (eds.),
''Quantum Fields and Strings: A Course for Mathematicians'', Vols 1 and 2,
Institute for Advanced Study, American Mathematical Society (1994);\ J.
Polchinski, ''String Theory'', Vols 1 and 2, Cambridge University Press
(1998).

\bibitem{qg} V. G. Drinfeld, ''Quantum groups'', in: Proceedings of the
International Congress of Mathematicians, Vol. 1, 2; Berkeley, California.,
1986, Amer. Math. Soc., Providence, RI, 1987, pp. 798--820;\ M. Jimbo, ''A
q--analogue of $U(gl(N+1)),$ Hecke algebra and Yang--Baxter equation'', Let.
Math. Phys. \textbf{11} (1986) 247--252;\ S. L. Worinowicz, ''Twisted $SU(2)$%
--group: an example of a noncommutative differential calculus'', Publ. R. I.
M. S. (Kyoto Univ.) \textbf{23} (1987) 117--181;\ Yu. I. Manin, ''Quantum
Groups and Noncommutative Geometry'', Universite de Montreal, Centre de
Recherches Mathematiques, Montreal, QC (1988).\

\bibitem{ll} L. Landau and E. M. Lifshits, \textit{The Classical Theory of
Fields}, vol. 2 (Nauka, Moscow, 1988) [in russian]; S. Weinberg, \textit{\
Gravitation and Cosmology}, (John Wiley and Sons, 1972).

\bibitem{landi} G.\ Landi, ''An Introduction to Noncommutative Spaces and
their Geometries'', Springer--Verlag (1997).\

\bibitem{majid} S. Majid, ''Foundations of Guantum Group Theory'', Cambridge
University Press, Cambridge (1995); ''A Quantum Group Primer'', L. M.\ S.
Lect. Notes. Series. \textbf{292} (2002); ''Meaning of Noncommutative
Geometry and the Planck--Scale Quantum Group'', Springer Lecture Notes in
Physics, \textbf{541} (2000), 227--276; ''Noncommutative Geometry and
Quantum Groups'', Phys. Trans. Roy. Soc. \textbf{A 358} (2000) 89--109.

\bibitem{fradkin} E. S. Frandkin and A. A. Tseytlin, ''Nonlinear
Eloctrodynamics from Quantized Strings'', Phys. Lett. B 163 \ (1985) 123-- \
\ ; C. G. Callan, C. Lovelace, C. R. Nappi and S.\ Yost, ''String Loop
Corrections to Beta Functions'', Nucl. Phys. B288 (1987) 525--550\ ; A.
Aboulesaood, C. G.\ Callan, C. R. Nappi and S.\ A. Yost, ''Open Strings in
Background Gauge Fields'', Nucl. Phys. B280 (1987) 599--624.

\bibitem{friedan} D. H. Friedan, ''Non--linear models in 2+$\varepsilon $
dimensions'', Ann. Physics 1963 (1985) 318--419; L. Alvarez--Gaume and D. Z.
Freedman, ''Geometrical structure and ultraviolet finitness in the
supersymmetric sigma model'', Comm. Math. Phys. 80 (1981) 443--451; T. L.
Curtright, L. Mezincescu and C. K. Zachos, ''Geometrostasis and torsion in
covariant superstrings'', Phys. Lett. B 161 (1985) 79--84; C. G. Callan, D.\
Friedan, E. J. Martinec and M. J. Perry, ''Strings in background fields'',
Nucl. Phys. B 262 (1985) 593--609\ .

\bibitem{gar} M. R. Garousi, ''Superstring scattering from D--branes bound
states'', JHEP 9812 (1998) 008.

\bibitem{kaw} A. Kawaguchi, ''On the theory of non--linear connections, I,
II'', Tensor, N. S. \textbf{2 }(1952) 123--142; (1956) 165--199; ''Beziehung
zwischen einer metrischen linearen Ubertragung und iener nicht--metrischen
in einem allgemeinen metrischen Raume'', Akad. Wetensch. Amsterdam, Proc.
\textbf{40} (1937) 596--601.

\bibitem{kern} J.\ Kern, ''Lagrange Geometry'', Arch. Math. 25 (1974)
438--443.

\bibitem{kir} E.\ Kiritsis, ''Introduction to Superstring Theory'', Leuven
Notes in Mathematical and Theoretical Physics. Series B: Theoretical
Particle Physics, 9. Leuven University Press, Leuven (1998); J. Scherk and
J. Schwarz, ''How to get masses from extra dimensions'', \ Nucl. Phys.
\textbf{B153} (1979) 61--88; J. Maharana and J.\ Schwarz, ''Noncompact
symmetries in string theory'', Nucl. Phys. \textbf{B390} (1993) 3--32.

\bibitem{lidsey} J. E. Lidsey, D. Wands and E. J. Copeland, ''Superstring
Cosmology'', Phys. Rept. \textbf{337} (2000) 343-492.

\bibitem{liu} H. Liu, ''*--trek II: *n operations, open Wilson lines and the
Seiberg--Witten map'', Nucl. Phys. \textbf{B614} (2001) 305-329; H. Liu and
J. Michelson, ''Ramond--Ramond couplings of noncommutative D--branes'',
Phys.Lett. \textbf{B518} (2001) 143-152; S. Mukhi and N. V. Suryanarayana,
''Gauge--invariant couplings of noncommutative branes to Ramond--Ramond
backgrounds'', JHEP \textbf{0105} (2001) 023; Y. Okawa and H.\ Ooguri, ''An
exact solution to Seiberg--Witten equation of noncommutative gauge theory'',
Phys.Rev. \textbf{D64} (2001) 046009; N. Nekrasov, ''Lectures on D--branes
and noncommutative gauge theories'', hep-th/0203109.

\bibitem{miron} R. Miron, ''The Geometry of Higher--Order Lagrange Spaces,
Application to Mechanics and Physics'', FTPH no. 82, Kluwer Academic
Publishers, Dordrecht, Boston, London (1997); R. Miron, ''The Geometry of
Higher--Order Finsler Spaces'',\ Hadronic Press, Palm Harbor, USA (1998);\
R. Miron, D. Hrimiuc, H. Shimada and V. S. Sabau, ''The Geometry of Hamilton
and Lagrange Spaces'',\ Kluwer Academic Publishers, Dordrecht, Boston,
London (2000).

\bibitem{ma} R. Miron and M. Anastasiei, ''The Geometry of Lagrange Spaces:
Theory and Applications'', Dordrecht, Kluwer (1994).

\bibitem{mtw} C. W. Misner, K. S. Thorne and J. A. Wheeler, ''Gravitation'',
\ Freeman (1973).

\bibitem{nahm} W. Nahm, ''Supersymmetries and their representations'', Nucl.
Phys. \textbf{B135} (1978) 149--166.

\bibitem{polch} J. Polchinski, ''String duality: A colloquium'', Rev. Mod.
Phys. 68 (1996) \ 1245--1258; ''TASI lectures on D--branes'',
hep--th/9611050.

\bibitem{risi} G. De Risi, G. Grignani and M. Orselli, "Space/Time
Noncommutativity in String Theories without Background Electric
Field", hep--th/0211056.

\bibitem{finsler} H. Rund, ''The Differential Geometry of Finsler Spaces'',
Springer--Verlag, Berlin (1959);\ G. S. Asanov, ''Finsler Geometry,
Relativity and Gauge Theories'', Reidel, Boston (1985);\ M. Matsumoto,
''Foundations of Finsler Geometry and Special Finsler Spaces'', Kaisisha,
Shigaken (1986).

\bibitem{salam} A. Salam and J. Strathdee, ''On Kaluza--Klein theory'', Ann.
Phys. (NY) \textbf{141} (1982) 316--352; R. Percacci and S. Randjbar--Daemi,
''Kaluza--Klein theories on bundles with homogeneous fibers'', J. Math.
Phys. \textbf{24} (1983) 807--814; J. M. Overduin and P. S. Wesson,
''Kaluza--Klein Theory'', Phys. Rep. \textbf{283} (1997) 303--378.

\bibitem{salamsezgin} A. Salam and E. Sezgin, eds., ''Supergravities in
Diverse Dimensions'', vols. I, II, World Scientific (1989).

\bibitem{cosm} Ashoke Sen, ''Magnetic Monopoles, Bogomol'nyi Bound and
SL(2,Z) Invariance in String Theory'', Mod.Phys.Lett. \textbf{A8} (1993)
2023-2036; A. Shapere, S. Trivedi and F. Wilczek, ''Dual dilaton dyons.
Modern Phys. Lett. \textbf{A6} (1991), 2677--2686; J. E. Lidsey, D. Wands
and E. J. Copeland, ''Superstring cosmology'', Phys. Rept. \textbf{337}
(2000) 343-492.

\bibitem{sw} N. Seiberg and E. Witten, ''String theory and noncommutative
geometry'', JHEP, 09 (1999) 032.

\bibitem{taylor} W. Taylor, ''D--brane field theory on compact spaces'',
Phys. Lett. B 394 (1997) 283--287.

\bibitem{vspinors} S. Vacaru, ''Clifford structures and spinors on spaces
with local an\-iso\-tro\-py'', Buletinul Academiei de \c{S}tiin\c{t}e a
Republicii Moldova, Fizica \c{s}i Tehnica [Izvestia Akademii Nauk Respubliki
Moldova, seria fizica i tehnika] \textbf{3} (1995) 53--62;\ ''Spinor
structures and nonlinear connections in vector bundles, generalized Lagrange
and Finsler spaces'', J. Math. Phys. \textbf{37} (1996) 508--523;\ ''Spinors
and Field Interactions in Higher Order Anisotropic Spaces'', JHEP\ \textbf{\
9809} (1998) 011.

\bibitem{vsuper} S. Vacaru, ''Nonlinear Connections in Superbundles and
Locally Anisotropic Supergravity'', gr-qc/9604016.

\bibitem{vstring} S. Vacaru, ''Locally Anisotropic Gravity and Strings'',\
Ann. Phys. (N. Y.), \textbf{256} (1997) 39--61.

\bibitem{vstr2} S. Vacaru, "Superstrings in Higher Order Extensions of
Finsler Superspaces", Nucl. Phys. B \textbf{434} (1997) 590--656.

\bibitem{vankin} S. Vacaru, ''Stochastic Processes and Thermodynamics on
Curved Spaces'', Ann. Phys. (Leipzig), \textbf{9} (2000), Special Issue,
175--176, gr--qc/0001057;\ ''Locally Anisotropic Kinetic Processes and
Thermodynamics in Curved Spaces'', Ann. Phys. (NY)\ \textbf{290} (2001)
83--123; S. Vacaru, ''Stochastic Differential Equations on Spaces with Local
Anisotropy'', Buletinul Academiei de Stiinte a Republicii Moldova, Fizica si
Tehnica [Izvestia Academii Nauk Respubliky Moldova, fizica i tehnika],
\textbf{3} (1996) 13-25;\ ''Stochas\-tic pro\-cesses and dif\-fu\-sion on
spaces with lo\-cal an\-i\-sot\-ropy'', gr-qc/9604014;\ ''Locally
ansiotropic stochastic processes in fiber bundles'', Proceedings of Workshop
on Global Analysis, Differential Geometry and Lie Algebra, Thessaloniki,
(1995), pp. 123--144.

\bibitem{vexsol} S. Vacaru, ''Anholonomic Soliton-Dilaton and Black Hole
Solutiions in General Relativity'', JHEP \textbf{04} (2001) 009;\ S. Vacaru,
D. Singleton, V. Botan and D. Dotenco, ''Locally Anisotropic Wormholes and
Flux Tubes in 5D Gravity'', Phys. Lett. B \textbf{519} (2001) 249-258;\ \ S.
Vacaru and O. Tintareanu-Mircea, ''Anholonomic Frames, Generalized Killing
Equations, and Anisotropic Taub NUT Spinning Spaces'', Nucl. Phys. \textbf{B}
\textbf{626} (2002) 239-264;\ \ S. Vacaru, P. Stavrinos and E. Gaburov,
''Anholonomic Triads and New Classes of (2+1)-Dimensional Black Hole
solutions'', gr-qc/0106068;\ S. Vacaru, P. Stavrinos and Denis Gontsa,
''Anholonomic Frames and Thermodynamic Geometry of 3D Black Holes'',
gr-qc/0106069;

\bibitem{vbel} S. Vacaru, ''Horizons and Geodesics of Black Ellipsoids'',
Int. J. Mod. Phys. D. [to be published], gr-qc/0206014; S. Vacaru,
''Perturbations and Stability of Black Ellipsoids'', Int. J. Mod. Phys. D.
[to be published], gr-qc/0206016.

\bibitem{vmon1} S. Vacaru, ''Interactions, Strings and Isotopies in Higher
Order An\-isot\-ropic Superspaces'', Hadronic Press, FL, USA (1998),
math-ph/0112056.

\bibitem{vnonc} S. Vacaru, ''Gauge and Einstein Gravity from Non-Abelian
Gauge Models on Noncommutative Spaces'', Phys. Lett. B \textbf{498} (2001)
74-82;\ S. I. Vacaru, I. Chiosa and N. Vicol, ''Locally Anisotropic
Supergravity and Gauge Gravity on Noncommutative Spaces'', in: NATO Advanced
Research Workshop Proceedings ''Noncommutative Structures in Mathematics and
Physics'', eds S. Duplij and J. Wess, September 23-27, Kyiv, Ukraine, Kluwer
Academic Publishers (2001), 229 - 243.

\bibitem{vmethod} S. Vacaru, ''Locally Anisotropic Black Holes in Einstein
Gravity, gr--qc/0001020;\ ''A New Method of Constructing Black Hole
Solutions in Einstein and 5D Dimension Gravity'', hep-th/0110250; ''Black
Tori in Einstein and 5D Gravity'', hep-th/0110284.

\bibitem{vncf} S.\ Vacaru, ''Noncommutative Finsler Geometry, Gauge Fields
and Gravity'', math--ph/0205023.

\bibitem{vgauge} S. Vacaru and Yu. Goncharenko, ''Yang--Mills fields and
gauge gravity on generalized Lagrange and Finsler spaces'', Int. J. Theor.
Phys. \textbf{34} (1995) 1955--1978;\ S. Vacaru, ''Gauge Gravity and
Conservation Laws in Higher Order Anisotropic Spaces'', hep-th/9810229;\
''Yang-Mills Fields on Spaces with Local Anisotropy'', Buletinul Academiei
de Stiinte a Republicii Moldova, Fizica si Tehnica [Izvestia Academii Nauk
Respubliky Moldova, fizica i tehnika], \textbf{3} (1996) 26-31;\ ''Gauge
Like Treatment of Generalized Lagrange and Finsler Gravity'', Buletinul
Academiei de Stiinte a Republicii Moldova, Fizica si Tehnica [Izvestia
Academii Nauk Respubliky Moldova, fizica i tehnika] \textbf{3} (1996) 31-34.

\bibitem{vsolsp} S. Vacaru and F. C. Popa, ''Dirac Spinor Waves and Solitons
in Anisotropic Taub-NUT Spaces'', Class. Quant. Gravity, \textbf{18} (2001)
4921-4938; Vacaru and D. Singleton, ''Warped, Anisotropic Wormhole / Soliton
Configurations in Vacuum 5D Gravity'', Class. Quant. Gravity, \textbf{19}
(2002) 2793-2811.

\bibitem{vsingl} S. Vacaru and D. Singleton, ''Warped Solitonic Deformations
and Propagation of Black Holes in 5D Vacuum Gravity'', Class. Quant.
Gravity, \textbf{19} (2002) 3583-3602.

\bibitem{vsingl1} S. Vacaru and D. Singleton, ''Ellipsoidal, Cylindrical,
Bipolar and Toroidal Wormholes in 5D Gravity'', J. Math. Phys. \textbf{43}
(2002) 2486-2504.\

\bibitem{vmon2} S. Vacaru and P. Stavrinos, ''Spinors and Space--Time
Anisotropy'', Athens Universtity Press, Greece (2002), gr-qc/0112028.

\bibitem{wit} B. de Wit, J. Hoppe and H. Nicolai, ''On the quantum mechanics
of supermembranes'', Nucl. Phys. \textbf{B305} (1988) 545--581; T. W. Banks,
S. H. Fischler, S. H. Snenker, and L. Susskind, ''M theory as a matrix
model: A conjecture'', Phys. Rev. \textbf{D55} (1997) 5112--5128; W. Tailor,
''M(atrix) theory: Matrix quantum mechanics as a fundamental theory'', Rev.
Mod. Phys. \textbf{73} (2001) 419-462.

\bibitem{strncg} E. Witten, ''Noncommutative Geometry and String Field
Theory'', Nucl. Phys. \textbf{B268} (1986) 253--294.

\bibitem{kad} V. E. Zakharov and A. B.\ Shabat, ''A plan for integrating the
nonlinear equations of mathematical physics by the method of the inverse
scattering problem. I.'', Funk. Analiz i Ego Prilojenia [in Russian] \textbf{%
8} (1974) 43--53; B. Harrison, Phys.\ ''B\"{a}cklund transformation for the
Ernst equation of general relativity'', \ Phys. Rev. Lett., \textbf{41}
(1978) 1197--1200.

\end{thebibliography}
\end{document}